\documentclass[a4paper,12pt]{report}
\usepackage{amssymb}
\usepackage{amsmath}
\usepackage{epsfig}
\usepackage{graphics}
\usepackage{fancyhdr}
\renewcommand{\headrulewidth}{0.5pt}
\renewcommand{\footrulewidth}{0pt}
\fancypagestyle{plain}{%
\fancyhead{} 
\renewcommand{\headrulewidth}{0pt} 
}

\usepackage{hyperref}
\usepackage{nomencl}
\renewcommand{\nomname}{List of Abbreviations}
\makenomenclature

\makeatletter

\title{Search for $\eta$-mesic $^{4}\mbox{He}$\\ with the WASA-at-COSY detector}
\author{Wojciech Krzemie\'n}
\date{10.11.2011}

\renewcommand{\maketitle}{\begin{titlepage}

\begin{center}\small
	Jagiellonian University\\
	Institute of Physics\\
\end{center}

\vspace{2cm}
\noindent

\begin{center}
	\LARGE \textsc{\@title}
\end{center}

\vspace{0.5cm}

\begin{center}
	\@author
\end{center}

\vspace{7cm}

\noindent A doctoral dissertation prepared at the Institute of Physics of the Ja\-gie\-llo\-nian University and at the Institute of
Nuclear Physics of the Research Centre Juelich, submitted to the Faculty of Physics, Astronomy and Applied Computer Science at the Jagiellonian University, conferred by prof. Jerzy Smyrski.

\vspace{2cm}

\begin{center}
   \@date
\end{center}

\end{titlepage}%
}
\makeatother

\newcommand{\scaleFactor}{0.25}

\begin{document}

\maketitle

\newpage \vspace*{13cm}
\pagestyle{empty}
\begin{flushright}
Dla mojej c\'{o}rki Natalki.
\end{flushright}
\vspace*{3cm} 

\begin{flushright}
{\footnotesize
{\em Alice laughted. ``There is no use trying'' she said: ``one can't believe impossible things.''\\
``I daresay you haven't had much practice,'' said the Queen. ``When I was your age,\\ I always did it for half-an-hour a day. Why, sometimes I've believed as many as six impossible things before breakfast.'' }\\
(L. Carroll  Alice's Adventures in Wonderland)
}
\end{flushright}
\newpage
\pagestyle{fancy}

\begin{abstract}
We performed a search for the ${^4\mbox{He}}-\eta$ bound state via  exclusive measurement
of the excitation function for the $dd \rightarrow {^3\mbox{He}} p \pi^-$ reaction,
where the outgoing $p-\pi^-$ pair originates from the conversion of the $\eta$ meson
on a neutron inside the ${^4\mbox{He}}$ nucleus.
The measurements were performed at the Cooler Synchrotron COSY-Juelich with the WASA-at-COSY detection system.
The internal deuteron beam of COSY was scattered on a pellet-type deuteron target.
The data were taken during a slow acceleration of the beam
from 2.185\,GeV/c to 2.400\,GeV/c crossing the kinematical threshold
for the $\eta$ production in the $dd \rightarrow {^4\mbox{He}}\,\eta$ reaction at 2.336~GeV/c.
The corresponding excess energy in the ${^4\mbox{He}}-\eta$ system varied from -51.4~MeV to 22~MeV.
Events corresponding to decays of the $\eta$-mesic ${^4\mbox{He}}$ were selected using cuts
on the ${^3\mbox{He}}$ momentum, $p$ and $\pi^-$ kinetic energies and the relative $p-\pi^-$ angle
in the center of mass system. The range of the applied cuts was inferred from simulations
of decay of the ${^4\mbox{He}}-\eta$ bound state proceeding via excitation the $N^*$ resonance.
The integrated luminosity in the experiment was determined using the $dd \rightarrow {^3\mbox{He}} n$ reaction
and the relative normalization of the points of the $dd \rightarrow {^3\mbox{He}} p \pi^-$ excitation
function was based on the quasi-elastic proton-proton scattering.
No signal of the ${^4\mbox{He}}-\eta$ bound state was observed in the excitation function.
An upper limit for the cross-section for the bound state formation and decay in the process
$dd \rightarrow ({{^4\mbox{He}}-\eta})_{bound} \rightarrow {^3\mbox{He}} p \pi^{-}$,
determined for the bound state width of
10, 20 and 30\,MeV equals to  28, 32 and 41\,nb, respectively.
The measured angular and momentum distributions of the reaction products are close
to those simulated under the assumption of uniform phase-space distribution.

\end{abstract}

\pagenumbering{roman}

\tableofcontents

\pagenumbering{arabic}
\setcounter{page}{1}

\chapter{Introduction}
\label{ch:intro}
Observation of bound states of hadrons and atomic nuclei such as hypernuclei or pionic atoms
opened new fields of research in nuclear physics and provided very fruitful results in the studies
of the hadron-nucleon interaction in a many-body environment.
Hypernuclei contain at least one hyperon in addition to nucleons.
Since their discovery by Danysz and Pniewski in 1952~\cite{danysz},
the study of their properties led to a considerable progress in understanding the $\Lambda N$
and $\Sigma N$ interactions.
In turns, pionic atoms are formed by a negatively charged pion  trapped in the Coulomb
field of the atomic nucleus.
Observation of shifts and broadening of the energy levels in pionic atoms
induced by the strong interaction allows for precise studies of this interaction.
One of the most interesting results originating from investigations of deeply bound pionic atoms
is the evidence for partial restoration of chiral symmetry in the nuclear medium~\cite{suzuki}.

It is also conceivable that neutral mesons such as $\eta, \bar{K},\omega,\eta'$ can form bound states
with atomic nuclei.
In this case the binding is exclusively due to the strong interaction and the bound state - {\em mesic nucleus}
- can be considered as a meson moving in the mean field of the nucleons in the nucleus.
Due to the strong attractive $\eta$-nucleon interaction, the $\eta$-mesic nuclei are ones of the most promising candidates for such states.
One expects that properties of $\eta$-mesic nuclei are strongly influenced by the excitation
of the $N^*$(1535) resonance which dominates the low energy $\eta-N$ interaction.
The $N^*$ decays with roughly equal probabilities in $\eta N$ and $\pi N$ channel.
In the second case it leads to the decay of the $\eta$-nucleus bound state.
The discovery of the $\eta$-mesic nuclei would be interesting on its own but it would be also
valuable for investigations of the $\eta-N$ interaction and for the study of the in-medium properties
of the $N^*$ resonance~\cite{jido} and of the $\eta$ meson~\cite{osetNP710}.
It could also help to determine the flavor singlet
component of the $\eta$ wave function~\cite{basssymposium}.

The existence of $\eta$-mesic nuclei was postulated in 1986 by Haider and Liu \cite{liu2},
and since then a search for such states  was conducted in many experiments.
However, up to now no firm experimental evidence for $\eta$-mesic nuclei was found.
A possible reason for this can be a high background observed in the experiments
and large predicted widths of the $\eta$-nucleus bound states
ranging from ~7 to 40\,MeV~\cite{osetPLB550,liu3,haider}.
One expects that the width should be smaller for light nuclei, where the absorption of the $\eta$ mesons
is weaker due to smaller number of nucleons.
A very strong final state interaction (FSI) \nomenclature{FSI}{Final State Interaction} observed
in the $dd \rightarrow {^4\mbox{He}} \eta$ reaction close to kinematical threshold
and interpreted as possible indication of ${^4\mbox{He}}-\eta$ bound state~\cite{Willis97}
suggests, that ${^4\mbox{He}}-\eta$ system is a good candidate for experimental study of possible binding.

Taking into account the above arguments, we proposed to perform a search for $\eta$-mesic $^{4}\mbox{He}$
by measuring the excitation function for the $dd \rightarrow {^3\mbox{He}} p\pi^-$  reaction
in the vicinity of the $\eta$ production threshold.
The outgoing particles in the proposed reaction correspond to the decay of $\eta$-mesic ${^4\mbox{He}}$
proceeding via  $\eta$ absorption on one of neutrons in  ${^4\mbox{He}}$ leading to excitation
of the $N^*$ resonance which subsequently decays in the $p-\pi^-$ pair.
The remaining three nucleons bind forming the ${^3\mbox{He}}$ nucleus which plays a role of a spectator
moving with relatively low momenta in the overall center of mass (CM) \nomenclature{CM}{Center-of-Mass frame} frame
corresponding to the Fermi momentum distribution in ${^4\mbox{He}}$.
If the $\eta$-mesic ${^4\mbox{He}}$ is produced as an intermediate state in the process
$dd \rightarrow ({^4\mbox{He}}-\eta)_{bound}\rightarrow {^3\mbox{He}} p \pi^{-}$
then we expect to observe a resonance-like structure in the corresponding excitation function at energy
below the ${^4\mbox{He}}-\eta$ threshold.

We proposed to perform a search for the $\eta$-mesic ${^4\mbox{He}}$
using the deuteron beam of the COSY accelerator scattered
on internal deuteron target.
The WASA-at-COSY detector was chosen for registration of the reaction products.
Our proposal was presented at the 34th meeting of the Program Advisory Committee (PAC) \nomenclature{PAC}{Program Advisory Committee}
of the COSY accelerator in November 2007 \cite{prop186}.
It received a positive opinion with the recommendation of nine days of deuteron beam time.
The measurements were performed in June 2008.
During the experimental run the momentum of the deuteron beam was varied continuously within each acceleration cycle
from  2.185~GeV/c to 2.400~GeV/c, crossing the kinematical threshold for the $\eta$ production in the $dd \rightarrow {^4\mbox{He}}\,\eta$ reaction at 2.336~GeV/c.
This range of the beam momenta corresponds to an interval of the excess energy in the $^4\mbox{He}-\eta$
system from -51.4~MeV to 22~MeV.
Unfortunately, out of nine days of the allocated beam time, due to failures in operation
of the deuteron target
and of the COSY beam we could collect the data only for one day.
Additionally, during this period, due to the failure of the cooling system
of the WASA-at-COSY solenoid, the measurement was performed without the magnetic field,
making impossible the momentum analysis of charged particles.
In spite of these difficulties, the  measurements delivered valuable data
for the search for the ${^4\mbox{He}}-\eta$ bound state.

The present work is devoted to the analysis of these data.
In Chapter~2, theoretical and experimental background of the search for the $\eta$-mesic nuclei is presented.
Chapter~3 describes the COSY accelerator and the WASA-at-COSY detection system.
A basic concept of the experiment and simulations, which were performed in order to validate
this concept, are presented in Chapter~4.
Chapter~5 is devoted to the data analysis. It describes the detector calibration,
the reconstruction of $dd \rightarrow {^3\mbox{He}} p \pi^-$ events and the luminosity determination.
Chapter~6 presents  the final results of the data analysis including the excitation function
 for the $dd \rightarrow {^3\mbox{He}} p\pi^-$ process and an upper limit for the cross-section
characterizing the production and decay of the $\eta$-mesic ${^4\mbox{He}}$.
Also, analysis of experimental background, based on the reconstructed momentum and angular
distributions of the final state particles, is included.
Chapter~7 summarizes the thesis and provides the outlook.

\chapter{Experimental and theoretical background}

In the first section of this chapter we gathered basic notions and formulas from the scattering theory which will be used in the further parts
of the dissertation. They concern mainly the description of bound and virtual states and their influence
on the scattering process.
For a systematic discussion of these topics the interested reader is referred to the cited literature.
In the second section, the basic information about the $\eta$ meson is presented.
The next section is devoted to the theoretical studies of the $\eta$-mesic nuclei.
Also, the physical motivation of the research is presented.
The last section contains a review of the experimental research of $\eta$-mesic nuclei in different experiments.

\section{Bound and virtual states in the scattering theory}

\subsection{Scattering matrix}
In the scattering theory the initial state of a system before collision
and the final state after collision are connected by the scattering operator $\hat{S}$:
\begin{equation}
 \mid \Psi_{out}\rangle=\hat{S}\mid \Psi_{in}\rangle.
\end{equation}
The $\hat{S}$ operator satisfies special conditions like the unitarity, the time reversal symmetry, the analyticity and,
in the case of relativistic theory, the Lorentz invariance.
The unitarity expresses the conservation of probability and requires
that the norm of a state before and after collision has to be preserved.
The time reversal invariance implies equality of probabilities for a direct
and inverse transition~\cite{landau2}. 
The analyticity properties are the consequences of the underlying locality of the interaction~\cite{white}.

The initial and final state can be expanded in a basis of orthonormal states~\cite{landau2}:
\begin{equation}
\mid\Psi_{out}\rangle =\sum_{f,i} \mid f\rangle \langle f\mid \hat{S}\mid i\rangle \langle i \mid \Psi_{in}\rangle .
\end{equation}
The  elements $ S_{fi}=\langle f\mid \hat{S}\mid i\rangle $ define the scattering matrix.
The squared module  $ |S_{fi}|^2$ is a probability of transition from the initial state $\mid i\rangle$
to the final state $\mid f\rangle$.

In the case of scattering of spinless particles the scattering matrix is diagonal in the basis of the
angular momentum states and it depends only on the absolute value of the  relative momentum $p$
of the colliding particles:
\begin{equation}
S_{ll'}(p) = S_l(p) \delta(l-l').
\end{equation}
Imposing the unitarity condition ($|S_l(p)|=1$) the scattering matrix can be expressed
in the following way:
\begin{equation}
S_l(p)=e^{2i\delta_{l}(p)},
\end{equation}
where $\delta_{l}(p)$ is the phase shift being a real number.
The corresponding scattering amplitude equals:
\begin{equation}
f_l=\frac{S_l-1}{2ik},
\end{equation}
where $k$ is the length of the wave vector equal to $p/\hbar$.
In the following, we set $\hbar=1$ and we use the relation $k=p$.

The scattering amplitude expressed in terms of $\delta_l$ reads:
\begin{equation}
f_l=\frac{e^{2i\delta_l}-1}{2ip}=\frac{1}{p\cot\delta_l-ip}.
\end{equation}

\subsection{Scattering length}

At low momenta the scattering proceeds mainly in the s-wave ($l=0$) and, therefore,
it is sufficient to take into account the $f_{l=0}$ amplitude:
\begin{equation}
f_{l=0}=\frac{1}{p\cot\delta_0-ip}.
\end{equation}
The phase shift $\delta_0$ can be approximated by the effective range expansion:
\begin{equation}
p\cot\delta_0=\frac{-1}{a}+\frac{r_0}{2}p^2,
\end{equation}
where $a$ is the scattering length and $r_0$ is the effective range.
For very small $p$:
\begin{equation}
p\cot\delta_0=\frac{-1}{a}.
\end{equation}
The above equation can be considered as a definition of the scattering length,
however, one should keep in mind, that also a different convention with
opposite sign of the scattering length is sometimes used.
With this definition, the scattering amplitude at low momenta can be written as:
\begin{equation}
f_{l=0}=\frac{1}{\frac{-1}{a}-ip}=\frac{a}{-1-iap}.
\label{eq:s_wave_amplitude}
\end{equation}

The scattering length is a quantity that describes the strength of the interaction potential
in the low energy region.
For repulsive potential the scattering length is larger than zero.
For attractive interaction, positive scattering length corresponds to a bound state
and negative one to lack of binding.
A large negative scattering length can be connected with the so called virtual state which is discussed
in the next subsection.

If besides the elastic scattering, there are also inelastic channels open,
then the scattering length has to  be a complex number:
\begin{equation}
a=a_{R} +i \cdot a_{I}.
\end{equation}
The real part of the scattering length can be interpreted as a measure of the elastic scattering of particles,
whereas the imaginary part corresponds to the losses in the inelastic channels.
Because of the unitarity, the imaginary part fulfils the condition:
\begin{equation}
a_{I} \ge 0.
\end{equation}
In order to have a bound state there is a commonly quoted necessary condition \cite{liu3}:
\begin{equation}
|a_R|>a_I.
\end{equation}

For a complex scattering length the $f_{l=0}$ amplitude has the form:
\begin{equation}
f_{l=0}=\frac{a_R+ia_I}{-1+a_Ip-ia_Rp}.
\end{equation}
The elastic cross-section is proportional to the squared scattering amplitude:
\begin{equation}
|f_{l=0}|^2=\frac{|a|^2}{1-2a_Ip+|a|^2p^2}.
\label{eq:squared_amplitude}
\end{equation}
From this formula it is clear that on the basis of a scattering experiment one can determine
the absolute value of the real part of the scattering length but not its sign.
Therefore it is not possible to answer the question if a bound state exists in the system
of the colliding particles.
The scattering length can be also determined on the basis of the FSI
between the produced particles.
According to Watson and Migdal~\cite{watson,migdal}, the energy dependence of the reaction cross-section
with strongly interacting particles in the final state can be approximated by the squared amplitude
for the final state as given by Eq.~\ref{eq:squared_amplitude}.

\subsection{Bound states, virtual states and resonances}

One of the most powerful techniques in the scattering theory 
is the analysis of properties of the scattering amplitude in the complex momentum or energy plane~\cite{taylor}.
In this approach we allow momentum to become a complex variable $p$
and  we treat the s-matrix elements $S_l(p)$ as analytic functions of complex incident momentum.
The analytical properties of the s-matrix for given reaction depend on the properties of the potential
e.g. on its asymptotic behavior in the limit $r \rightarrow \infty$.
For example, if the potential falls exponentially at large $r$ and it is an analytic function of $r$ in $\Re {r} >0$
then the region of the analyticity of the s-matrix can be extended to whole complex plane
except for finite number of singular points ~\cite{taylor}. 
Alternatively, we consider the s-matrix elements as a function of the complex energy
$E=\frac{p^2}{2 m}$.
However, the mapping from $p$ to $E$ is a two-to-one relation.
For a one-to-one relation the complex energy plane has to be extended to two-sheeted Riemann surface~\cite{taylor}.

\begin{figure}[htb]
\begin{center}
      \scalebox{0.5}
         {
         \includegraphics{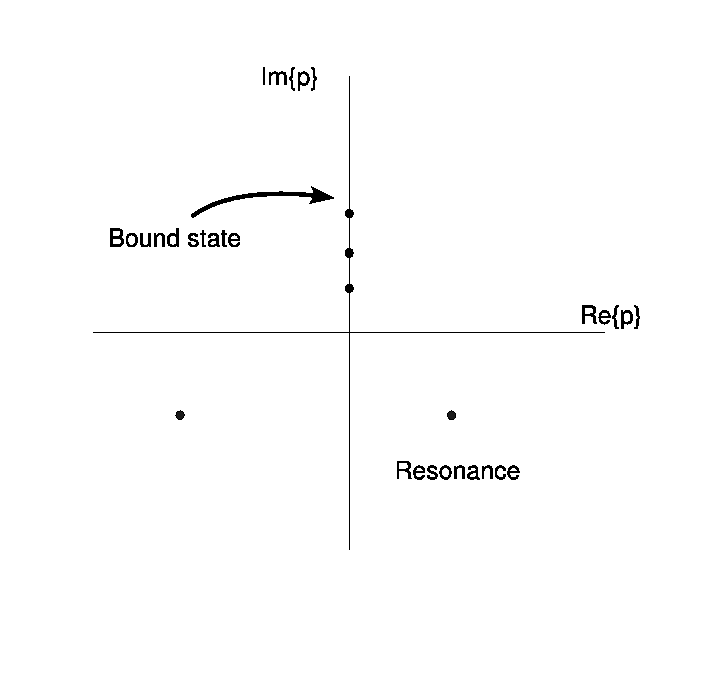}
	
  	 }
\caption[Positions of the poles corresponding to resonances and bound states]{\label{resonance_p_plane_p}
Positions of poles of the scattering matrix
corresponding to bound states and resonances.}
\end{center}
\end{figure}

If the interaction is described by a real potential (no absorption channel) then the poles can only lay
either on the imaginary $\Im p$ axis or on the lower open half plane of the complex momenta~\cite{Cassing}.
Existence of a pole of the scattering matrix in the complex momentum plane can be observed in experiment as a resonance, a bound state or a virtual state. The character of this physical phenomenon is determined by the position of the pole in the complex plane
(see Fig.~\ref{resonance_p_plane_p}).
A pole $p$ which lays on the upper part of the $\Im {p}$ axis ($\Im {p} >0$) corresponds to a bound state which is the proper eigenvalue of the Hamiltonian with energy defined as $E=\frac{p^2}{2m}=-\frac{\alpha^2}{2m}$ where $p=i \alpha$.
If the pole $p$ lays on the lower half of the complex plane ($\Im {p} <0$) then it can correspond to a resonance state.
However, the correspondence between resonances and the poles in $\Im p <0$ is less exact than between  bound states and poles in $\Im {p}>0$. There can exist poles located far from the real axis that do not lead to observable resonance effects~\cite{taylor}. 
The resonance state lays on the second Riemann sheet in the $E$ plane and is not the proper eigenfunction of the Hamiltonian with the energy defined as $E=\frac{p^2}{2m}=E_{R}-\frac{i\Gamma}{2}$.
The resonance is represented by a non-normalizable state.

The simplest form of the scattering matrix which fulfils the unitarity condition  for real $p>0$
with a pole corresponding to a bound state is:
\begin{equation}
S_{l=0}=\frac{-p-ip_0}{p-ip_0},
\end{equation}
where $p_0$ is real and larger than zero and $p=ip_0$ is the position of the pole
on positive part of the imaginary axis in the complex $p$ plane.
The corresponding scattering amplitude equals:
\begin{equation}
f_{l=0}=\frac{1}{-p_0-ip}.
\end{equation}
Comparing to Eq.~\ref{eq:s_wave_amplitude} the following relation
between the position of the pole and the scattering length can be found:
\begin{equation}
p_0=\frac{1}{a}.
\end{equation}
If the scattering length is much larger than the range of the potential, then one can derive the following
relation between the binding energy and the scattering length:
\begin{equation}
-E_{bound}=\frac{p_0^2}{2m}=\frac{1}{2ma^2},
\end{equation}
where $m$ is the reduced mass.
This means that we can infer the binding energy from scattering experiments at low energies.

A pole of the scattering matrix on the negative imaginary axis of the momentum plane corresponds
to a virtual state. The proximity of this pole results in large scattering cross-sections
at low energies.

An example of an attractive interaction leading to the formation of a bound state is the neutron-proton
interaction in the triplet state and the corresponding bound state is the deuteron.
In turns, the neutron-proton scattering in the singlet state is strongly influenced by a virtual state.

A resonance can be described by the following expression for the scattering matrix
as a function of the complex energy:
\begin{equation}
S_l(E)=\frac{E-E_R-i\Gamma/2}{E-E_R+i\Gamma/2},
\end{equation}
where $E_R$ is the resonance energy and $\Gamma$ is the width of the resonance.
The corresponding scattering amplitude has the form:
\begin{equation}
f_l(E)=\frac{-1}{p}\frac{\Gamma/2}{E-E_R+i\Gamma/2},
\end{equation}
and the resonance cross-section (the Breit-Wigner formula) reads:
\begin{equation}
\sigma_l(E)=\frac{4\pi}{p^2}\frac{(2l+1)(\Gamma/2)^2}{(E-E_R)^2+(\Gamma/2)^2}.
\end{equation}

In the case of several competing decay channels the spin averaged formula for the resonance cross-section
reads as follows:
\begin{equation}
\sigma_l(E)=\frac{2J+1}{(2S_1+1)(2S_2+1)} \cdot \frac{\pi}{p^2} \cdot \frac{\Gamma_{i} \cdot \Gamma_{f}}{(E-E_R)^2+(\Gamma/2)^2},
\end{equation}
where $J$ is the spin of the resonance, $S_1$ and $S_2$ are the spins of colliding particles, $ \Gamma_{i}$ and $ \Gamma_{f}$ are the partial widths of the resonance to decay into the entrance and exit channels, respectively, $\Gamma$ is the total width.

\begin{figure}[htb]
\begin{center}
      \scalebox{0.5}
         {
         \includegraphics{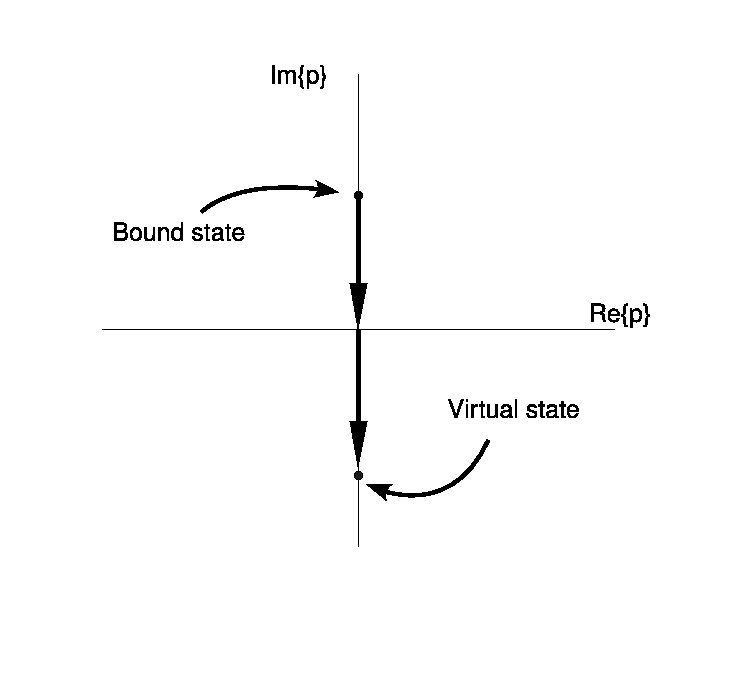}
  	 }
\caption[Transition of the s-wave bound state into the "virtual state"]{\label{bound_virtual_plane_p} Transition of the s-wave bound state into virtual state.}
\end{center}
\end{figure}

The presence of a resonance or a bound state can be observed  as a rapid change in the phase of $ S(p)$  and as a sharp peak in the total cross-section as a function of energy. As we see both, resonance and bound state are related. To understand the  connection between resonances and bound states let consider the limit situation where the pole lays exactly at the boundary between upper and lower half plane and let consider a potential in the form of $\lambda V$, where $\lambda$ is a real coupling parameter. If we change  $\lambda$ to obtain a more attractive potential then the pole moves up and becomes a bound state. If we make the potential less attractive then the pole will move down to the lower half plane ($\Im {p}$) and the resonance state will appear. In case of the s-wave limit ($l=0$) the crossing of the pole to the lower half plane corresponds to a "virtual state" (see Figure \ref{bound_virtual_plane_p}). Therefore, the strength of potential can have a capital importance in differentiating between bound states and resonances.

\begin{figure}[htb]
\begin{center}
      \scalebox{0.5}
         {
         \includegraphics{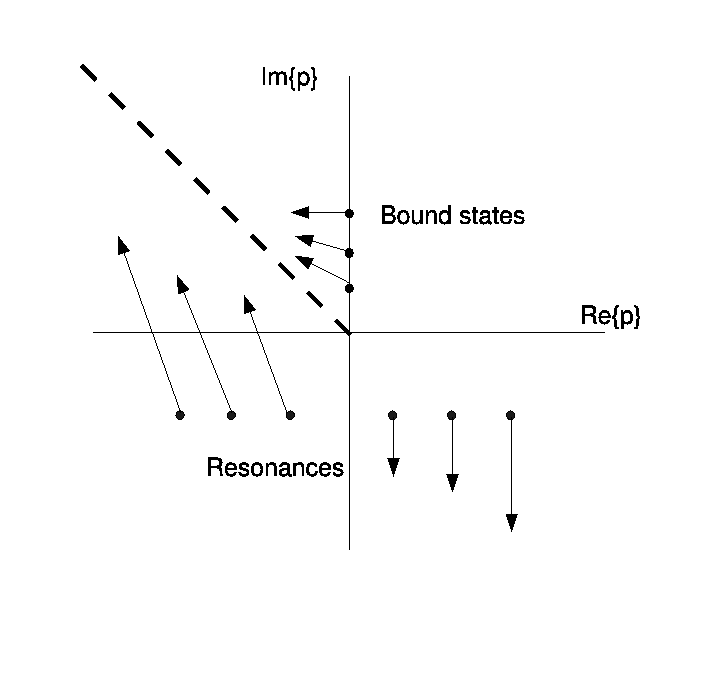}
	
  	 }
\caption[Motion of poles]{\label{poles_motion}Motion of poles of scattering matrix in the complex momentum plane as a function of the imaginary coupling strength $\lambda$ of the interaction potential.
The idea of the picture was taken from ~\cite{sofianos1}.}
\end{center}
\end{figure}

The situation is more complicated if we consider also the inelastic channels. In that case, the potential carries the imaginary part, and thus, the coupling parameter $\lambda$ is a complex number.  As it is shown in ~\cite{Cassing} the positions of the poles in the s-matrix are shifted and the boundary limit between bound states and resonances is no longer the $\Re$ axis. The poles and zeros of the s-matrix are "moving" as a function of the parameter $\lambda$ (see Figure \ref{poles_motion}).

\label{ch:definitions}

\section{Basic properties of the \texorpdfstring{$\eta$}{eta} meson}

\subsection{Quantum numbers and decay channels}
\label{quantum_numbers}
\begin{figure}[!ht]
\begin{center}
\includegraphics[width=8cm]{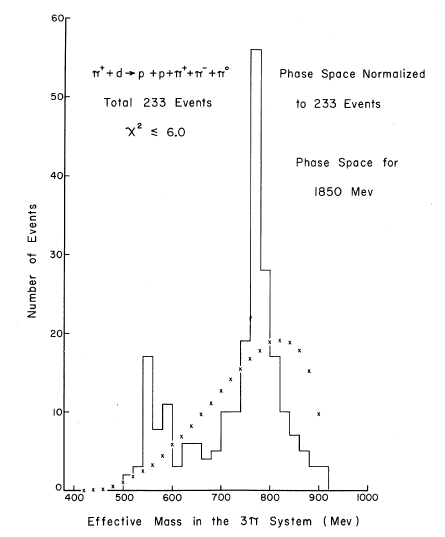}
\caption[Invariant mass spectrum of three pions- discovery of the $\eta$ meson]{Invariant mass spectrum of three pions as measured by Pevsner and collaborators \cite{pevsner}.
Besides $\omega$ meson peak at mass of around 780\,MeV/c$^2$, a signal from $\eta$ is visible
at around 550\,MeV/c$^2$.
The figure was adopted from Ref.~\cite{pevsner}.}
\label{eta_discovery}
\end{center}
\end{figure}

The $\eta$ meson was discovered in 1961 by Pevsner et al. ~\cite{pevsner}
in the Lawrence Radiation Laboratory.
It was observed as a three pion resonance in the reaction $\pi^+ + d \rightarrow p+p+\pi^+ +\pi^- +\pi^0$.
The experiment was performed using a 1.23\,GeV/c pion beam from the Bevatron
scattered in a bubble chamber filled with deuterium.
The trajectories of the charged pions and protons were measured in the chamber
and the neutral pions were identified using the missing mass method.
The $\eta$ meson was observed in the invariant
mass spectrum of the three produced pions as a resonance with a mass of about 550\,MeV/c$^2$
(see Fig.~\ref{eta_discovery}).

During the last 50 years since its discovery, the $\eta$ meson was
intensively studied both experimentally and theoretically.
The basic properties of this meson are now well established.
According to the Particle Data Group \cite{pdg2010} the $\eta$ mass is equal to $547.853\pm0.024$\,MeV/c$^2$.
The full width equals  $1.3\pm0.07$\,keV and is very small compared to
other mesons with higher masses.
The values of the spin $J=0$ and an odd parity $P=-1$ define $\eta$ as a pseudoscalar meson.
$\eta$ is a neutral meson with isospin $I$ equal to zero, even charge parity ($C=+1$) and G-parity ($G=+1$).

The very small total width of the $\eta$ meson results from the fact that
the decay into two pions is forbidden due to the parity and angular momentum conservation.
In turns, the  strong decay into three pions is forbidden by the G-parity conservation.
Dominant are second order electromagnetic two gamma decay and isospin violating three pion decay
(see Table~\ref{TabEtaProp}).
Many $\eta$ decay channels, which are energetically possible but are forbidden by the conservation
of the $C$, $P$ or $CP$ symmetry are used for precise tests of these symmetries
\cite{pdg2010,kupsc2}.

{
\renewcommand{\arraystretch}{1.2}
\begin{table}[!ht]
 \begin{center}
 \begin{tabular}{|l|l|}
	\hline
	{\bf Decay modes} & Fraction ($\Gamma_j /\ \Gamma$) \\
	\hline
	Neutral modes & (71.90 $\pm$ 0.34) \% \\
	\hline
	$ \gamma \gamma $   & (39.31 $\pm$ 0.20) \%  \\
	$ \pi^{0} \pi^{0}\pi^{0}$ & (32.57 $\pm$ 0.23) \% \\
	Other neutral modes & $\sim$ 0.02 \%\\
	\hline
	Charged modes & (28.10 $\pm$ 0.34) \%\\
	\hline
	$\pi^{+} \pi^{-}\pi^{0}$ & (22.74 $\pm$ 0.28) \% \\
	$\pi^{+} \pi^{-}\gamma$ & (4.60 $\pm$ 0.16) \% \\
	Other charged modes& $\sim$ 0.76 \%\\
	\hline
 \end{tabular}
 \caption{Basic decay modes of $\eta$ meson~\cite{pdg2010}.} \label{TabEtaProp}
 \end{center}
\end{table}
}

\subsection{\texorpdfstring{$\eta$}{eta} meson in quark model}
\label{quark_model}
\begin{figure}[!htb]
\begin{center}
\includegraphics[width=0.65\textwidth]{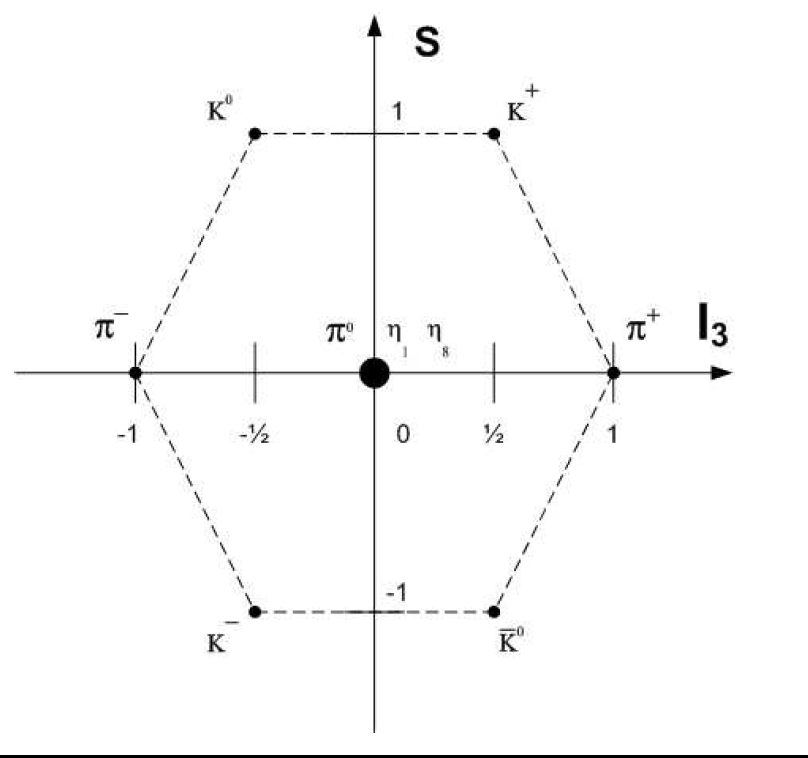}
\caption[Nonet of pseudoscalar mesons]{\label{pseudoscalar_p} Nonet of pseudoscalar mesons ( $J^{P}=0^{-}$). The third component of the isospin $I_{3}$ is on the $x$-axis  and the strangeness $S$
is on the $y$-axis.}
\end{center}
\end{figure}

In the same year as the $\eta$ meson was discovered, Gell-Mann published the famous {\it The Eightfold Way} where he proposed classification of hadrons assuming that they are built of elementary species named quarks.
In his model, quarks appear in three flavors: up ($u$), down ($d$) and strange ($s$).
Barions consist of three quarks ($qqq$) and mesons are quark-anti-quark ($q-\bar{q}$) pairs.
Due to assumed symmetry of the strong interactions between quarks in the flavor space, barions and mesons can
be classified in multiplets, grouping particles with the same $J^{PC}$ quantum numbers
and with similar properties.
According to the quark model, the $\eta$ meson can be classified as a component of the SU(3)-flavour nonet
of the lightest pseudoscalar mesons~\cite{halzen} (see  Appendix \ref{app:pseudoscal}).
The components of this nonet are plotted in Fig.~\ref{pseudoscalar_p} in terms of the
third component of the isospin $I_3$ and of the strangeness $S$.
The $\eta_{1}$ and  $\eta_{8}$ is a flavor singlet and flavor octet state, respectively,
with the following quark content:
\begin{equation*}
\begin{split}
|\eta_8>=(d \bar{d}+u \bar{u}-2s \bar{s})/\sqrt{6},\\
|\eta_1>=(d \bar{d}+u \bar{u}+s \bar{s})/\sqrt{3}.\\
\end{split}
\end{equation*}
In nature, the $\eta_{1}$ and  $\eta_{8}$ are not observed
and the existing $\eta$ and $\eta'$ mesons are their superposition:
\begin{equation*}
\begin{split}
|\eta> = \cos\theta |\eta_{8}> - \sin\theta |\eta_{1}>, \\
|\eta'>= \sin\theta |\eta_{8}> + \cos\theta |\eta_{1}>,
\end{split}
\end{equation*}
where $\theta$ is the mixing angle equal to about -15$^{\circ}$~\cite{bramon,pham}.
Because of a relatively small value of the mixing angle,
the real observed $\eta$ can be treated as a flavor-octet state
with a small admixture of a flavor-singlet component.

One presumes that the flavor-singlet component $\eta_1$ can mix with pure gluonic states. This effect should be more pronounced in the $\eta'$ case, because of a greater fraction of the flavor-singlet component.
However, it  can be also important in the $\eta$ case. According to the suggestions of Bass and Thomas it
can significantly influence  properties of the $\eta$-meson embedded in nuclear matter~\cite{bass}.

\subsection{\texorpdfstring{$\eta$}- nucleon interaction}\label{etanucl}
Due to the short lifetime of the $\eta$ meson (t$\sim 10^{-18}$ s) it is not feasible to create an $\eta$ beam.
Therefore, its interaction with nucleon or nuclei must be studied via the observation of final states
of nuclear reactions including the $\eta$-nucleon (or $\eta$-nuclei) pair. As it will be discussed in the next
section, the FSI between produced particles
can strongly influence the production cross-sections and, in this way,
can be used for studies of the interaction itself.

\begin{figure}[!htb]
\begin{center}
\includegraphics[width=0.95\textwidth]{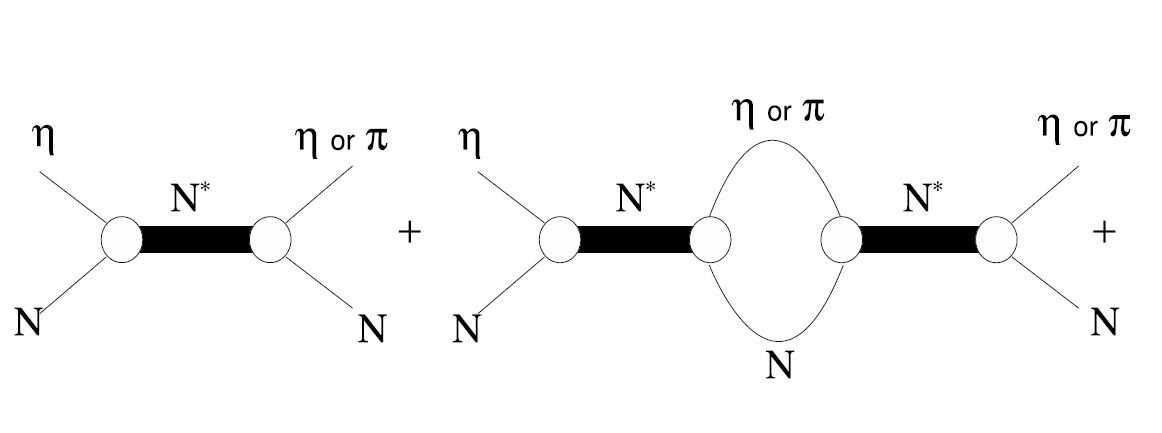}
\caption[$N-\eta$ interaction scheme]{\label{coupled_channel_p} $N-\eta$ interaction presented
as a series of formation and decays of the $N^*$ resonance: $\eta N\rightarrow N^* \rightarrow \pi N$
and $\eta N\rightarrow N^* \rightarrow \eta N$. The idea for this picture was taken from~\cite{sofianos1}.}
\end{center}
\end{figure}

In the low energy region, where the $\eta-N$ pairs are produced in the s-wave,
the $\eta$ meson interaction with nucleons is dominated by excitation of the $S_{11}$ resonance $N^*(1535)$.
The mass of this resonance is equal to 1535\,MeV/c$^2$ and the full width is of about 150\,MeV~\cite{pdg2010}.
It is the first excited state of the nucleon with odd parity  $P=-1$.
As a nucleon excitation state it has the  spin $J$ and isospin $I$
equal to $\frac{1}{2}$.
Because of its large width, it influences the whole low energy $\eta-N$ interaction region.
$N^*(1535)$ decays predominately into $N-\eta$  and  $N-\pi$ channels with roughly equal probability.
This feature suggests that $N-\eta$  and  $N-\pi$ pairs should be treated as strongly coupled
systems and that the coupled channel formalism is an appropriate tool to describe these systems.
In 1985 Bhalerao and Liu ~\cite{liu1} performed coupled-channel calculations including
$N-\eta$,  $N-\pi$ and $\Delta-\pi$ channel and showed that s-wave $\eta-N$ interaction is of strong and attractive nature.
This result has raised a question whether the total interaction in a nucleus-$\eta$ system is strong enough
to form  a bound-state.
In the bound state the $\eta$ meson undergoes multiple elastic scattering
$\eta N \rightarrow N^* \rightarrow \eta N \rightarrow N^* ...$
until it annihilates after the interaction with a nucleon
in the process $\eta N \rightarrow N^* \rightarrow \pi N$ (see Fig.~\ref{coupled_channel_p}).


\label{ch:etameson}
\section{Theoretical studies of \texorpdfstring{$\eta$}{eta}-mesic nuclei}
\subsection{Predictions for bounding \texorpdfstring{$\eta$}{eta} mesons in nuclei}

The existence of $\eta$-mesic nuclei was suggested by Haider and Liu in 1986~\cite{liu2}.
They investigated the $\eta$-nucleus interaction using the attractive
$\eta-N$-scattering length $a_0=0.28+i0.29$\,fm  or $a_0=0.27+i0.22$\,fm,
obtained by Bhalerao and Liu \cite{liu1}.
With this value of the scattering length $\eta$-mesic nuclei could be formed
for nuclei with $A\ge12$.

In the most of the contemporary estimations the real part of the ${\eta N}$-scattering length is larger
with its real part lying in the range from 0.5 to 1\,fm
and the imaginary part equal to about 0.3\,fm~\cite{green}.
With larger scattering length, a bound state can be formed in lighter nucleus.
This effect can be studied using the optical potential of the $\eta$-nucleus interaction.
In the first order in density, this potential can be written as~\cite{Tryasuchev}:
\begin{equation}
V(r)=-\frac{4\pi}{2\mu}(1+\frac{m_{\eta}}{m_N}) \rho(r)a_0,
\end{equation}
where, $m_{\eta}$, $m_N$ are the meson and nucleon masses,
$\mu$ is the reduced meson-nucleus mass, $a_0$ is the $\eta N$-scattering length and
$\rho(r)$ is the nuclear density.
Assuming $\mu=m_{\eta} = 547$\,MeV/c$^2,$ $m_N = 939$\,MeV/c$^2$, and $\rho_0 = 0.17$\,fm$^{-3}$
we obtain V=120\,MeV for $a_0 = 1$\,fm which is strongly attractive.

\begin{figure}[!ht]
\begin{center}
\includegraphics[width=10cm]{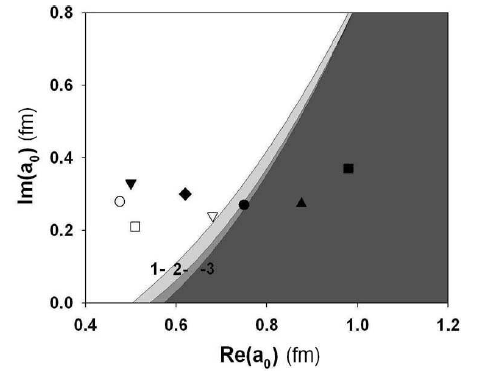}
\caption[Predictions for quasi-bound state in ${^4\mbox{He}}-\eta$ systems]{Curves show the boundaries of quasi-bound states in ${^4\mbox{He}}-\eta$ system.
The darkened areas are areas of quasi-bound state formation of $\eta$-meson with ${^4\mbox{He}}$
nucleus in the complex plane of $\eta-N$-scattering length for different diffuseness $a/R_c$ parameters:
1~-~0.25, 2~-~0.15, 3~-~0.05. Open and full symbols correspond to values of the $\eta N$-scattering length
taken from various works.
The figure was adopted from Ref.~\cite{Tryasuchev}.}
\label{he4bound}
\end{center}
\end{figure}

Tryasuchev and Isaev performed  calculations of the binding of $\eta$-meson in
${^3\mbox{He}}$ and ${^4\mbox{He}}$ nuclei
using the optical potential with the nuclear density parametrized   with the Fermi form~\cite{Tryasuchev}:
\begin{equation}
\rho(r)=\frac{\rho_0}{1+exp(\frac{r-R_c}{a})},
\end{equation}
where $R_c$ is the half-density radius, $a$ is the thickness of the nucleus diffusion surface
and $\rho_0$ is the nucleon density of the nucleus in the center.
They calculated the binding energy of $\eta$-mesons based on the exact solution of Schroedinger equation
with the optical potential.
They investigated the binding as a function of the real and the imaginary parts
of the $\eta N$-scattering length.
They concluded that for the ${^3\mbox{He}}-\eta$ system the binding is not possible,
at least for eight different values of $\eta N$-scattering length taken from the literature.
For the ${^4\mbox{He}}-\eta$ system they found that the bound state occurs in the case
of three out of the eight considered values of the $\eta-N$-scattering length (see Fig.~\ref{he4bound}).
These three values of the $\eta N$-scattering length have the real part larger than 0.7\,fm
and the imaginary part is of about 0.3\,fm.

\subsection{Motivation for a search for \texorpdfstring{$\eta$}{eta}-mesic nuclei}

With the present knowledge, the existence of $\eta$-mesic nuclei is not clear
and observation of such state would be interesting on its own.
In a more fundamental sense, its discovery would have many important implications
for the $\eta$ meson physics.

The determination of the binding energy and the decay of the $\eta$-mesic nucleus
would provide important information about the $\eta N$ interaction.
It would allow to verify the current values of the $\eta N$-scattering length
which is only very poorly known due to the difficulties in extracting it from experimental data.

Investigations of the $\eta$-nucleus bound states can  be also important for studies
of the properties of the $N^*(1535)$ resonance in nuclear matter.
This resonance dominates the $\eta N$ interaction at low energies and if its properties change inside nucleus,
then it also influences the $\eta$-nucleus optical potential~\cite{jido}.
The experimental investigation of $\eta$-mesic nucleus can help to distinguish
between different models describing the $N^*$ structure.
In chiral doublet models, $N^*(1535)$ as the first excited state with odd parity
is treated as chiral partner of the nucleon. In this picture, the in-medium $N^*$ mass shift
will be reduced due to the partial restoration of chiral symmetry.
Consequently, the $\eta$-nucleus will have a repulsive core with the attractive part at the nuclear surface.
On the other hand, in chiral unitary approach $N^*$ is treated as a dynamically generated object
in meson-baryon scattering.
Here the reduction of the mass shift is expected to be considerably smaller
and the potential stays basically attractive inside the nucleus~\cite{nagahiro2,jido}.

Studies of the $\eta$-mesic nuclei can also help to learn about the structure of the $\eta$ meson.
Bass and Thomas have shown that the binding energy of the $\eta$-nucleus system is sensitive
to the flavor-singlet component in the $\eta$ meson~\cite{basssymposium}.
Increasing the singlet component at the cost of the octet component results in a greater binding.
The same happens when the gluon content contributing to the singlet component increases.

The wave function of the $\eta$ meson in the bound state largely overlaps with the one from the nuclei.
In such compact system, large medium effects influencing properties of mesons  are expected~\cite{nagahiro2}.
Since the $\eta$-nucleus optical potential is dominated by the s-wave part,
the spectroscopic studies of $\eta$-mesic nuclei
can provide the precise information about the s-wave potential, which manifest itself in the mass shift
of the $\eta$ meson in the nucleus~\cite{hayano}.
The estimate of the mass shift would be a very interesting result in the context of studies
of the spontaneous chiral symmetry breaking.
There are theoretical predictions ~\cite{nagahiro1} which claim that the effective restoration
of the U$_{A}$(1) anomalous symmetry should be observable as an $\eta$ meson mass shift
at the finite density in $\eta$-mesic nuclei.

\label{ch:theory_pred}
\section{Experimental search for \texorpdfstring{$\eta$}{eta}-mesic nuclei}
\subsection{Heavy nuclei region}

Haider and Liu  have shown that, with the value of the $\eta N$-scattering length determined
by Bhalerao and Liu \cite{liu1}, the $\eta$ can be bound  in nuclei with $A\ge12$~\cite{liu2}.
Their estimations were  confirmed by the calculations of Li~\emph{et al.}~\cite{li}.
Consequently, the first experimental efforts were concentrated on the heavy nuclei region.

\begin{figure}[!ht]
\begin{center}
\includegraphics[width=8cm]{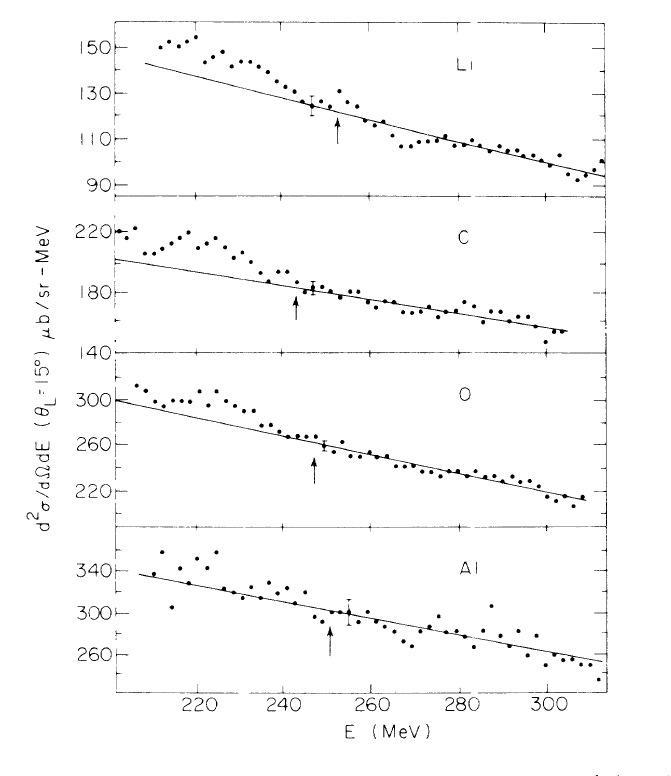}
\caption[BNL search for $\eta$-mesic nuclei]{The proton kinetic energy spectra measured for four targets in the BNL search
for $\eta$-mesic nuclei in reactions of the type ($\pi^+,p$).
The arrows indicate the eta production
threshold for each target. In each case, a Maxwellian function was fitted to part of spectrum
above the $\eta$ production threshold.
The figure was adopted from Ref.~\cite{chri}.}
\label{chrien_search}
\end{center}
\end{figure}

The first experiment dedicated to the search for $\eta$-mesic nuclei was performed in 1988 at BNL~\cite{chri}
in nuclear reactions of the type $^A{\mbox{X}}(\pi^+,p) ^{A-1}{\mbox{X}}-\eta$.
The measurements were conducted using 800\,MeV/c $\pi^+$ beam and four different targets: lithium, carbon,
oxygen and aluminium. The momenta of the outgoing protons were measured
with the Moby Dick spectrometer set at an angle of 15$^{\circ}$.
Signals from formation of $\eta$-mesic nuclei were searched in the energy spectra of the registered protons.
In the case of formation of such nuclei,
e.g. in the case of the oxygen target in the reaction $^{16}{\mbox{O}}(\pi^+,p)^{15}{\mbox{O}}-\eta$,
an enhancement in the proton energy spectrum at an energy close to the $\eta$ production threshold
was expected. No such a signal was observed in the measured spectra (see Fig.~\ref{chrien_search}).
As suggested recently in Ref.~\cite{nagahiro}, the kinematic conditions chosen
in the BNL experiment were not optimal for the production of eta mesic nuclei due to relatively
large kinetic energies of the $\eta$ mesons with respect to the target nuclei.
Therefore, further studies of the production of eta mesic nuclei in the ($\pi,N$) reaction
with optimized kinematic conditions are planned at J-PARC~\cite{fujioka}.

Five years after the BNL experiment, a search for $\eta$-mesic $^{18}$F was performed at LAMPF
in pion double charge exchange (DCX) reaction $^{18}$O($\pi^+,\pi^-$)~\cite{john}.
The $\eta$ mesons are produced in this case in collisions of the $\pi^+$ beam with neutrons
inside the $^{18}$O nucleus: $\pi^+ n \rightarrow \eta p$, leading to the appearance
of a bound $\eta$-$^{18}$F state.
One of the possible decay channels of this state proceeds via absorption of the $\eta$
on one of the protons in the $^{18}$F nucleus, leading to emission of negatively charged pions:
$\eta n \rightarrow \pi^- p$.
No clear signal from the $\eta$ mesic $^{18}$F nucleus was found in the measured excitation curves
for the registered DCX process.

In 1998 Hayano, Hirenzaki and Gillitzer~\cite{gillitzer} proposed to use the recoilless
($d,^{3}$He) reaction to produce $\eta$ mesic nuclei.
The idea of the recoilless reaction is based on a choice of kinematic conditions
corresponding to production of the $\eta$ meson at rest in the laboratory (LAB) \nomenclature{LAB}{laboratory frame}  frame
so, that it can be easily bound by the target nucleus.
The momentum of the beam is fully taken over by the outgoing $^3\mbox{He}$ nucleus.
The production of the $\eta$ mesons proceeds via the elementary process $p(d,{^3\mbox{He}})-\eta$,
for which the cross-section is relatively large even very close to threshold.
Following this idea, series of experiments were performed at GSI, including the reactions:
$^7{\mbox{Li}}(d,{^3\mbox{He}}){^6\mbox{He}}-\eta$
and $^{12}{\mbox{C}}(d,{^3\mbox{He}})^{11}{\mbox{Be}}-\eta$
\cite{gillitzer1}.
Analysis of data from these measurements is in progress.

Besides experimental searches for the $\eta$ mesic nuclei concluded with negative results,
there are also measurements claiming the discovery of such states.
One of them originates from photo-production measurements on carbon target performed by Sokol and Pavlyuchenko \cite{sokol}.
The studied reaction was:
\begin{equation}
\gamma + {^{12}{\mbox{C}}} \rightarrow p(n) + {^{11}{\mbox{B}}}-\eta ({^{11}{\mbox{C}}}-\eta) \rightarrow \pi^+ + n +X.
\end{equation}
In this process the incoming photon produces a slow $\eta$ meson that is bound in the nucleus
and a fast nucleon that leaves the nucleus.
The decay of the $\eta$ mesic nucleus proceeds via the excitation of the $N^{\star}$(1535) resonance and its subsequent decay
in a $\pi N$ pair.
In the experiment the $\pi^+ n$ pairs were registered in a two arm TOF scintillation spectrometer.
The observed shift by $90 \pm 15$\,MeV/c$^2$ of the invariant mass of $\pi^+ n$  pairs
with respect to the $N^*(1535)$ mass
was interpreted by the authors as an effect of binding the $\eta$ meson in nucleus.

\begin{figure}[!ht]
\begin{center}
\includegraphics[width=10cm]{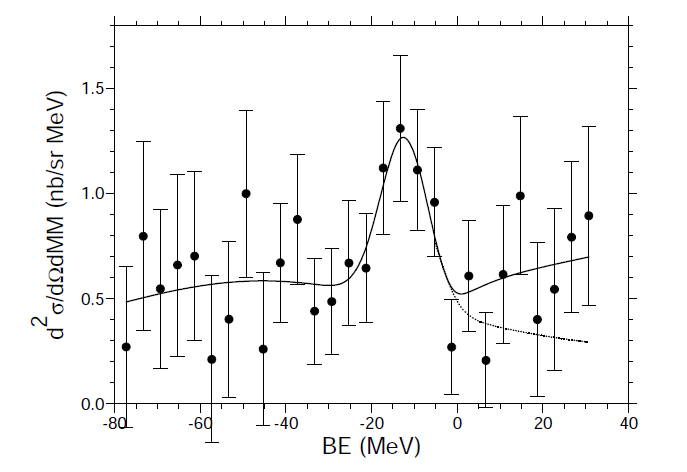}
\caption[GEM-COSY result]{Missing mass spectrum converted to binding energy BE of a bound state ${^{25}\mbox{Mg}}-\eta$ obtained
by the GEM-at-COSY group. The figure was adopted from~\cite{gem}.}
\label{machner}
\end{center}
\end{figure}

Recently, the GEM-at-COSY group reported an indication of the $\eta$-mesic magnesium produced
in the reaction p$+{^{27}\mbox{Al}} \rightarrow {^{3}\mbox{He}}+{^{25}\mbox{Mg}}-{\eta}$ ~\cite{gem}.
In their experiment, they used recoilless kinematics which means that the $\eta$ mesons were produced at rest
with respect to the target.
The momenta of the outgoing $^3$He ions were registered with the BIG KARL magnetic spectrometer
and, additionally, $\pi^- p$ pairs originating from the $\eta$ absorption on a neutron were detected.
Fig.~\ref{machner} shows the missing mass spectrum of the $^3$He ions with a peak around -20\,MeV interpreted
as a signal from ${^{25}\mbox{Mg}}-{\eta}$ production.
In our opinion, it would be important to confirm this result with higher statistics.

\subsection{Light nuclei region}


The current estimations of the $\eta N$-scattering length are mostly larger
compared to the one used by Haider and Liu in their first predictions of $\eta$-mesic nuclei.
Consequently, there exist suggestions that the binding of $\eta$ mesons is possible
in light nuclei including  ${^{3,4}\mbox{He}}$~\cite{wilkin1,wycech1,wycech2} and even deuteron~\cite{green}.
In our opinion the light nuclei are  better suited for a search of $\eta$ binding than the heavy ones,
since due to the smaller $\eta$ absorption, the bound states are expected to be narrower compared to
the case of the heavy nuclei.
Besides, there are no problems with interpretation of the results connected to the excitation
of higher nuclear levels.

Strong indications of existence of $^3{\mbox{He}}-\eta$ and $^4{\mbox{He}}-\eta$ bound states originate
from studies of the final state interaction in those systems.
The early measurements of the $dp\to {^3\mbox{He}}\eta$ reaction close to threshold
performed at SATURNE with the SPES-4~\cite{berger} and SPES-2~\cite{mayer} spectrometers
revealed a strong enhancement of the production cross-section due to an attractive
$^3{\mbox{He}}-\eta$ FSI.
The enhancement was  interpreted by Wilkin as a possible indication
of the $^3{\mbox{He}}-\eta$ bound state \cite{wilkin1}.
A similar effect was observed in the $d d \rightarrow {^4\mbox{He}} \eta$ cross-sections
measured at SPES-4~\cite{frascaria} and SPES-3~\cite{Willis97}
and it was interpreted as possible manifestation of $\eta$-mesic ${^4\mbox{He}}$~\cite{Willis97}.

\begin{figure}[htb]
\begin{center}
\includegraphics[scale=0.4]{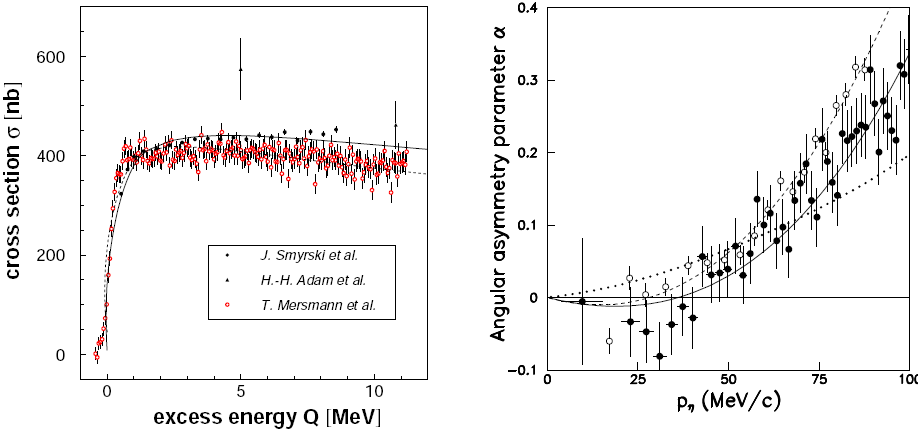}
\caption[Total cross-section and angular assymetry parameter for  $dp\to ^{3}\!\!\!He\eta$ reaction based on ANKE and COSY-11 data]{Left: Close-to-threshold total cross-section for the $dp\to ^{3}\!\!\!He\eta$ reaction
plotted as a function of the excess energy Q. Shown are the measurements performed
by ANKE collaboration~~\cite{timo} (open circles) and COSY-11 group:~~\cite{jurek-he3}
(full dots) and~~\cite{adam} (triangles). The solid line represents the scattering length fit
to the COSY-11 data~~\cite{jurek-he3}, while the dashed line is the analogous fit to the data set of Ref.~~\cite{timo}.
Right: Angular asymmetry parameter $\alpha$ plotted as a function of CM momentum.
Closed circles are the experimental data
from ANKE~~\cite{timo}, whereas open circles represent the data set of COSY-11 group~~\cite{jurek-he3}.
The dashed and solid lines are the theoretical parametrizations~\cite{wilkin2}
explained in the text. The Figure is adapted from Ref.~~\cite{wilkin2}.
}
\label{fig1}
\end{center}
\end{figure}

Recently, high precision measurements of the $dp\to {^3\mbox{He}}\eta$ reaction
have been performed at COSY-Juelich by the COSY-11 collaboration~\cite{jurek-he3} and independently,
by the ANKE collaboration~\cite{timo}.
Both measurements were done using deuteron beam slowly accelerated in a momentum interval
covering the $\eta$ production threshold.
The total cross-sections as measured in both experiments are shown in Fig.~\ref{fig1} (left panel).
The measurements confirm a rapid increase of the cross-section within an excess energy
of about 1\,MeV above the threshold to a plateau of about 400\,nb.
The presented data points were parametrized with the s-wave scattering length formula~\cite{wilkin1,jurek-he3,timo}
and a value of the ${^{3}{\mbox{He}}} \eta$ scattering length of $a_0~=~[\pm(2.9\pm 0.6)+(3.2\pm 0.4)i]$~fm has been extracted
from the fit to the COSY-11 data ~\cite{jurek-he3}.
The performed analysis did not allow to answer the question if
the real part of the scattering length is larger than the imaginary part due to a strong
correlation of these two parameters in the performed fit.
Therefore, the necessary condition for the existence of the bound
state  could not be checked.

The COSY-11 and ANKE collaborations measured also angular distributions
of the near threshold $dp\to {^{3}{\mbox{He}}\eta}$ cross-section.
Close to threshold, these distributions can be very well described
by a linear function of $\cos\theta_{\eta}$~~\cite{jurek-he3}:
\begin{equation}
\frac{d\sigma}{d\Omega}=\frac{\sigma_{tot}}{4\pi}[1+\alpha \cos\theta_{\eta}],
\end{equation}
where $\alpha$ is an asymmetry parameter, which changes with energy as shown in Fig.~\ref{fig1} (right panel).
As pointed out by Wilkin~\cite{wilkin2}, this behaviour results from a very strong variation
of the s-wave amplitude indicating the proximity of a pole in the $^{3}{\mbox{He}}-\eta$
scattering matrix.
However, information whether the pole lies
on the bound state or the virtual part of the complex energy plane cannot be accessed.

\begin{figure}[htb]
\begin{center}
\includegraphics[scale=0.29]{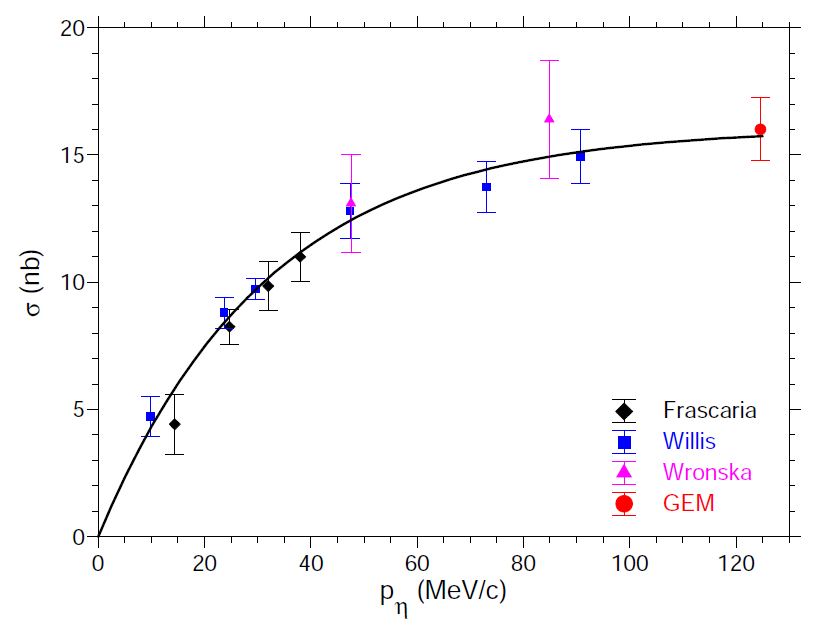}
\caption[Total cross-section for the $dd \to ^{4}\!\!\!He\eta$ reaction]{\label{4Heeta}Total cross-section for the $dd \to ^{4}\!\!\!He\eta$ reaction
plotted as a function of the CM momentum. Shown are the measurements of Frascaria et al.~\cite{Frascaria94}
(diamonds), Willis et al.~\cite{Willis97} (squares), Wronska et al.~\cite{wronska} (triangles)
and Budzanowski et al.~\cite{budzan} (circle).
The solid line represents a fit in the scattering length approximation.
The figure is adapted from Ref.~\cite{machner}.}
\label{gem1}
\end{center}
\end{figure}

The near threshold cross-sections for the $dd\to {^{4}{\mbox{He}}}\eta$ reaction
are by about one order of magnitude smaller than the $dp\to {^{3}{\mbox{He}}}\eta$ cross-sections (see Fig.~\ref{gem1}),
but they show a similar energy dependence
indicating for  a strong ${^{4}{\mbox{He}}}-\eta$ interaction in the final state.
A fit of the cross-section data using the scattering length approximation results in a ${^{4}{\mbox{He}}}-\eta$
scattering length of: $a_0 = [\pm (3.1 \pm 0.5) + i( 0.0 \pm 0.5 ]$~fm~\cite{budzan}.
This result can be converted into a pole position in the complex energy plane of $|W| \approx 4$\,MeV.
Since the sign of the real part of the scattering length is not known, the pole position
corresponds either to a virtual state or to a bound state. In the later case the $W$ equals to the binding energy.

\begin{figure}[!ht]
\begin{center}
\includegraphics[width=13cm]{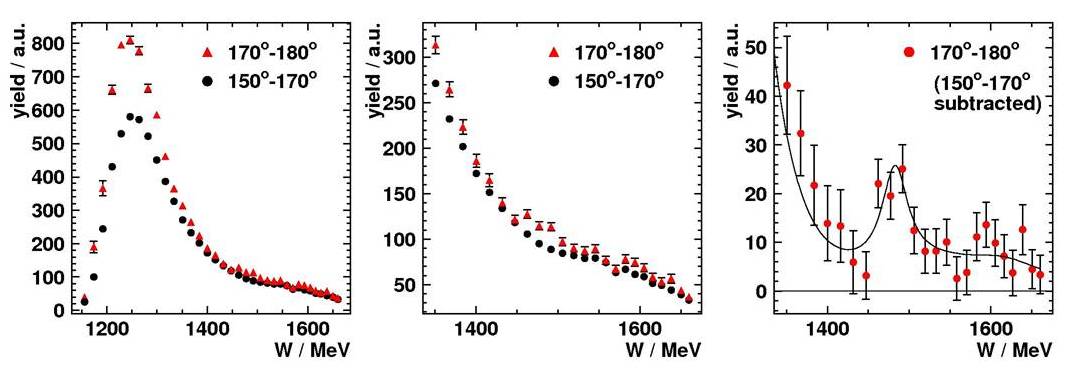}
\caption[Excitation function of the $\pi^0-p$ production obtained at MAMI]{Left and center: Excitation function of the $\pi^0-p$ production
for opening angles of 170$^{\circ}$-180$^{\circ}$ (triangles) compared to opening angles
150$^{\circ}$-170$^{\circ}$ (circles) in the $\gamma-{^3\mbox{He}}$ center of momentum system.
Right: Difference of both distributions with a Breit-Wigner distribution plus background fitted
to the data.
The figure was adopted from Ref.~\cite{mami}.}
\label{pfeiffer}
\end{center}
\end{figure}

The first direct experimental indication of a bound state of the $\eta$ meson and a light nucleus was
reported by the TAPS collaboration from their measurements of photo-production of $\eta$ mesons
on ${^3\mbox{He}}$ target~\cite{mami}.
Besides registration of $\gamma {^3\mbox{He}} \rightarrow {^3\mbox{He}} \eta$ events
they measured also production of $\pi^0 p$ pairs in the process
$\gamma {^3\mbox{He}} \rightarrow \pi^0 p X$.
According to their expectations, production of such pairs with opening angle close to $180^{\circ}$
in the $\gamma-{^3\mbox{He}}$ center of momentum frame for gamma energies below the $\eta$ threshold
can indicate a decay of ${^3\mbox{He}}-\eta$ bound state.
And indeed, the difference between excitation functions for two ranges
of the $\pi^0 p$ relative angle showed a structure which was interpreted
as a possible signature of the bound state (see Fig.~\ref{pfeiffer}).
From a fit of a Breit-Wigner distribution to the observed structure, a binding energy of $-4.4\pm4.2$\,MeV
and a width of $25.6\pm6.1$\,MeV was deduced for the $\eta$-mesic state in ${^3\mbox{He}}$.
As pointed out by Hanhart~\cite{Han}, due to the limited statistics the TAPS results
could also be interpreted in terms of a virtual ${^3\mbox{He}}-\eta$ state.
Recently, the TAPS collaboration has repeated the measurements  with higher statistics
and after a preliminary analysis Krusche {\em et al.}~\cite{krusche} suggest that the structure observed
in the $\pi^0 p$ excitation function is most likely
an artefact from the quasi-free $\pi^0$ production.

A search for the ${^3\mbox{He}}-\eta$ bound state was also performed at COSY, independently by
the COSY-TOF and COSY-11 collaboration.
They measured the excitation function for the reaction $d p \rightarrow ppp \pi^-$ close to the $\eta$
production threshold.
One expects that $ppp \pi^-$ is one of the favourable decay channels of the ${^3\mbox{He}}-\eta$ bound state
corresponding to the $\eta$ absorption on the neutron inside the ${^3\mbox{He}}$ nucleus leading
to creation of the $p \pi^-$ pair in the $\eta n \rightarrow N^*(1535) \rightarrow p \pi^-$ reaction .
The COSY-11 collaboration observed only 9 events which could originate from the decay
of the ${^3\mbox{He}}-\eta$ bound state and an upper limit of 270\,nb was derived
for the production of such state in the $d p$ collisions \cite{smyrs,moskalsymposium}.
The analysis of the COSY-TOF data is still in progress.

\label{ch:etaboundstate}
\chapter{Experimental setup}
\label{ch:wasa}

The ${4\pi}$ detector facility WASA (Wide Angle Shower Apparatus) \nomenclature{WASA}{Wide Angle Shower Apparatus}
was designed for studies of production and decays of light mesons, especially the rare $\eta$ meson decays.
Originally, it was installed and operated at the CELSIUS storage ring at the TSL in Uppsala,
Sweden~\cite{zabierowski}~\cite{calen}, offering beams of protons and light ions
with momenta up to 2.1~GeV/c.
After CELSIUS shut down in 2006 the detector was moved and mounted at the COSY accelerator.
In comparison with CELSIUS, the COSY offers higher momenta of proton and deuteron beams,
up to 3.7~GeV/c, which allows to extend the studies of light mesons to a higher mass region
including the $\phi$ meson.
The new, upgraded version of the WASA detector operating at COSY is called
WASA-at-COSY~\cite{Wasa1}~\cite{WasaComm}.

In the first section of the present chapter basic characteristics of the COSY accelerator are given.
The second section contains description of the WASA-at-COSY facility including the pellet target,
the forward detector and the central detection system.
The last two sections are devoted to the data acquisition system and the data analysis software.

\section{The COSY accelerator}

\begin{figure}[htp]
\begin{center}
 \scalebox{0.75}
 {
        \includegraphics{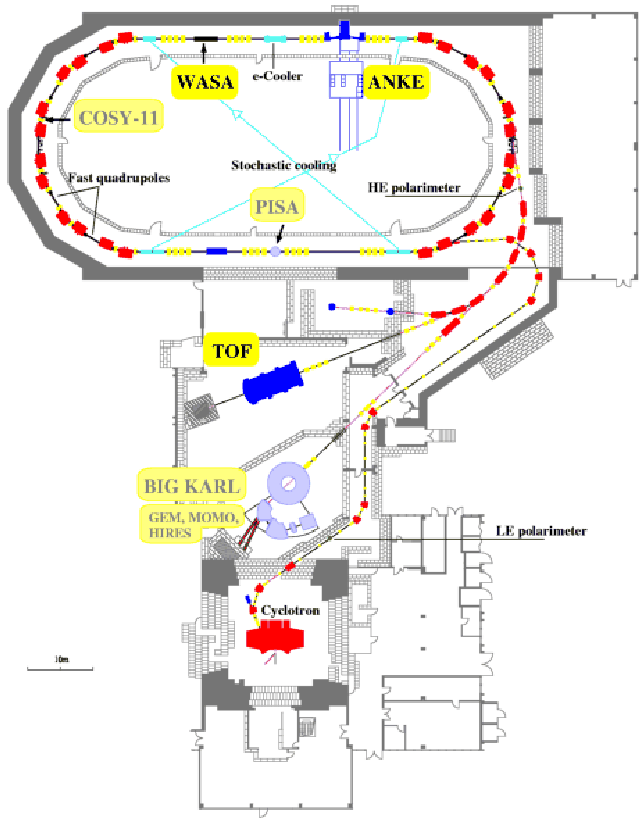}
 }
\caption[Plan of the COSY accelerator complex]{\label{cosy}Plan of the COSY accelerator complex. The beam of H$^-$ or,  alternatively, D$^-$ ions is preaccelerated in the isochronous cyclotron JULIC and after passing 100 m long transfer beamline, is injected into the COSY synchrotron storage ring via stripping injection. The operating detector systems (ANKE, WASA-at-COSY and TOF) are highlighted in yellow. The places occupied by previous experiments (GEM, MOMO, HIRES, PISA and COSY-11) are marked as shaded. WASA-at-COSY is installed next to the electron cooling facility.  }
\end{center}
\end{figure}

The Cooler Synchrotron COSY\cite{cosy1,cosy2} \nomenclature{COSY}{Cooler Synchrotron} is an accelerator and storage ring
equipped with  phase-space cooling, operated at the Research Center J\"{u}lich in Germany since 1993. A schematic view of the COSY accelerator complex is presented in Fig.~\ref{cosy}.
The accelerator facility consists of the isochronous cyclotron JULIC used as an injector, the cooler synchrotron ring with a circumference of 184 m, as well as internal and external target stations. Currently, there are three internal beam experiments: ANKE, WASA-at-COSY and  EDDA,
and one detector system using extracted beam: the TOF facility.

COSY delivers beams of polarized or unpolarized protons and deuterons in the momentum range
from 0.3 to 3.7 GeV/c. The number of stored unpolarized particles reaches a value of 10$^{11}$.  Two beam cooling techniques, the electron cooling at injection energies  and the stochastic cooling
at higher energies, are applied to reduce the beam emittance and to decrease losses of luminosity due to heating of the beam when interacting with targets of the internal experiments. The typical luminosity achieved
with the internal cluster target used by the ANKE experiment is of about $10^{31}$\,cm$^{-2}$ s$^{-1}$
and with the WASA-at-COSY pellet target it is by one order of magnitude higher.
Typical beam preparation time, including injection, accumulation and acceleration, is of the order
of a few seconds and the beam lifetime with the pellet target is of the order of several minutes.
One of the advantages of COSY is the possibility of conducting measurements during a slow
acceleration (ramping) of the internal beam within a given momentum interval.
This method permits to reduce significantly a number of systematic errors occurring in the case
when the beam is set up for each momentum separately and it has been successfully used in  past experiments \cite{jurek-he3},\cite{jureketa},\cite{moskal-prl}.

\section{The WASA-at-COSY detector}

\begin{figure}[htp]
\begin{center}
 \scalebox{0.5}
 {
  \includegraphics{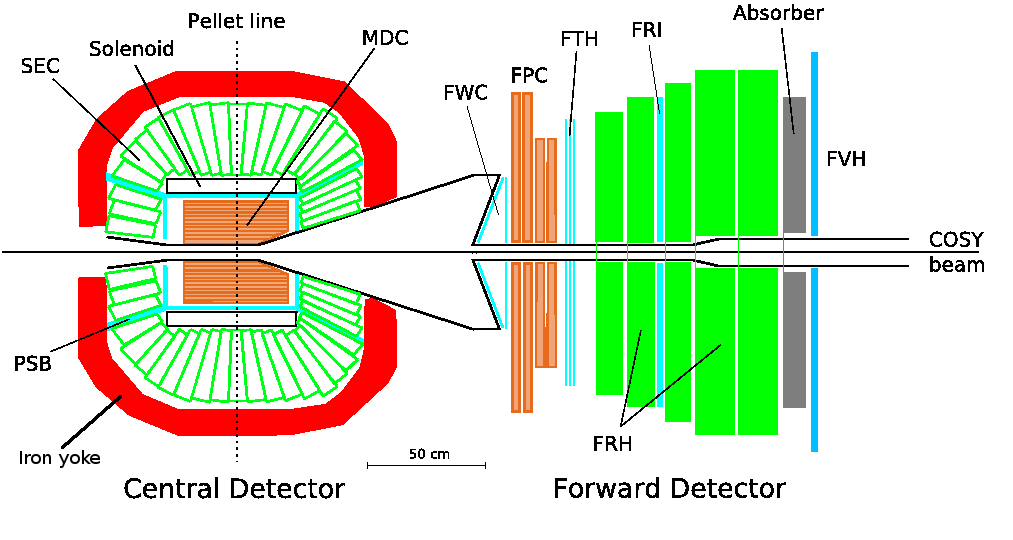}
 }
\caption[Schematic view of the WASA detector setup]{\label{wasa} Schematic view of the WASA-at-COSY detector setup. The COSY beam comes from the left side. The abbreviations are explained in the text.}
\end{center}
\end{figure}

The WASA-at-COSY detector is installed at one of the two straight sections of the COSY ring, right
in front of the electron cooler (see Fig.\ \ref{cosy}).
The WASA-at-COSY detector is depicted in the Fig.\ \ref{wasa}.
It consists of two main parts: the Forward Detector dedicated to the measurement
of scattered projectiles and target-recoils and the Central Detector optimized for registering of photons, electrons and pions originating from decays of mesons and excited baryonic states.
The forward part consists of several layers of plastic scintillators allowing for
particle identification on the basis of the $\Delta$E-E and $\Delta$E-$\Delta$E information
and of a proportional drift chamber providing track coordinates.
The Central Detector is composed of the electromagnetic calorimeter used for the energy measurement of the charged and neutral particles, the cylindrical drift chamber, the superconducting solenoid providing a magnetic field for momentum determination of the tracks of charged particles measured in the drift chamber and the barrel of plastic scin\-ti\-lla\-tors which provides fast signals for the first level trigger, and together with the drift chamber and  the calorimeter, is used for charged particle identification via $\Delta$E-p and $\Delta$E-E methods. WASA-at-COSY uses
an internal target system which provides pellets of frozen hydrogen or deuterium.
In the following subsections the individual components of the WASA-at-COSY detector are described.


\subsection{Pellet target}
\label{pellet_target}

\begin{figure}[htp]
\begin{center}
 \scalebox{0.7}
 {
  \includegraphics{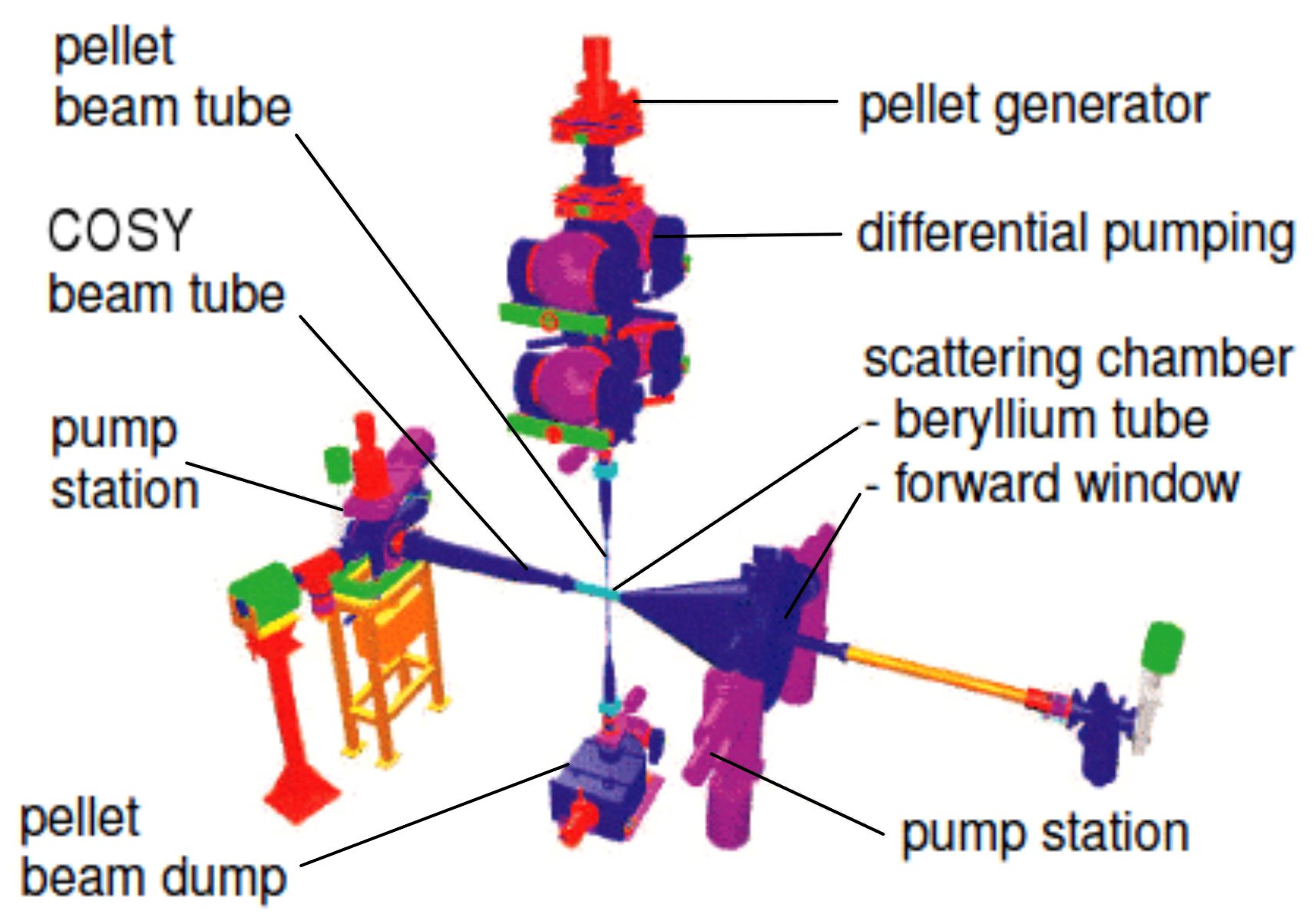}
 }
\caption{\label{pellet_p} Schematic view of the WASA-at-COSY pellet target system.}
\end{center}
\end{figure}

The WASA-at-COSY target system provides a stream of droplets (pellets) of frozen hydrogen or deuterium.
The main parts of the system are shown in Fig.~\ref{pellet_p}.

Production of the pellets starts in the pellet generator with formation of droplets from a jet of high purity liquid hydrogen (H$_2$) or deuterium (D$_2$) using a vibrating nozzle. The typical frequency of the nozzle vibrations is 70 kHz. The average diameter of the droplets is around 35 $\mu$m. The droplets are then frozen by evaporation process in a droplet chamber. Afterwards, the frozen pellets enter a 7 cm long vacuum-injection capillary and obtain the speed of 60-80 m/s due to gas pressure difference at the ends of the capillary. After the collimation process the pellets are directed through a  2~m long pipe into the scattering chamber and further down to the pellet beam dump.  The basic properties of the pellet target at the interaction point are summarized in the Table \ref{pellet_t}.
\begin{table} [htp]
 \begin{center}
 \begin{tabular}{|l|l|}
\hline
 Pellet diameter & 35 $\mu$m\\
 Pellet frequency & 8-10 kHz \\
 Pellet velocity &  60-80 m/s \\
 Pellet stream diameter at the COSY beam & 2-4 mm \\
 Pellet stream divergence & $0.04^{\circ}$ \\
 Effective target thickness & $10^{15}$ -  $10^{16}$ atoms cm$^{-2}$ \\
 \hline
 \end{tabular}
 \caption{\label{pellet_t} Parameters of the pellet stream at the interaction point.}
 \end{center}
\end{table}

\subsection{Forward Detector}
The Forward Detector (FD) \nomenclature{FD}{Forward Detector} provides information about charged hadrons like protons, deuterons or $\mbox{He}$ ions scattered in the forward direction within the polar angle range from 2.5 to 18${^\circ}$.
Also neutrons and charged pions can be measured.
The FD comprises a proportional chamber of the straw tube type designed for tracking charged particles
and a few layers of segmented plastic scintillation detectors used for the measurement of energy losses
of charged particles.
The particle identification in the FD is based on the $\Delta$E-E and $\Delta$E-$\Delta$E information
from the scintillation detectors.  The signals from scintillators provide also the information for the first level trigger logic. The amount of sensitive material of 50 g cm$^{-2}$ corresponds to values of 0.6 radiation lengths and 0.4  nuclear interaction lengths.
The FD is placed directly behind the vacuum chamber, having a conical shape opening in the forward direction.
In the forward part, the chamber contains an exit window for particles made of stainless steel with a thickness
of 0.4 mm.
The individual components of the FD are described in the following subsections.

\subsubsection{The Forward Window Counter}
\label{fwc}
\begin{figure}[htb]
\begin{center}
 \scalebox{0.5}
 {
  \includegraphics{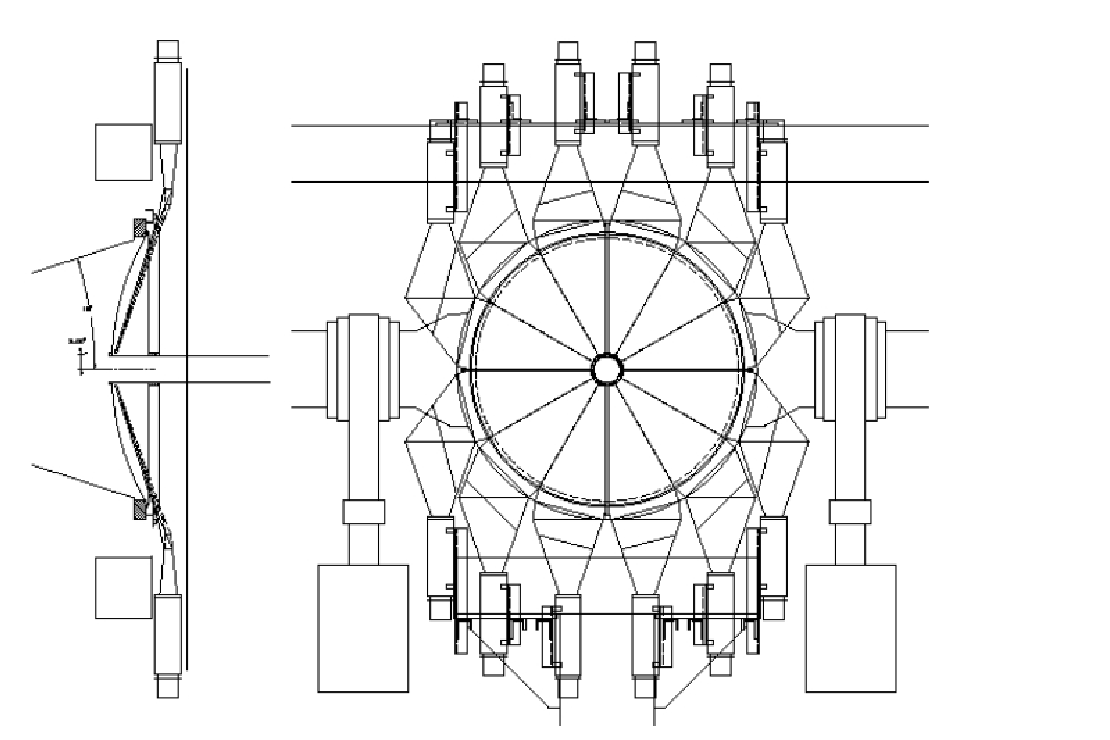}
 }
\caption[Forward Window Counter]{\label{fwc_p} Front view of a one layer of Forward Window Counter containing 12 segments of scintillation detectors (right panel) and the side view of the detector (left panel). The picture is taken from \cite{olena}.}
\end{center}
\end{figure}

The Forward Window Counter (FWC) \nomenclature{FWC}{Forward Window Counter} is a thin scintillation hodoscope mounted directly after the vacuum chamber.
It consists of two layers of 12 plastic scintillators, each 5\,mm thick.  The components are mounted on the paraboloidal stainless steel vacuum window, and they are inclined by $10^{\circ}$ with respect to the plane perpendicular to the beam direction (see Fig.~\ref{fwc_p} (left panel)).
The FWC is used in the first level trigger logic to reduce the background caused by particles scattered
downstream the target.
The information about the energy loss in individual FWC layers combined with the total energy
deposited in the Forward Detector can be used to identify charged particles and, in particular,
to select the ${^3\mbox{He}}$ ions.

\subsubsection{The Forward Proportional Chamber}
\label{fpc_l}
\begin{figure}[htb]
\begin{center}
 \scalebox{0.3}
 {
    \includegraphics{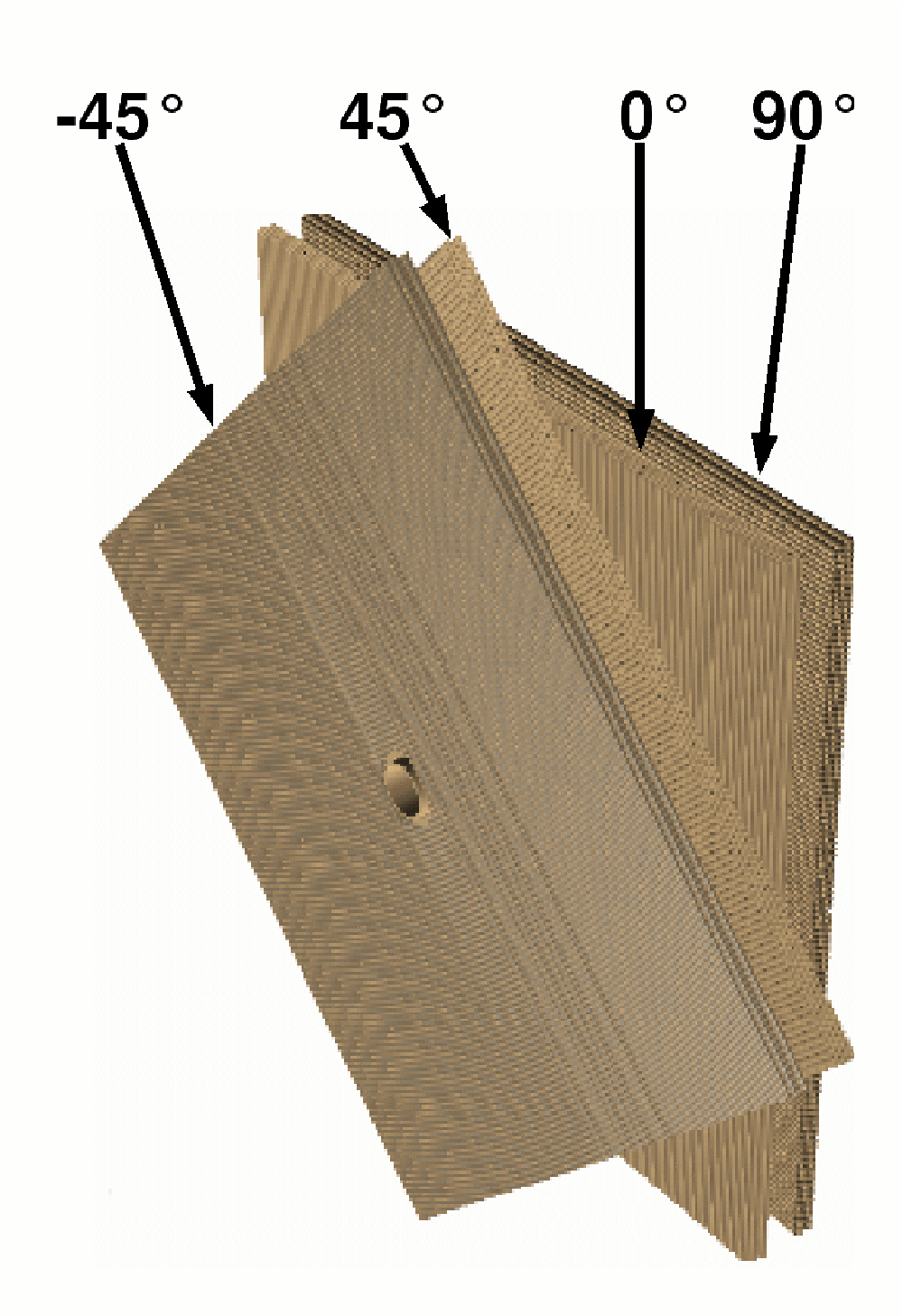}
 }
\caption[Scheme of the Forward Proportional Chamber]{\label{fpc_p}Forward Proportional Chamber 3-D layout.}
\end{center}
\end{figure}

The Forward Proportional Chamber (FPC) \nomenclature{FPC}{Forward Proportional Chamber} is a tracking device placed after the FWC. It provides precise track coordinates (up to $\sim$ 0.2${^\circ}$ angular resolution~\cite{janusz}) of charged particles passing through.
It consists of 4 quadratic modules with a circular opening for the beam pipe at the center.
Each module has 4 layers of 122 proportional straw tubes.
For a 3-dimensional reconstruction of multi-track events the straw tubes in consecutive modules are oriented
at +45$^{\circ}$, -45${^{\circ}}$, 0${^{\circ}}$ and 90${^{\circ}}$ with respect to the vertical direction
(see Fig. \ref{fpc_p}).
The straws have 8\,mm diameter and are made of thin (25\,$\mu$m) mylar foil coated with 0.1\,$\mu$m aluminium on the inner side only. The 20\,$\mu$m diameter anode wire made of gold plated tungsten is placed in the center of each straw.
The FPC works with Argon-Ethane 20\%-80\% gas mixture at atmospheric pressure.

\subsubsection{The Forward Trigger Hodoscope}

\begin{figure}[htp]
\begin{center}
 \scalebox{0.8}
 {
    \includegraphics{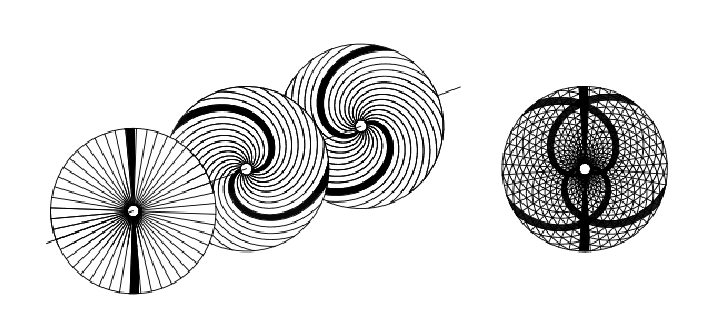}
 }
\caption[Schematic view of the Forward Trigger Hodoscope]{\label{fth_p} Left: schematic view of the Forward Trigger Hodoscope.
Right: example of pixel determination for two particle tracks.}
\end{center}
\end{figure}

The Forward Trigger Hodoscope (FTH) \nomenclature{FTH}{Forward Trigger Hodoscope} is made up of three layers of 5 mm thick plastic scintillators.
The first and the second layer consist of 24 Archimedian spiral shaped segments oriented clock-wise and counterclock-wise, respectively. The third layer has 48 scintillator segments with a torte-like shape (see Fig.~\ref{fth_p}, left part).
The  overlap of  hit segments in the three layers allows to localize tracks with the FTH (see right part of Fig.~\ref{fth_p}). This provides  fast information about the track polar angle for the triggering system. The FTH supplies also the information about the track multiplicity and the energy losses for the trigger.

\subsubsection{The Forward Range Hodoscope}

\begin{figure}[htp]
\begin{center}
 \scalebox{0.5}
 {
  \includegraphics{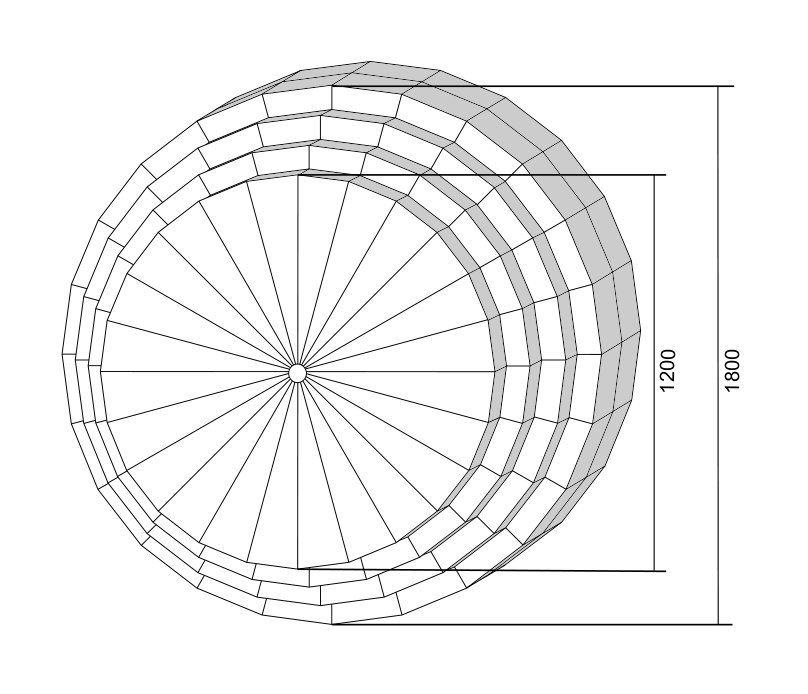}
 }
\caption[Scheme of the Forward Range Hodoscope]{\label{frh_p} Scheme of the Forward Range Hodoscope. The specified diameter of the 1st and 4th layer is given in [mm].}
\end{center}
\end{figure}

The Forward Range Hodoscope (FRH),\nomenclature{FRH}{Forward Range Hodoscope} which is situated behind the FTH, consists of five layers with 24 plastic scintillator elements each. The first three layers are 11 cm thick, while the scintillator elements in last two layers are 15 cm thick (see Fig. \ref{frh_p}).
The information from FRH combined with FTH and FWC is used for energy determination of charged particles and for particle identification by means of  $\Delta$E-E method. The reconstruction of the kinetic energy and the identification of charged particle are based on the pattern of deposited energy in the different detector layers. The maximum kinetic energy  for particles stopping in the FRH is listed in  Table \ref{frh_t}. The energy resolution for protons, deuterons, and alpha particles stopped in the detector is approximately 3$\%$.

\begin{table}[htp]
 \begin{center}
 \begin{tabular}{|l|l|}
\hline
Particle & Maximum stopping energy [MeV]\\
\hline
$\pi^{\pm}$ &  200  \\
 p & 360 \\
 d & 450 \\
${^3\mbox{He}}$ & 1000 \\
${^4\mbox{He}}$ & 1100 \\
 \hline
 \end{tabular}
 \caption{\label{frh_t} Maximum stopping energies in the Forward Range Hodoscope.}
 \end{center}
\end{table}

\subsubsection{The Forward Range Intermediate Hodoscope}
The Forward Range Intermediate Hodoscope is an additional scintillator hodoscope, which provides two-dimensional position sensitivity. It can be mounted between the third and fourth layer of the FRH.
This detector was not used during the experiment reported in this dissertation.

\subsubsection{The Forward Veto Hodoscope}
\begin{figure}[htp]
\begin{center}
 \scalebox{0.4}
 {
    \includegraphics{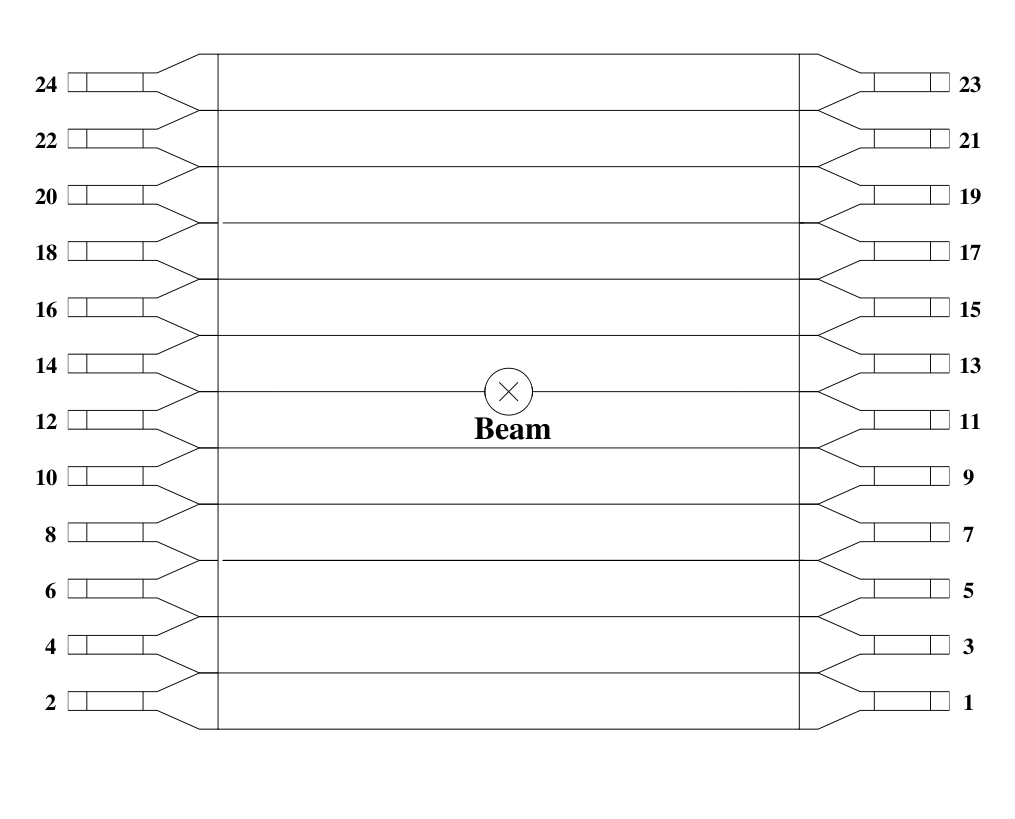}
 }
\caption{\label{fvh_p} Scheme of the Forward Veto Hodoscope.}
\end{center}
\end{figure}

The Forward Veto Hodoscope (FVH) \nomenclature{FVH}{Forward Veto Hodoscope} is the last detector layer of the FD. It consists of twelve horizontal plastic scintillator bars equipped with photomultipliers on both sides forming a wall (see Fig. \ref{fvh_p}).
The information from the FVH is used in the first level trigger logic to select or reject particle that punched through the FRH.

\subsubsection{The Forward Absorber}
The Forward Absorber (FRA) \nomenclature{FRA}{Forward Absorber} is a layer of iron which can be introduced between the last layer of FRH and the FVH, to discriminate the fast protons from elastic scattering from slower protons originating from investigated reactions. The thickness of the absorber can be chosen between 5 cm and 10 cm.
It has been used for example in the $pp \rightarrow pp\eta$ measurement. In this case, the fast protons
from elastic scattering penetrated FRA, whereas the protons associated with the $\eta$ production were stopped.
This absorber was not used during the present experiment.

\subsection{Central Detector}
The Central Detector (CD) \nomenclature{CD}{Central Detector} surrounds the interaction point and is designed mainly for detection and identification of decay products of $\pi^{0}$ and $\eta$ mesons: photons, electrons and charged pions. The CD consists of the Mini Drift Chamber, the Superconducting Solenoid, the Plastic Scintillator Barrel
and the Scintillation Electromagnetic Calorimeter.

\subsubsection{The Superconducting Solenoid}
The Superconducting Solenoid (SCS) \nomenclature{SCS}{Superconducting Solenoid} is installed inside the calorimeter and encloses the MDC and the PSB detectors. It provides an axial magnetic field along the beam for the momentum determination of tracks of charged particles measured in the MDC. It shields also the detector parts against the high flux of low energy delta electrons produced in the interaction region. The SCS provides a field of 1.3\,Tesla at the interaction point. In order to reduce the probability of electromagnetic showers caused by  gamma conversion in the material, the wall thickness of SCS is minimized to only 0.18 radiation lengths. The flux of the magnetic field outside the solenoid is closed by means of an iron yoke. The yoke serves also as support for the calorimeter crystals.  In normal operating mode the SCS is cooled using liquid helium to the temperature $\sim$4.5\,K.

During the present experiment the cooling system of the SCS was broken and, therefore, no magnetic field was provided. Thus, the momentum analysis of charged particles registered in the MDC was not possible.

\subsubsection{The Mini Drift Chamber}
\label{mdc_l}
\begin{figure}[htp]
\begin{center}
 \scalebox{0.5}
 {
 \includegraphics{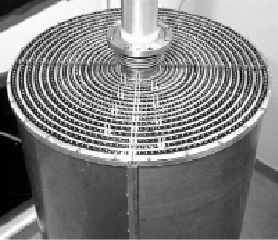}
 }
 \scalebox{0.5}
 {
  \includegraphics{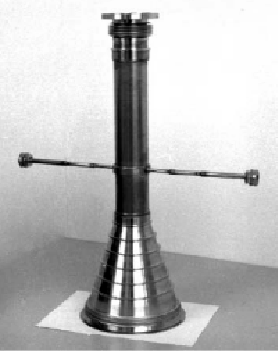}
 }
\caption[Photos of the MDC and the beam pipe]{\label{mdc_p} Left: The assembled MDC inside the Al-Be cylinder. Right: Cross formed by the beam pipe (vertical element) and part of the target pipe (horizontal tubes).}
\end{center}
\end{figure}

The Mini Drift Chamber (MDC) \nomenclature{MDC}{Mini Drift Chamber} surrounds the beam--target interaction region. It is used for the determination of charged particle momenta and the reaction vertex. It covers scattering angles from 24${^\circ}$ to 159${^\circ}$. The angle resolution provided by the MDC is about 1.2${^\circ}$.
The MDC is composed of 17 cylindrical layers with 1738 straw tubes in total. The diameter of the straw tubes in the 5 inner layers is 4 mm, 6 mm in the 6 intermediate layers and 8 mm in the 6 outer layers.  The straws are made of 25\,$\mu$m mylar foil coated with 0.1 $\mu$m aluminium on the inner side. A 20 $\mu$m diameter anode wire made of gold plated tungsten is placed in the center of each straw.
The straws in the nine inner layers are parallel to the beam axis (z-axis). The next 8 layers have small skew angles (6-9${^\circ}$) with respect to the z-axis. These layers form a hyperboloidal shape. The MDC is fitted inside a cylindrical cover made of 1 mm thick Al-Be. The straws in each layer are mounted between 5\,mm thick Al-Be end-plates. The layers are assembled around 60\,mm diameter beryllium beam pipe. The wall thickness of the beam pipe is  1.2\,mm (see Fig.~\ref{mdc_p}).
The MDC works with Argon-CO$_2$ 50\%-50\% gas mixture.
The front end electronics for the MDC is based on the CMP16 amplifier-discriminator chip originally developed for the  CMS experiment at CERN~\cite{cms}. A detailed description of the MDC can be found in~\cite{jacewicz}.

\subsubsection{The Plastic Scintillator Barrel}
\begin{figure}[htp]
\begin{center}
 \scalebox{0.5}
 {
   \includegraphics{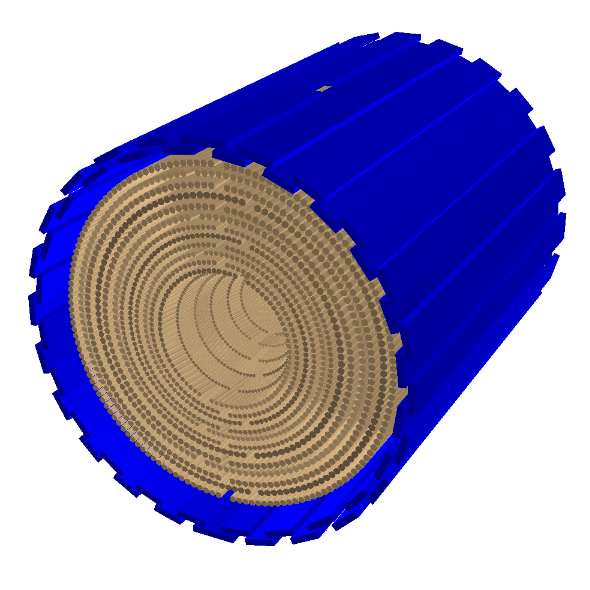}
 }
\caption[Schematic view of MDC and PS]{\label{psf_p} Schematic view of the Mini Drift Chamber (brown) enclosed by the Plastic Scintillator Barrel (blue). On the top part, the hole for the target tube is visible.}
\end{center}
\end{figure}

The Plastic Scintillator Barrel (PSB) \nomenclature{PSB}{Plastic Scintillator Barrel} is located inside the SCS coil and surrounds the MDC.
It provides fast signals for the first level trigger and together with the MDC and the calorimeter it is used for charged particle identification by the $\Delta$E-p and $\Delta$E-E method, respectively.
It serves also as a veto counter for $\gamma$ identification.
In total PSB contains 146 elements of fast plastic scintillator, each 8\,mm thick.
PSB is composed of  cylindrical part and two end-caps.
The cylindrical part consists of 50 scintillator bars arranged in two layers.
The bars are 550\,mm long and 38\,mm wide (see Fig.~\ref{psf_p}). The neighbouring bars overlap by 6\,mm
to avoid that particles pass without registration.
Two bars in the top and in the bottom part of the cylinder are split to leave space for the target tube. The end-caps contain 48 "cake-piece" shaped elements each. The front end-cap is flat while the rear end-cap forms conical surface. Both end-caps have a central hole for the beam pipe. Scintillators are glued to light guides coupled to the photomultiplier tubes. The photomultipliers are installed outside of the iron yoke to shield them against the magnetic field.
The length of the light-guides is about 500\,mm.

\subsubsection{The Scintillation Electromagnetic Calorimeter}

The Scintillation Electromagnetic Calorimeter (SEC) \nomenclature{SEC}{Scintillation Electromagnetic Calorimeter} is placed outside the SCS magnet. It is used to measure the energy of charged and neutral particles in the CD.
The energy resolution for 0.1 GeV photons is about of 8$\%$ and for stopped charged particles  is about 3$\%$.
The SEC can provide also  angular information with a scattering angle resolution of about 5${^\circ}$.
The energy threshold for detection of photons is about 2\,MeV.
The basic SEC parameters are presented in the Table~\ref{sec_t}.

\begin{figure}[htp]
\begin{center}
 \scalebox{0.5}
 {
  \includegraphics{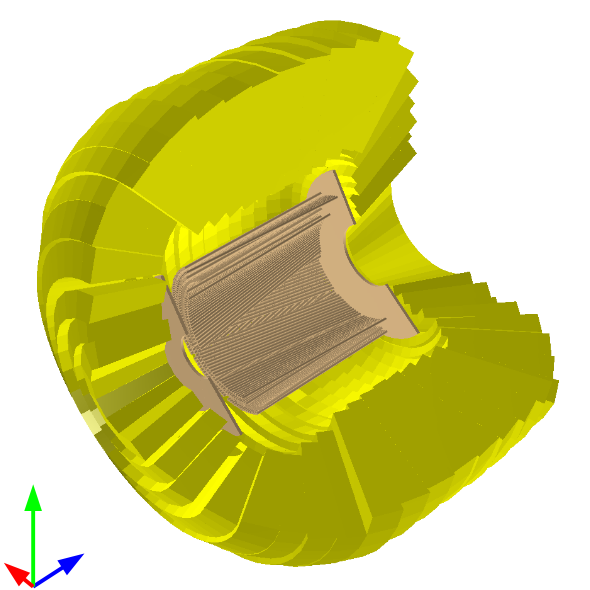}
 }
\\
 \scalebox{0.6}
 {
  \includegraphics{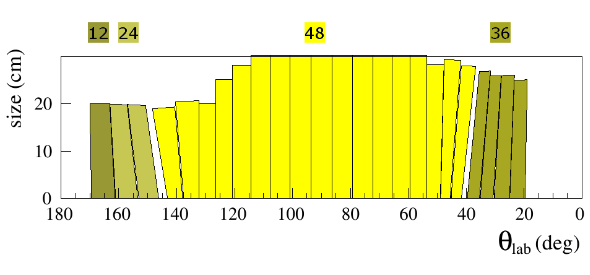}
 }
\caption[Cross-section and the angular coverage of the SEC]{\label{sec_p} (Top) Cross-section of Scintillation Electromagnetic Calorimeter. (Bottom) The angular coverage of the SEC. The 17 layers of central part are marked in bright colour. The first three layers on the left (dark colour) belong to the backward part and the last 4 layers on the right (dark colour)
belong to the forward part.
The length of the crystals are marked on the $y$-axis.
The number of elements in appropriate layer is given above the plot.}
\end{center}
\end{figure}

 The SEC is composed of 1012 sodium-doped CsI scintillating crystals placed between the superconducting solenoid and the iron yoke. It covers the scattering angles in the range from 20${^\circ}$ to 169${^\circ}$ and nearly 100$\%$  of the azimuthal angle providing about 96$\%$ of geometrical acceptance. The crystals are placed in 24 layers along the beam. There are three main parts in the SEC: forward, central and backward. The forward part covers a scattering angle range from 20${^\circ}$ to 36${^\circ}$ and  consists of  4 layers with 36 elements each. The central part covers scattering angle range from 36${^\circ}$ to 150${^\circ}$ and consists of 17 layers with 48 elements each. The backward part covers the region from  150${^\circ}$ to 169${^\circ}$ and consists of 3 layers. The layer closest to the beam pipe has 12 elements and the two outer layers consists of 24 elements each. The angular coverage of the SEC is presented in the Fig. \ref{sec_p}.
The crystals have the shape of a truncated pyramid.
The length of the crystals varies from 30 cm (central) to 25 cm (forward) and 20 cm (backward part).
The length of the crystals corresponds to the value of $\sim$16 radiation lengths and $\sim$0.8 nuclear interaction lengths.
The crystals are connected by plastic light guides with the photomultipliers placed outside of the iron yoke.
More details about the SEC can be found in~\cite{sec}.

\begin{table}
 \begin{center}
 \begin{tabular}{|l|l|}
\hline
Amount of sensitive material & 135 g cm$^{-2}$\\
   $[$radiation lengths$]$ & 16 \\
   $[$nuclear interaction lengths$]$ & 0.8 \\
Geometric coverage (4$\pi) $& 96$\%$ \\
    polar angle &  20${^\circ}$-169${^\circ}$\\
    azimuthal angle & 0${^\circ}$-360${^\circ}$\\
Maximum stopping kinetic energy &\\
    $\pi^{\pm}$/p/d& 190/400/450 MeV\\
Angular resolution & 5${^\circ}$ \\
Time resolution&\\
    photons & 40 ns \\
    charged particles & 5 ns \\
Relative energy resolution (FWHM) &\\
    photons (0.1 GeV) & 8$\%$ \\
    stopped charged particles & 3$\%$ \\
 \hline
 \end{tabular}
 \caption{\label{sec_t} Basic parameters of the Scintillator Electromagnetic Calorimeter.}
 \end{center}
\end{table}

\section{Data Acquisition System}

\begin{figure}[htp]
\begin{center}
 \scalebox{0.5}
 {
   \includegraphics{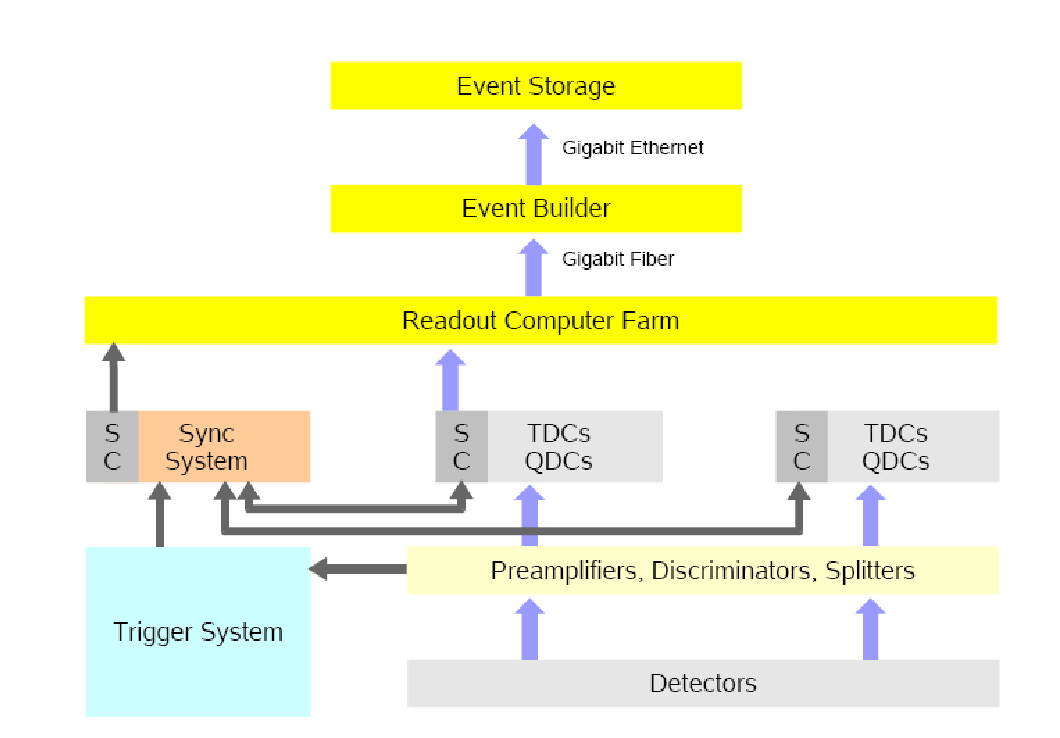}
 }
\caption[Schematic view of the data acquisition system]{\label{daq_p}Schematic overview of the data acquisition system. The figure is taken from~\cite{redmer}.}
\end{center}
\end{figure}

The Data Acquisition (DAQ) \nomenclature{DAQ}{Data Acquisition} for the WASA-at-COSY detector is based on the third generation
of the DAQ systems used in experiments at COSY~\cite{daq}. The overview of the DAQ is presented in Fig.~\ref{daq_p}.

The DAQ  is based on FPGA-controlled read out boards used for digitization and buffering of data.
It permits to conduct measurements at average accepted event rate
of  $\sim$10000 1/s ~\cite{vlasov} with a life time of 80\% of the system.
The limiting factors are the events size and the writing speed to disk storage of approximately
80\,MB/s.
The DAQ system runs in "common stop mode" which means that the trigger signal is coming after the data
have been digitized.

The analogue and digital signals from the detector front-end cards are digitized
by means of Charge-to-Digital Converter (QDC) \nomenclature{QDC}{Charge-to-Digital Converter}
and Time-to-Digital Converter (TDC) \nomenclature{TDC}{Time-to-Digital Converter} read out boards.
There are two types of the QDC boards: SlowQCD and FastQDC as well as two types of TDC boards: SlowTDC and FastTDC.
The SlowQDC is a 16 channel Flash ADC board designed for the readout of the SEC photomultipliers.
It provides time stamps along with the charge integration. The sampling frequency is 80 MHz.
The FastQDC is a 16 channel Flash ADC board with the higher sampling frequency of 160 MHz.
It is designed for the readout of the plastic scintillator detectors.
The SlowTDC is a 64 channel F1 ASIC board ~\cite{slowTDC}.
It is designed for the readout of straw tubes from the MDC and the FPC detectors.
Finally, the FastTDC is a 64 channel GPX ASIC board designed for time stamping of the plastic scintillator
signals~\cite{fastTDC}.

The synchronization system is used to control and synchronize the data flow.
When the trigger electronics generates a trigger signal, the synchronization system issues
an event number with a time stamp, which is distributed to all QDC and TDC  boards.

Read out boards collect and store the digitized signals
and mark them with time stamps. They work in self-triggering mode.
When the trigger arrives, the signals inside a given time interval are selected based on the trigger's time stamp.
The data are then sent via high speed optical links  to the PC computer farm and further to the event builder.
The events are finally written to the disk storage.

More information about the DAQ system can be found in~\cite{daq}.

\section{Analysis software}
For simulations of measurements with the WASA-at-COSY detector  Wasa Monte-Carlo (WMC) \nomenclature{WMC}{Wasa Monte Carlo}
package based on the Geant3 software has been developed.
In turns the data analysis is performed with the RootSorter which is based on the Root framework.
Both packages work in the Linux environment.

\chapter{Concept of experiment}
\label{ch:concept_exp}
Present chapter describes the basic idea of our experimental search for the
${^4\mbox{He}}-\eta$ bound state and results of simulations which were performed
to validate this idea.

\section{Basic idea}
\label{basic_idea_l}

\begin{figure}[!ht]
\begin{center}
      \scalebox{0.5}
         {
              \includegraphics{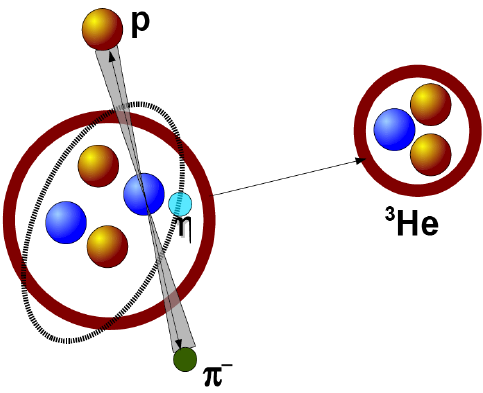}
         }
\caption[Bound state decay scheme]{\label{decay_scheme_p}Schematic picture of the  $({^4\mbox{He}}-\eta)_{bound} \rightarrow {^3\mbox{He}} p \pi^{-}$
decay. In the first step the $\eta $ meson is absorbed on one of the neutrons and the $N^{\star} $ resonance is formed. Next, the N$^{\star}$ decays into a $p-\pi^{-}$ pair. The $^3\mbox{He}$ plays the role of a spectator. }
\end{center}
\end{figure}

In our experimental studies, we use the deuteron-deuteron collisions at energies below the $\eta$ production threshold for production of the $\eta-{^4\mbox{He}}$ bound state.
We expect, that the decay of such state proceeds via absorption of the $\eta$ meson on one of the nucleons
in the ${^4\mbox{He}}$ nucleus leading to excitation of the  $N^{\star}$(1535)  resonance which subsequently
decays in pion-nucleon pair.
The remaining three nucleons play a role of spectators and they are likely to bind forming ${^3\mbox{He}}$
or ${^3\mbox{H}}$ nucleus.
This scenario is schematically presented in the Fig.~\ref{decay_scheme_p}.

According to the discussed scheme, there exist four equivalent decay channels
of the $({^4\mbox{He}}-\eta)_{bound}$ state:
\begin{itemize}
\item   $({^4\mbox{He}}-\eta)_{bound} \rightarrow {^3\mbox{He}} p \pi^{-}$
\item   $({^4\mbox{He}}-\eta)_{bound} \rightarrow {^3\mbox{He}} n \pi^{0}$
\item   $({^4\mbox{He}}-\eta)_{bound} \rightarrow {^3\mbox{H}} p \pi^{0}$
\item   $({^4\mbox{He}}-\eta)_{bound} \rightarrow {^3\mbox{H}} n \pi^{+}$
\end{itemize}

In our experiment we concentrated on the first one out of the listed decay modes
due to the highest acceptance of the WASA-at-COSY detector in this case.
The outgoing $^3\mbox{He}$ nucleus plays the role of a spectator and, therefore,
we expect that its momentum in the CM frame is relatively low and can be described
by the Fermi momentum distribution of nucleons in the $^4\mbox{He}$ nucleus.
This signature  allows to suppress background from reactions leading to the
${^3\mbox{He}} p \pi^{-}$ final state but proceeding without formation of the intermediate
$({^4\mbox{He}}-\eta)_{bound}$ state and, therefore, resulting on the average in much higher
CM momenta of $^3\mbox{He}$.
A kinematic variable correlated with the $^3\mbox{He}$ CM momentum is the relative angle  of the outgoing
$p-\pi^{-}$ pair.
In the limit of $^3\mbox{He}$ produced at rest in the CM frame this angle is exactly equal to ${180^{\circ}}$
but due to the presence of the Fermi motion it is smeared by about ${30^{\circ}}$.
Quantitative predictions for the $^3\mbox{He}$ CM momenta and for the $p-\pi^{-}$ relative angle are
given in the next section.

The principle of the present experiment is based on the measurement of the excitation function
of the $dd \rightarrow {^3\mbox{He}} p \pi^{-}$ reaction for energies in the vicinity of the $\eta$ production
threshold and on the selection of events with low ${^3\mbox{He}}$ CM momenta.
In the case of existence of the ${^4\mbox{He}}-\eta$ bound state we expect to observe
a resonance-like structure in the excitation function at CM energies below the $\eta$ threshold.
From the central energy of the observed structure $E^{CM}$ one can determine the binding energy
of the $({^4\mbox{He}}-\eta)$ system:
\begin{equation}
E_{BE}=m_{He}+m_{\eta} - E^{CM}.
\end{equation}
The width of the structure is equal to the width of the bound state.

\section{Simulations of the \texorpdfstring{$dd \rightarrow ({^4\mbox{He}}-\eta)_{bound}\rightarrow {^3\mbox{He}} p \pi^{-}$}{dd->(4He-eta)bound->3He p pi-} process}
\begin{figure}[!ht]
\begin{center}
      \scalebox{\scaleFactor}
         {
              \includegraphics{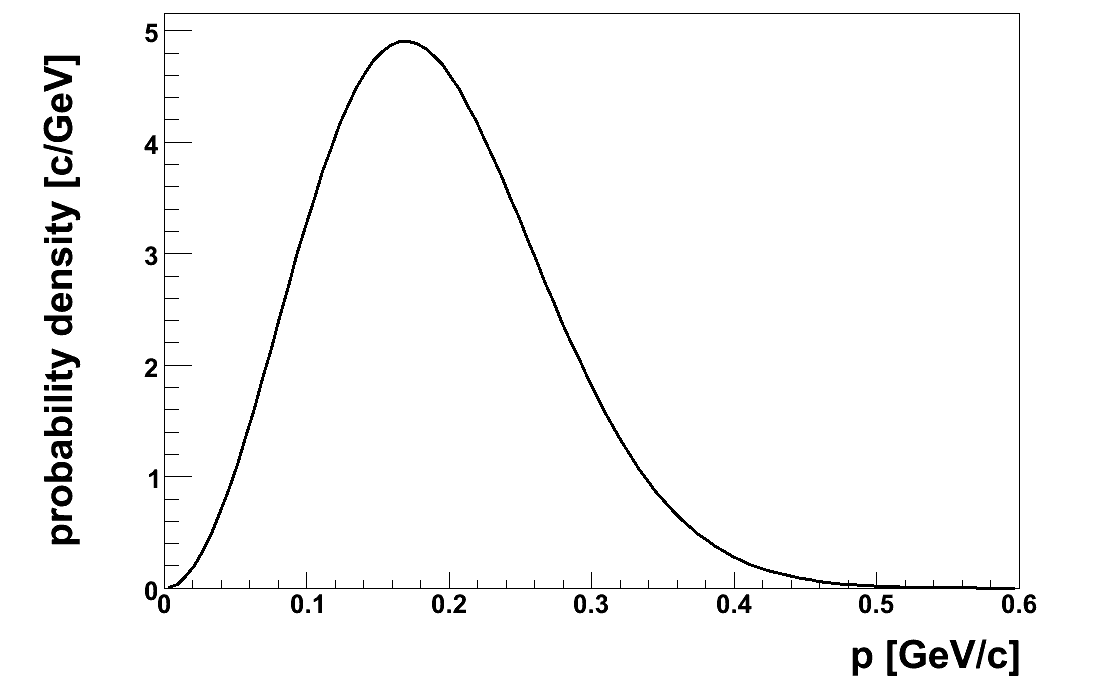}
         }
\caption[Probability density distribution of Fermi momentum in $^4{\mbox{He}}$]{\label{fermi_he4_p} Probability density distribution of the Fermi momentum in the $^4{\mbox{He}}$ nucleus taken from ~\cite{VH}. }
\end{center}
\end{figure}

In order to check the feasibility of the search for the $^4\mbox{He} \eta$ bound state with the WASA-at-COSY
detector we performed simulations
of the $dd \rightarrow ({^4\mbox{He}}-\eta)_{bound}\rightarrow {^3\mbox{He}} p \pi^{-}$ process.
These simulations were also used in preparatory phase of the experiment for setting adequate triggering conditions
of the detector read out and, further on, during the data analysis for choosing
optimal selection criteria and cuts.
We assumed that the decay of  $^4\mbox{He}-\eta$ bound state proceeds via absorption of the $\eta$ meson
on one of neutrons in the ${^4\mbox{He}}$ nucleus leading to excitation of the $N^{\star}$ resonance which subsequently
decays into the $p-\pi^{-}$ pair.
The remaining three nucleons bind forming the ${^3\mbox{He}}$ nucleus.
In the simulation, the direction of the ${^3\mbox{He}}$ momentum vector is selected assuming
its isotropic distribution in the CM frame.
The length of the momentum vector is randomized  according to the probability density distribution
of the Fermi momentum  in $^4\mbox{He}$ taken from \cite{VH} based on~\cite{McCarthy}:
\begin{equation}
f(p)=\frac{p^2}{a_1} \cdot \exp ^{\frac{p^2}{a_2}},
\end{equation}
 where the values of the parameters are: $a_1=0.0001989184519$ (GeV/c)$^3$, $a_2=0.0028615450879$ (GeV/c)$^2$.
This distribution has a maximum at around $p = 0.18$\,GeV/c (see Fig.~\ref{fermi_he4_p}).

The four-momentum of the intermediate $N^{\star}$  state is calculated using the energy-momentum conservation principle:
\begin{equation}
{p_{N^{\star}}}={p_{total}}-{p_{He}}=(\sqrt{s}-E_{He}^{ CM}, -{\vec{p}_{He}}^{ CM}).
\end{equation}

\begin{figure}[!ht]
\begin{center}
      \scalebox{\scaleFactor}
         {
              \includegraphics{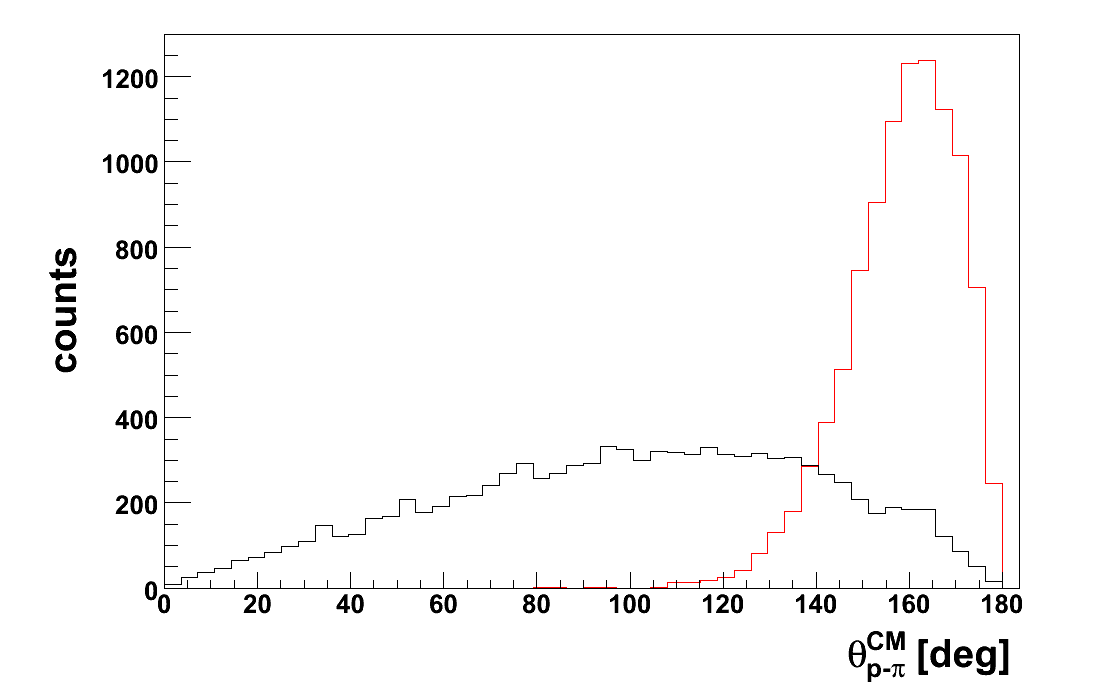}
         }

\caption[Distribution of the relative $p-\pi^{-}$ angle seen in CM]{\label{proton_pion_openAngleCM_p}
Distribution of the  $p-\pi^{-}$ opening angle in the CM system obtained in simulation of the processes
leading to the creation of the ${^4\mbox{He}}\eta$ bound state:
$dd\to({^4\mbox{He}}\eta)_{bound}\to{^3\mbox{He}}p\pi^{-}$ (red line) and
of the direct $dd\to{^3{\mbox{He}}}p\pi^{-}$ reaction (black line).
The simulation was done for momentum of the deuteron beam of 2.307~GeV/c.}
\end{center}
\end{figure}

The simulation of the $N^{\star}$ decay into a  $p \pi^{-}$ pair is performed under assumption of its isotropic
angular distribution in the $N^{\star}$ rest frame.
The opening angle between the outgoing proton and pion is equal to 180$^{\circ}$
in the $N^{\star}$ reference frame,
whereas in the CM frame its maximum is shifted to smaller values and it is smeared by about 30$^{\circ}$
due to  Fermi motion of the nucleons inside the $^4\mbox{He}$ nucleus (see Fig.~\ref{proton_pion_openAngleCM_p}).

As a model for background processes in the present studies
we consider the reaction $dd \rightarrow {^3\mbox{He}} p \pi^{-}$ for which we assume
a uniform distribution of the reaction products over the available phase space.
Further on, we refer to this process as a {\em direct production}.
In contrast to the $dd\to(^4{\mbox{He}}\eta)_{bound}\to{^3\mbox{He}}p\pi^{-}$ reaction, in the direct production
the distribution of the relative proton-pion angle in the CM system $\Theta_{p-\pi}^{CM}$ covers the full angular range
(see Fig.~\ref{proton_pion_openAngleCM_p}).

\begin{figure}[!ht]
\begin{center}
      \scalebox{\scaleFactor}
         {
              \includegraphics{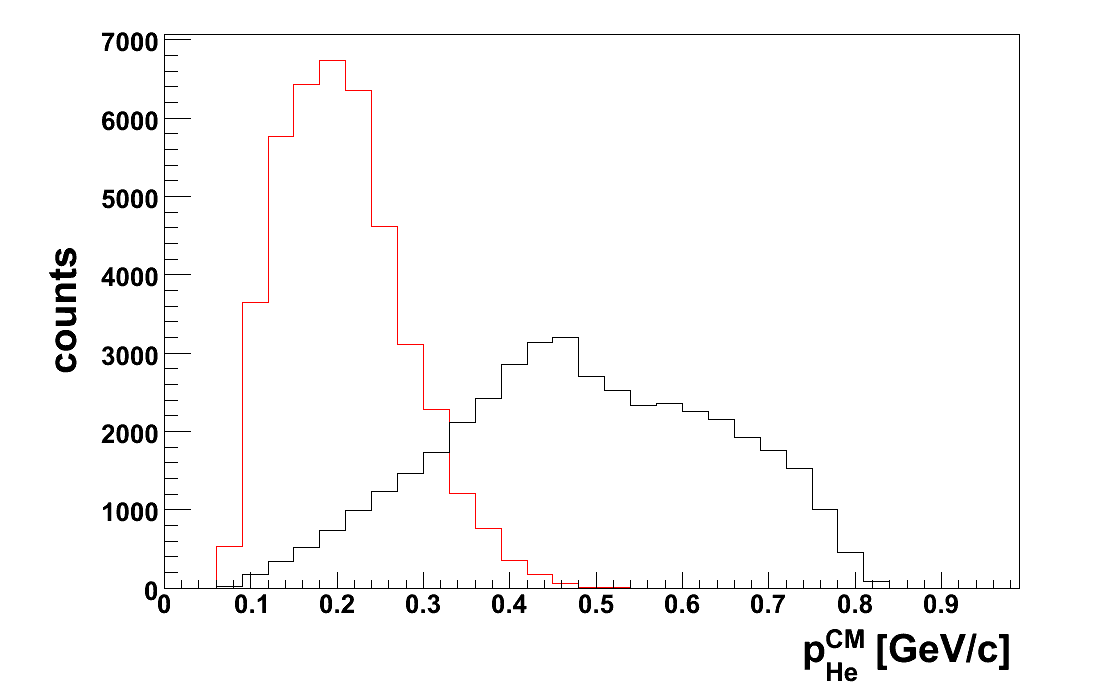}
         }

\caption[Distribution of the ${^3\mbox{He}}$ momentum in CM]{\label{pmomCM_p} Distribution of the ${^3\mbox{He}}$ momentum in the CM system obtained in simulation of the processes leading to the creation
of the  ${^4\mbox{He}}\eta$ bound state:
$dd \rightarrow (^4{\mbox{He}} \eta)_{bound} \rightarrow {^3{\mbox{He}}}p\pi^{-}$ (red line)
and of the direct $dd \rightarrow {^3{\mbox{He}}} p\pi^{-}$ decay (black line).
The simulation was done for momentum of the deuteron beam of 2.307~GeV/c.}
\end{center}
\end{figure}

The relative $p-\pi^{-}$ angle $\Theta_{p-\pi}^{CM}$ is strongly correlated
with the $^3\mbox{He}$ momentum $p_{He}^{CM}$.
In the case of $^3\mbox{He}$ produced at rest in the CM system, $\Theta_{p-\pi}^{CM}$ equals 180$^{\circ}$.
The Fermi motion results in smearing of both the  $\Theta_{p-\pi}^{CM}$ and $p_{He}^{CM}$.
For the direct production the $p_{He}^{CM}$ distribution is much wider than for the ${^4\mbox{He}}-\eta$ bound state
decay (see Fig.~\ref{pmomCM_p}). As it will be discussed in the next chapter devoted to the data analysis,
a cut on the $^3\mbox{He}$ momentum in the CM frame is used as a basic mean to suppress the background.

\begin{figure}[hpt]
\begin{center}
      \scalebox{\scaleFactor}
         {
              \includegraphics{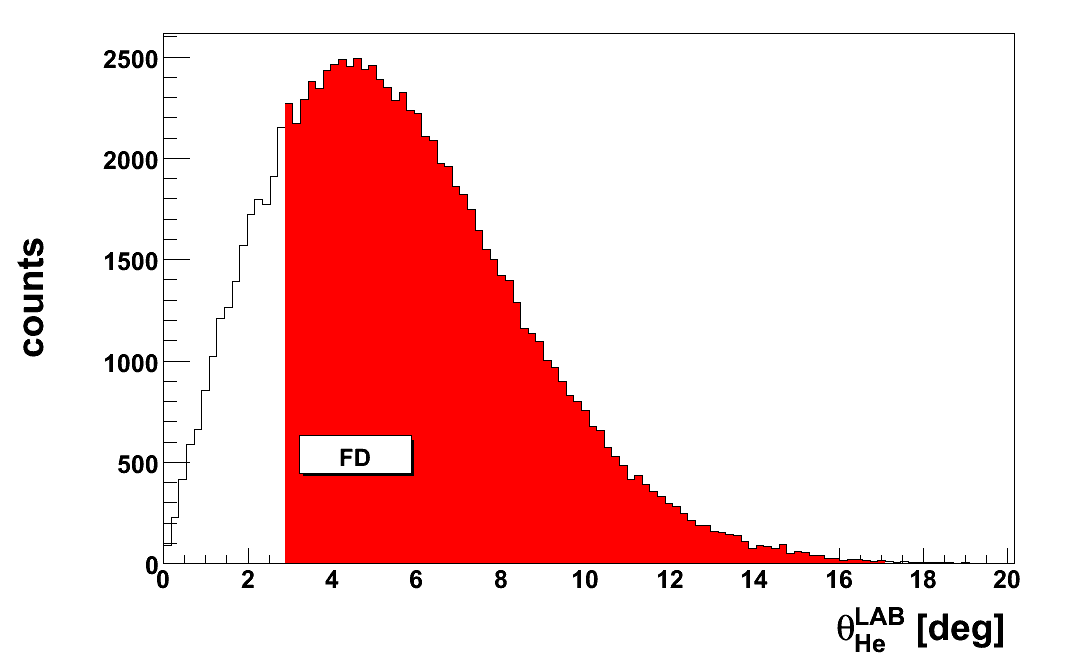}
         }
\caption[$^3{\mbox{He}}$ scattering angle in the LAB frame]{\label{he3_theta_p} Distribution of the $^3{\mbox{He}}$ scattering angle seen in the LAB frame as simulated for the processes leading to the creation of the  ${^4\mbox{He}}-\eta$ bound state:  $dd\to(^4{\mbox{He}}\eta)_{bound}\to{^3\mbox{He}}p\pi^{-}$. The red area represents the WASA-at-COSY Forward Detector acceptance.}
\end{center}
\end{figure}

\begin{figure}[!ht]
\begin{center}
      \scalebox{\scaleFactor}
         {
              \includegraphics{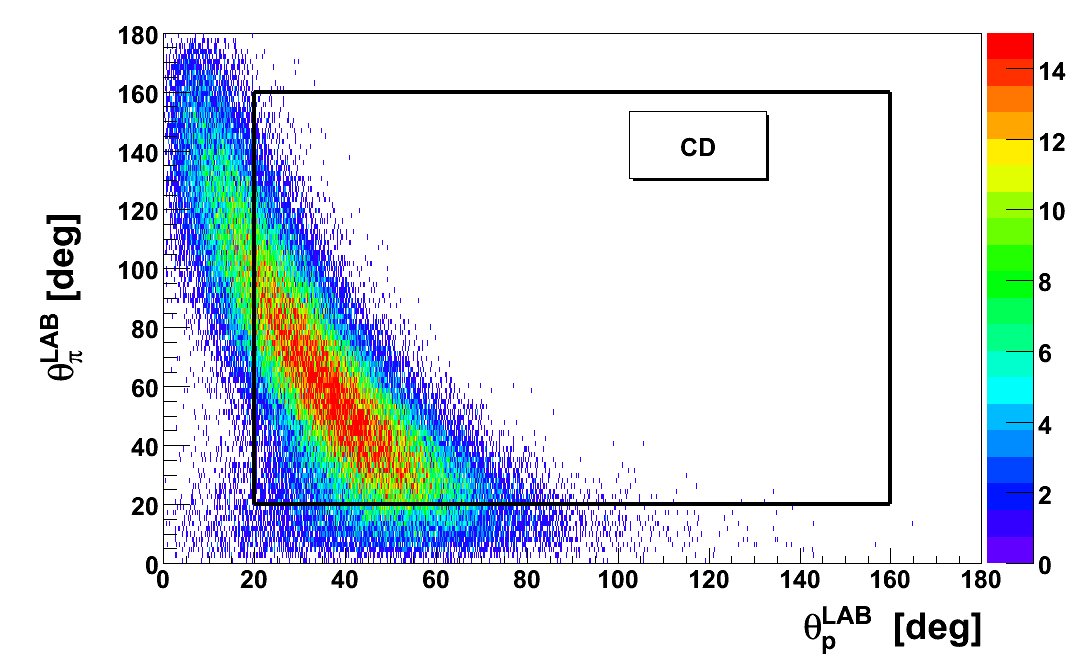}
         }
\caption[p-$\pi$ scattering angle in the LAB]{\label{ppim_theta_p} Distribution of $p$ ($x$-axis) versus $\pi^{-}$
($y$-axis) scattering  angle seen in the LAB frame as simulated for the processes leading to the creation of the ${^4\mbox{He}}-\eta$ bound state:  $dd\rightarrow({^4\mbox{He}}\eta)_{bound}\to{^3\mbox{He}}p\pi^{-}$. The black rectangle corresponds to the range of the WASA-at-COSY Central Detector acceptance.}
\end{center}
\end{figure}

For preparation of experimental trigger we performed estimation of angular acceptance of the WASA-at-COSY
for registration of the decay products of the ${^4\mbox{He}}-\eta$ bound state.
As shown in the Fig.~\ref{he3_theta_p} most  of the ${^3\mbox{He}}$ ions (81 \%) are emitted in the forward
direction and are confined in the acceptance of the Forward Detector.
In turns, the outgoing protons and negatively charged pions are registered at much larger angles and
about 71\% of all $p-\pi^{-}$ pairs is registered it the Central Detector (see Fig.~\ref{ppim_theta_p}).
Therefore, in the experiment the main trigger dedicated to the registration of
the $dd \rightarrow ({^4\mbox{He}}-\eta)_{bound}\rightarrow {^3\mbox{He}} p \pi^{-}$ process
was constructed under assumption that the   ${^3\mbox{He}}$ ions are detected in the Forward Detector
and the $p-\pi^{-}$ pairs are registered in the Central Detector.
The overall geometrical acceptance is equal to about 60\%.

\chapter{Data analysis}
\label{ch:analysis}
The present search for the ${^4\mbox{He}}-\eta$ bound state was performed in a dedicated experiment in June 2008.
During the measurements, a scan of the excitation function for the process
$dd \rightarrow ({^4\mbox{He}}-\eta)_{bound}\rightarrow {^3\mbox{He}} p \pi^{-}$
was conducted using  slow acceleration of the COSY deuteron beam.
The integrated luminosity in the experiment was determined using
the $dd \rightarrow {^4\mbox{He}} n$ reaction
and the relative normalization of the data points measured as a function of the beam momentum
was based on the quasi-elastic proton-proton scattering, which was registered with relatively high statistics.

The basic settings of the COSY acceleration cycle and the trigger applied for the detector read out
are presented in the first section of the present chapter.
The calibration of the WASA-at-COSY detectors is presented in the second section.
The main part of the data analysis is devoted to selection and reconstruction of the
$dd \rightarrow {^3\mbox{He}} p \pi^{-}$ events with subsequent application of cuts reducing the counts
not originating from the searched $({^4\mbox{He}}-\eta)_{bound}\rightarrow {^3\mbox{He}} p \pi^{-}$ decay.
This part of the analysis is described in the third section.
The luminosity determination is presented in the section four.

\section{Accelerator cycle and experimental trigger settings}
The presented experiment was planned for 9 days of the beam-time
which were granted by the COSY Programme Advisory Committee,
however, due to severe problems with the COSY beam
and with the WASA-at-COSY deuteron target, the data was taken during one day only.
The analysed data set consists of 66 runs  (run numbers: 9163-9228) and corresponds
to an effective measurement time of about 16.5 hours.

Setting of the accelerator cycle is specified in Tab.~\ref{beam_cycle_t}.
Duration of the cycle was 120\,s. In the first 3.431\,s the beam was accelerated in a routine way during a fast ramping
of the COSY dipole magnets to the momentum of 2.185\,GeV/c. After this, a slow ramping phase taking 107.6\,s
followed. During this phase the beam momentum was increased linearly in time to upper limit of 2.400\,GeV/c.
Events containing the detector response were stored together
with a content of a precise clock representing the information about the instantaneous beam momentum.
The data was taken during the slow acceleration phase in the momentum range
from  2.192~GeV/c to 2.400~GeV/c.
During the data analysis, this range was divided into 20 intervals of equal width and the excitation function was determined on the basis of events selected in each interval.

{
\renewcommand{\arraystretch}{1.2}
\begin{table}[!h]
 \begin{center}
 \begin{tabular}{|l|l|}
        \hline
        beam cycle time & 118 s \\
        \hline
        start DAQ & 7 s \\
        \hline
        start slow ramping &3.431 s\\
        \hline
        slow ramping time & 107.6 s  \\
        \hline
        lowest beam momentum & 2.185 GeV/c \\
        \hline
        highest beam momentum & 2.400 GeV/c \\
        \hline
        effective measurement time in the cycle & 88 \%\\
        \hline
 \end{tabular}
 \caption{Setting of accelerator cycle.}\label{beam_cycle_t}
 \end{center}
\end{table}
}

Several hardware triggers were used in the experiment for the read out of events.
The main trigger dedicated to the study
of the  $dd \rightarrow ({^4\mbox{He}}-\eta)_{bound}\rightarrow {^3\mbox{He}} p \pi^{-}$ reaction
required at least two charged tracks in the Central Detector for registration of the $p-\pi^-$ pairs
and at least one charged track in the Forward Detector for registration of the ${^3\mbox{He}}$ ions.
Additionally, a high energy threshold was set
in the first FWC layer and the first FRH layer in order to suppress signals
from fast protons and deuterons in the Forward Detector.
Besides,  in order to increase a selectivity for detection
of ${^3\mbox{He}}$ ions which are expected to stop in the first two layers of the FRH,
a veto was set on the third FRH layer.


For the luminosity monitoring two additional triggers were used.
The $dd \rightarrow {^3\mbox{He}} n$ reaction was registered for determination of absolute value
of the integrated luminosity. The measurement was based on detection
of the outgoing ${^3\mbox{He}}$ ions in the Forward Detector.
Therefore, the applied hardware trigger required at least one charged particle in the Forward Detector and,
in addition, a high energy threshold in the FWC.
The prescaling factor for this trigger was equal to $\frac{1}{150}$.
For determination of integrated luminosity as a function of the beam momentum the quasi-elastic $p-p$ scattering
was measured.
For this, at least one charged particle in the Forward Detector
and also at least one charged track in the Central Detector was required.
The prescaling factor was equal to $\frac{1}{4000}$.

\section{Detector calibration}
\label{calib_ch}
Registration of the reactions $dd \rightarrow {^3\mbox{He}} p\pi^-$ and $dd \rightarrow {^3\mbox{He}} n$,
which are the main processes measured in the present experiment,
is based on the energy deposits  of  the $^3\mbox{He}$ ions in the Forward Range Hodoscope.
Therefore, a precise energy calibration of the FRH is of high importance for the analysis of theses channels
and, in particular, for a clean identification of the  $^3\mbox{He}$ ions on the basis of the energy losses
in the FRH layers.
The calibration of the FRH is described in details in the first subsection.
The  second subsection presents the energy calibration procedure
of the Electromagnetic Calorimeter which was used for identification of the protons and pions
from the $dd \rightarrow {^3\mbox{He}} p\pi^-$ reaction on the basis of the energy losses.
The position calibration of the Straw Tube Chambers (FPC and MDC)
which were used for reconstruction of trajectories of charged ejectiles including
${^3\mbox{He}}$, protons and pions is described in the third subsection.
The calibration for other detectors and are not discussed here. Their settings were taken from the previous measurements.

\subsection{Forward Range Hodoscope}

The calibration of the plastic scintillator detectors in the FRH comprises translation from
the ADC channels to the energy.
The calibration is performed in two steps.
First, a deviation from the uniform light collection efficiency over the area of the scintillator detector
is determined and corresponding corrections to the ADC channels are calculated.
In the second step, the conversion of the corrected ADC values to the deposited energy is performed.

\begin{figure}[!ht]
      \centering
      \scalebox{\scaleFactor}
         {
         \includegraphics{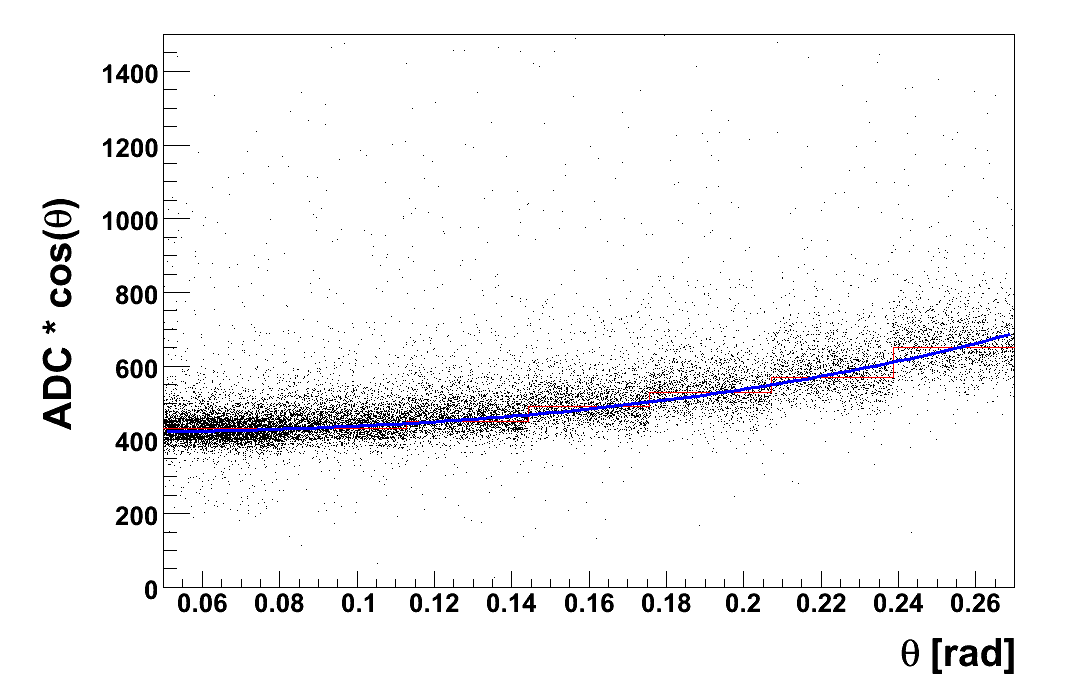}
         }
         \caption[Example of the nonuniformity corrections for one element of FRH]{'Raw' ADC value multiplied by $\cos(\theta)$ as a function of $\theta$ for the third element of the first layer of FRH. The fitted third order polynomial is shown as a solid blue line.}
         \label{FRH1_el3_nonuniform_v2_p}
\end{figure}

The non-uniformity of the light collection is caused by the fact that the photomultipliers
reading out the scintillator segments in the FRH are attached to these segments
on the outer rim of the FRH layers.
Therefore, particles scattered at small angles interact with the segments close to the center of the layers,
and the produced scintillation light has to pass the longest way to the photomultipliers. This leads
to the largest light losses.
In turns the particles scattered at the maximum angle accepted by the FRH ($18^{\circ}$) produce scintillations
close to the photomultipliers and thus the light losses are relatively small.
The non-uniformity of the light collection is parametrized
as a function of the scattering angle $\theta$.
To correct for the non-uniformity effect we select fast particles, such as protons
from the quasi-elastic scattering, demanding exactly one reconstructed charged track
reaching the fifth layer of the FRH.
These particles are expected to be close to the ionization minimum
which means that the energy deposit per unit length does not depend on the scattering angle.
The ADC values are multiplied by  $\cos\theta$ in order to correct the light output for track
inclination  with respect to direction perpendicular to the scintillator segment area.
For each scintillator segment in the FRH  we plot the values of $\mbox{ADC} \cdot \cos\theta$
as a function of the scattering angle $\theta$ (see Fig.~\ref{FRH1_el3_nonuniform_v2_p}).
We parametrize this dependence using a fit with the third order polynomial $f(\theta)$.
The ADC values for the analysed events are then corrected in the following way:
\begin{equation}
 \mbox{ADC'} =\frac{\mbox{ADC }\cdot \cos\theta}{f(\theta)}.
\end{equation}

\begin{figure}[!ht]
      \centering
      \scalebox{\scaleFactor}
         {
         \includegraphics{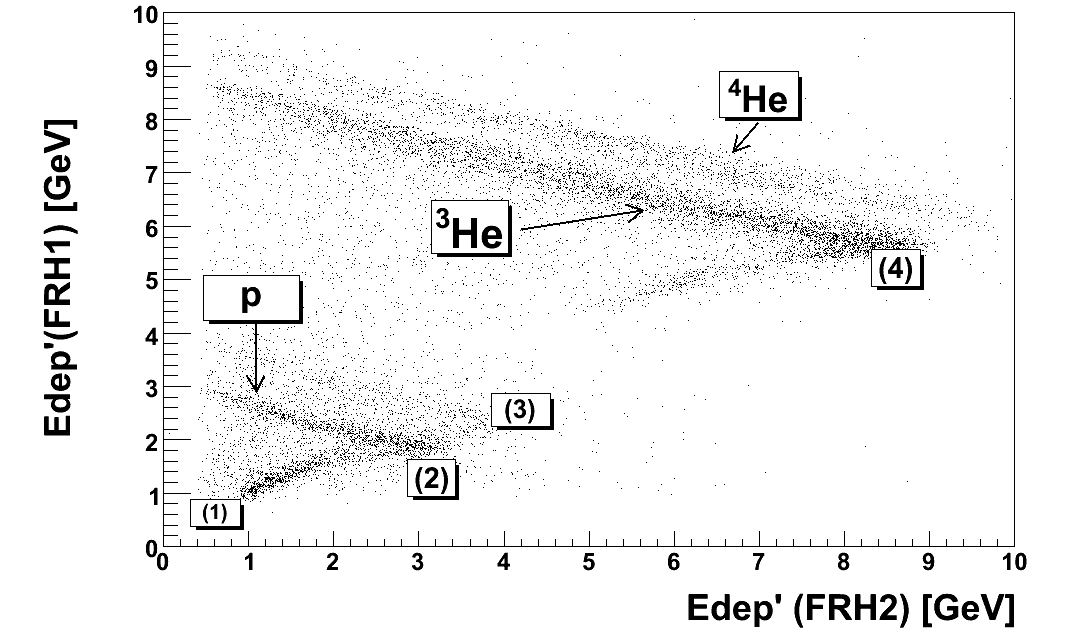}
         }
         \caption[Determination of nonlinearity corrections in one element of FRH]{Distribution of corrected ADC values from one element of the first layer of FRH (y-axis)
and adjacent element of the second layer of FRH (x-axis). The indicated characteristic points (1)...(4) are
explained in the text.}
         \label{first_layer_one_element_singlehe3_data_p}
\end{figure}

For determination of the relation between the corrected ADC values and the deposited energy
we compare some characteristic points visible in the experimental ADC spectra with results of simulations.
For this, we create two dimensional plots of the corrected ADC values for pairs of overlapping segments
from consecutive FRH layers.
Figure \ref{first_layer_one_element_singlehe3_data_p} shows an example of such  spectrum created
for a pair of segments from the FRH1 and FRH2 layer.
The characteristic `banana` bands corresponding to protons, deuterons,
${^3\mbox{He}}$ and ${^4\mbox{He}}$ ions are visible.
The following characteristic points are well defined in this plot:
\begin{itemize}
\item  minimum of ionization for protons (1),
\item  punch through point for protons (2),
\item  punch through point for deuterons (3),
\item  punch through point for ${^3\mbox{He}}$ (4).
\end{itemize}
The punch through points correspond to situation where particles are almost stopped
in the two scintillator layers.

\begin{figure}[!ht]
      \centering
      \scalebox{\scaleFactor}
         {
         \includegraphics{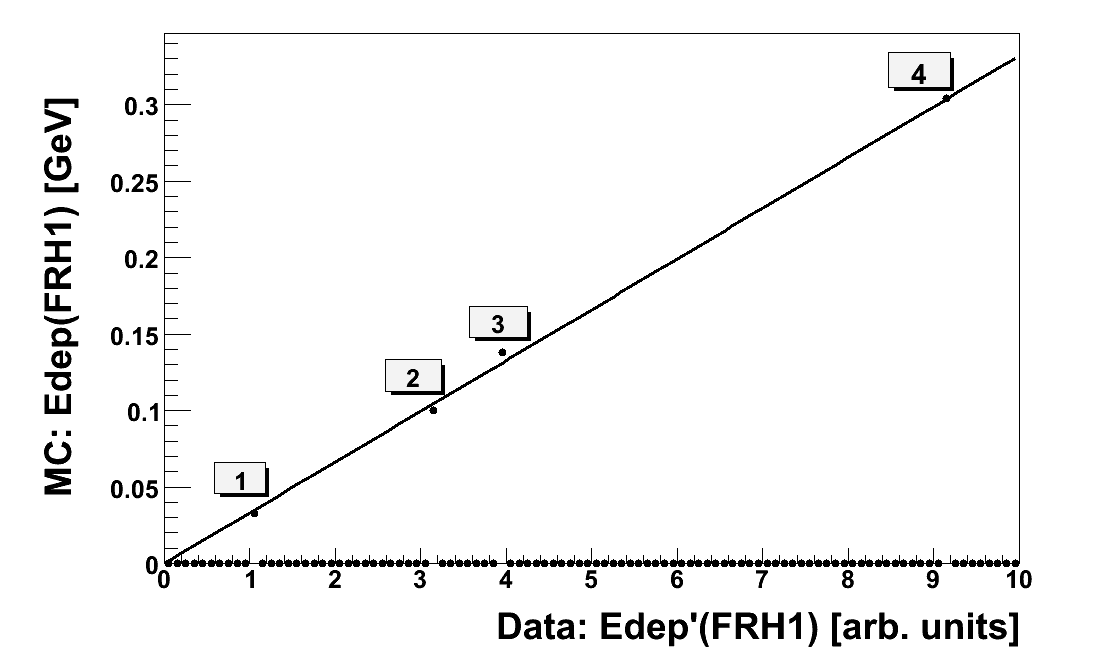}
         }
         \caption[MC vs corrected ADC values for an element of FRH]{Deposited energy obtained from MC simulations (y-axis)
versus corrected ADC value for the third element of the first layer of FRH (x-axis).
The marked points correspond to: (1) minimum ionizing point for protons,
(2) punch through point for protons,
(3) punch through point for deuterons,
(4) punch through point for ${^3\mbox{He}}$.
The fitted linear function represents the calibration relation.}
         \label{first_layer_third_element_p}
\end{figure}

The set of the characteristic points is compared to corresponding values obtained from the Monte Carlo simulations,
and a linear fit is then used as a calibration relation (see Fig.~\ref{first_layer_third_element_p}).
The deviations of the characteristic points from the linear fit are of about 5\,MeV and this value was
taken as an uncertainty of the energy calibration for the first three layers of the FRH.
In the case of the 4th layer, the punch through point  for the ${^3\mbox{He}}$ ions couldn't be determined
due to a very poor statistics. Therefore, the calibration, which was based on deuteron and proton bands only,
is less accurate than the ones for the first three layers.


\subsection{Scintillating Electromagnetic Calorimeter}
The main purpose of the Scintillating Electromagnetic Calorimeter is measurement of energies and emission angles
of photons and, therefore, the energy calibration is performed to optimize the reconstruction of these particles.
The initial calibration of the SEC crystals was done before the installation of the WASA detector at COSY, using cosmic muons and radioactive sources \cite{benedykt}\cite{sec}.
The standard calibration method which corrects possible variations of the gain of individual SEC detectors
is based on reconstruction of the invariant mass of neutral pions $\pi^{0}$ from two detected gammas
originating from the $\pi^{0} \rightarrow \gamma \gamma$ decay.
In the calibration process events with exactly two registered neutral particles are selected.
The particles are regarded as gammas and the invariant mass is calculated according to the following formula~\cite{redmer}:

\begin{equation}
\begin{split}
M_{\gamma1\gamma2}=\sqrt{(E_{\gamma1}+E_{\gamma2})^2 -(\vec{p}_{\gamma1} +\vec{p}_{\gamma2})^2}= \\
= \sqrt{2 \cdot k_{1} \cdot k_{2} \cdot E_{\gamma1}\cdot E_{\gamma2} \cdot(1-\cos(\theta_{1,2}))},
\end{split}
\end{equation}
where:$ E_{\gamma1}$ and$ E_{\gamma2}$ are the measured energies of the gammas
based on preliminary calibration constants,
$\vec{p}_{\gamma1}$ and $\vec{p}_{\gamma1}$ are their momentum vectors, $\theta_{1,2}$ is their opening angle
and $k_{1}$ and $k_{2}$ are calibration correction factors for two crystals with the highest energy deposited
by the pair of the gammas.
The deviation from the actual invariant mass of $\pi^{0}$ is used to optimize the values of the calibration factors
$k$ in an iterative way according the formula:
\begin{equation}
\begin{split}
k=\frac{M_{\pi^0}^2}{M_{\gamma\gamma}^2}.
\end{split}
\end{equation}

This procedure is applied to all crystals in the calorimeter. It corrects for differences in the gain
of individual detectors as well as for shower losses and border effects.

\subsection {Straw Tube Chambers}
The calibration of the Straw Tube Chambers FPC and MDC
means determination of the relation between
the measured drift time of electrons and  the track distance to anode wire.
This relation can be derived from the drift velocity and it depends on the gas mixture used in the chambers,
the voltage on the anode wires and the magnetic field.
The calibration is performed using the so called uniform irradiation method which is based on an assumption,
that the number of particles $dN$ passing through a single straw tube within
a distance interval $dr$ from the anode wire depends only on the width of this interval
and not on the distance to the wire.
This can be written as:
\begin{equation}
dN = c \cdot dr,
\end{equation}
where $c$ is a constant.
Division of both sides of the above equation by the drift time interval $dt$ corresponding to the drift distance $dr$
gives:
\begin{equation}
\frac{dN}{dt} = c \cdot \frac{dr}{dt} = c \cdot v(t),
\end{equation}
where $v(t)$ is the drift velocity being a function of the drift time $t$.
Integration of this equation over drift time interval from $t_{min}$ to $t$ gives:
\begin{equation}
\int_{t_{min}}^{t} \frac{dN}{dt'}dt' = \int_{t_{min}}^{t}c \cdot v(t') dt' = c [r(t)-r(t_{min})].
\label{dtint}
\end{equation}
The $t_{min}$ is chosen as a minimum drift time corresponding to drift distance equal to zero ($r(t_{min})=0$).
In this way the searched distance-drift time relation can be written as:
\begin{equation}
r(t)=\frac{\int_{t_{min}}^{t} \frac{dN}{dt'}dt'}{c}.
\label{rt}
\end{equation}
The constant $c$ can be calculated with Eq.~\ref{rt} using requirement that the maximum drift time $t_{max}$
corresponds to the radius of the straw tube $R$:
\begin{equation}
c = \frac{\int_{t_{min}}^{t_{max}} \frac{dN}{dt'}dt'}{R}.
\end{equation}

\begin{figure}[!ht]
      \centering
      \scalebox{0.5}
         {
         \includegraphics{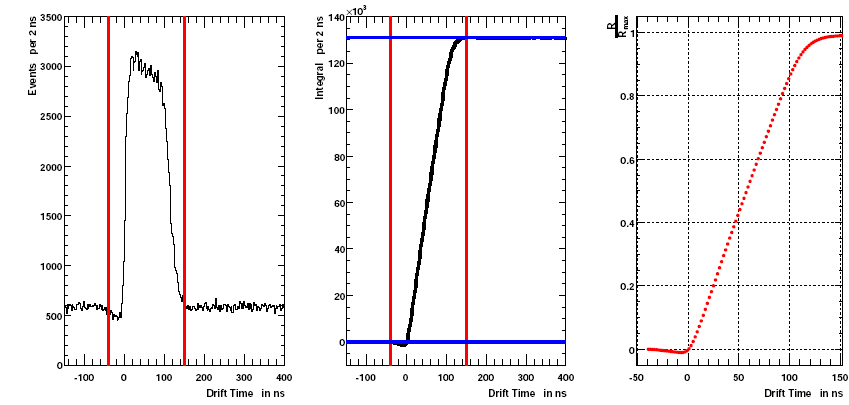}
         }
         \caption[Example of the MDC calibration drift time spectra.]{Drift time spectrum (left), integrated drift time spectrum after subtraction of background (middle)
and final drift time-distance relation (right).
The vertical red lines indicate the range of drift time taken for integration and
the blue horizontal lines indicate the range of integrals of the counts. The figure is adopted from~\cite{redmer}.}
         \label{drift_time_CFR_p}
\end{figure}

Left panel of Fig.~\ref{drift_time_CFR_p} presents drift time spectrum used for the calibration.
The middle panel presents result of integration of the drift time spectrum over the time interval
from $t_{min}$ to $t$ according Eq.~\ref{dtint}, and the right panel shows
the resulting distance-drift time relation $r(t)$.
As one can see from the figure, this relation is close to linear dependence
except for the region close to the wire and the one close to the straw tube wall.

\section{Analysis of the \texorpdfstring{${^3\mbox{He}} p \pi^{-}$}{3He p pi-} events}
\subsection{Scheme of analysis}

\begin{figure}[!hpt]
\begin{center}
      \scalebox{0.5}
         {
         \includegraphics{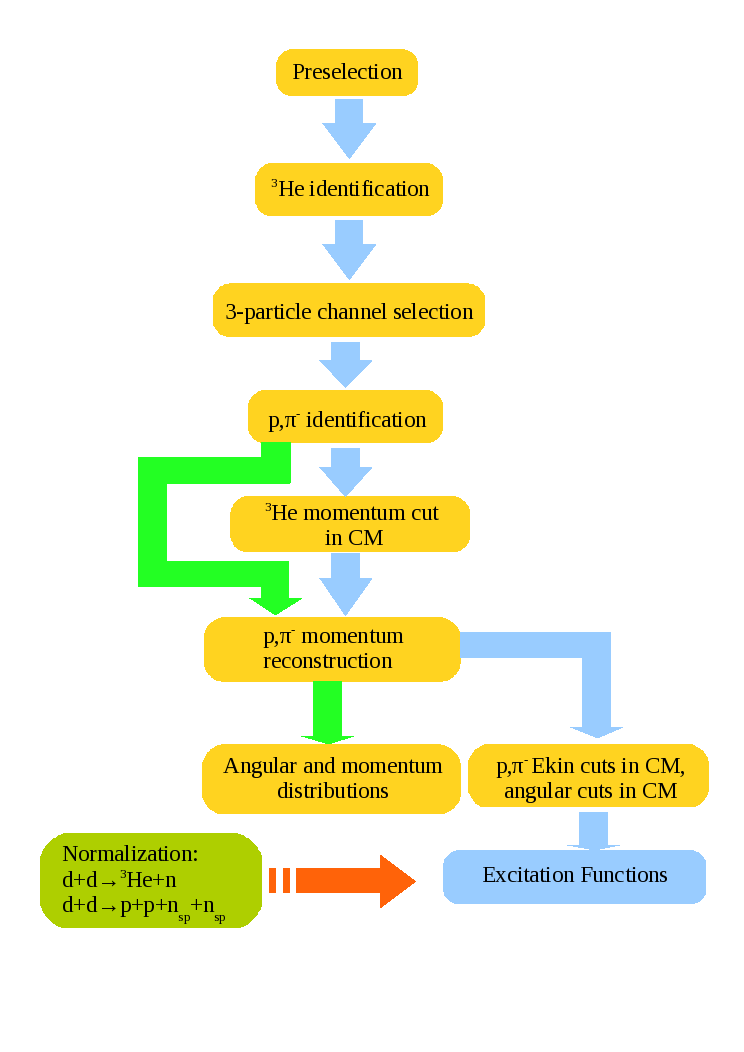}
	
  	 }
         \caption[Scheme of analysis.]{\label{analiza_schemat_p} Scheme of analysis of $dd \rightarrow {^3\mbox{He}} p \pi^{-}$ events leading to determination of excitation function as indicated with blue arrow as well as of the angular and momentum distributions
of the ejectiles marked with green arrows. Details of the presented scheme are explained in the text.}
\end{center}
\end{figure}

A general scheme of the analysis
of $dd \rightarrow {^3\mbox{He}} p \pi^{-}$
events is presented in Fig.~\ref{analiza_schemat_p}.
In the first step, a preselection of events was
performed by application of conservative cuts
in order to reduce the data sample and to speed up the analysis.
The next three steps aimed at selection of the $dd \rightarrow {^3\mbox{He}} p \pi^{-}$ events
and included (i) the identification of the ${^3\mbox{He}}$,
(ii) selection of three particles  final states and
(iii) $p, \pi^-$ identification.
For reconstruction of the $p$ and $\pi^-$ momenta which were not measured in the current experiment
we used the momentum conservation
combined with the information about the $p$ and $\pi^-$ scattering angles in the LAB system.
Selection of events associated with the decay of the ${^4\mbox{He}}-\eta$ bound state was performed
by setting an upper limit for the CM momenta of ${^3\mbox{He}}$ ions.
Additionally, appropriate cuts based on the simulations, were applied to the $p$ and $\pi^-$ kinetic energies
and the $p-\pi^-$ opening angle in the CM frame.
After normalization to the integrated luminosity, these events were used for determination
of the excitation function which is a basis of our search for the ${^4\mbox{He}}-\eta$ bound state.
For study of the background processes leading to the ${^3\mbox{He}} p \pi^{-}$ final state
we determined the angular and momentum distributions for the final state particles before application
the above-mentioned cuts on the ${^3\mbox{He}}$, $p$, $\pi^{-}$ kinematic variables.

The consecutive steps of the analysis are described in the next subsections.
The final results of the analysis including the excitation function and the angular and momentum distributions
are presented in the next chapter.

\subsection{Preselection of events}

\begin{figure}[!h]
\begin{center}
      \scalebox{\scaleFactor}
         {
         \includegraphics{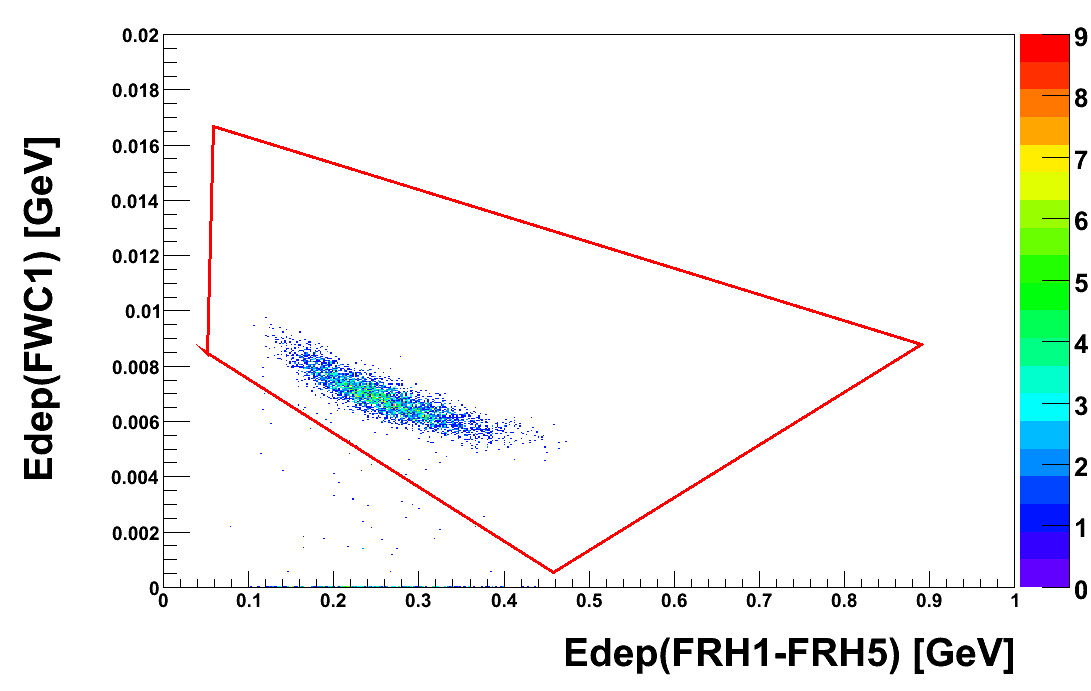}

         }
      \scalebox{\scaleFactor}
         {
         \includegraphics{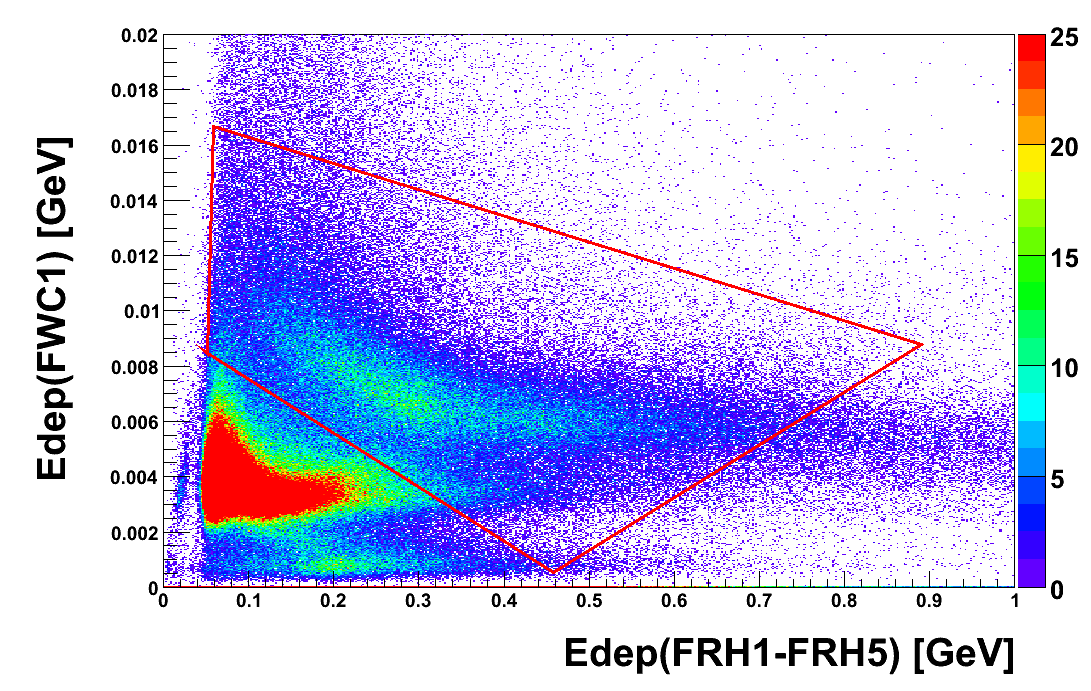}

         }
\caption[Helium selection based on the FWC energy losses]{\label{he3selection_p} Comparison of the Monte Carlo simulation (top) and experimental spectrum (bottom) of the energy loss in first layer of Forward Window Counter (y-axis) combined with energy deposited in all layers of Forward Range Hodoscope (x-axis). The preselection region comprising the ${^3\mbox{He}}$ ions
is limited by the red solid line.}
\end{center}
\end{figure}

In order to speed up the analysis, a preselection of  the raw data
was done using the following general conditions which are fulfilled by the ${^3\mbox{He}} p \pi^{-}$ events:
\begin{enumerate}
 \item exactly one track from a charged particle in the Forward Detector,
 \item exactly two tracks from charged particles  in the Central Detector,
 \item conservative graphical cut on two-dimensional $\Delta E-E$ plot representing the energy deposited in the first layer of the Forward Window Counter versus energy deposited in all layers of Range Hodoscope in order to reduce
     background from protons and charged pions  (see Fig.~\ref{he3selection_p}),
 \item conservative cut on the energy deposited in the first layer of the FRH: Edep(FRH1)$\in (0.05-0.4)$\,GeV
as expected for the ${{^3}\mbox{He}}$ ions,
\item scattering angle for tracks reconstructed in the Forward Straw Tracker in the range
$\theta_{FD} \in (3-18){^{\circ}}$ for eliminating tracks laying outside of the Forward Detector acceptance.
\end{enumerate}

%
%

\subsection{\texorpdfstring{$^{3}{\mbox{He}}$}{3He} identification}
\begin{figure}[!ht]
\begin{center}
      \scalebox{\scaleFactor}
         {
         \includegraphics{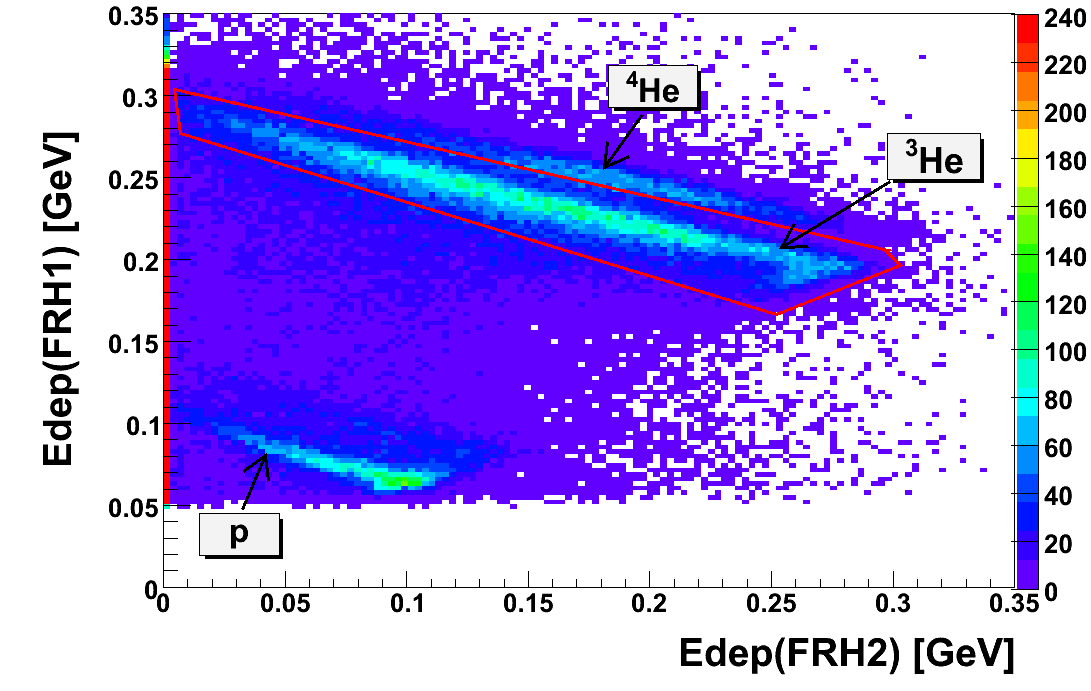}
  	 }

\caption[Helium selection based on the $\Delta E -\Delta E$ method in FRH.]{\label{he3selection_second_p} Experimental spectrum  of the energy losses in first two layers of the Forward Range Hodoscope. The selected area for $^{3}{\mbox{He}}$ ions is limited by the red line.
The empty area below 0.05\,GeV in the Edep(FRH1) distribution is due to the preselection cut.}
\end{center}
\end{figure}

Identification of the $^{3}{\mbox{He}}$ ions was carried out by means of the  $\Delta E- \Delta E$ techniques. According to the simulations of the reaction $dd \rightarrow ({^4\mbox{He}} \eta)_{bound} \rightarrow {^3\mbox{He}} p \pi^{-}$ , in 98\% of all cases the $^{3}{\mbox{He}}$ are stopped in the second layer of the FRH.
Therefore, we identify the $^{3}{\mbox{He}}$ ions on the basis of the energy losses in the first two layers of the FRH. As on can see in Fig.~\ref{he3selection_second_p}, a very clean separation of $^{3}{\mbox{He}}$ and protons
and even of $^{4}{\mbox{He}}$ ions is possible.

\subsection{Three-body channels}
\label{three_particle_coplanarity_cut_l}
The requirement of two reconstructed tracks in the Central Detector and one reconstructed track in the Forward Detector selects events from three-body channels like ${^3\mbox{He}} p \pi^{-}$ but it does not fully exclude events with more than three particles in the final state where some of particles are not detected. An example can be the  $d d \rightarrow {^3\mbox{He}} p \pi^{-} \pi^{0}$ reaction where the neutral pion
in the final state remains undetected.

In order to achieve a more restrictive selection of the three-body final states we
use the momentum conservation principle which can be written as:
\begin{equation}
\vec{p}_{beam}=\vec{p}_{He} +\vec{p}_{1}+\vec{p}_{2},
\label{pconserv}
\end{equation}
where $\vec{p}_{He}$ denotes the reconstructed momentum of the ${^3\mbox{He}}$ detected in the Forward Detector, and $\vec{p}_{1}$, $\vec{p}_{2}$ denote the momenta of two charged particles detected in the Central Detector.
This equation can be rewritten in a slightly changed form which defines a new vector $\vec{p}_k$ being a difference
of the beam momentum and the ${^3\mbox{He}}$ momentum vector:
\begin{equation} \label{coplanar_eq}
\vec{p}_{1} +\vec{p}_{2} = \vec{p}_{beam}-\vec{p}_{He} \equiv \vec{p}_k.
\end{equation}
The vectors $\vec{p}_k, \vec{p}_{1}, \vec{p}_{2}$ must lay in one plane 
and we use this requirement as a necessary condition for observation of the three-body final state.
For this, we define a measure of the coplanarity as an angle between the vector $\vec{p}_k$
and the cross product of the vectors $\vec{p}_{1}$ and $\vec{p}_{2}$:
\begin{equation}
\theta_{k,1x2} =  \angle(\vec{p}_k,\vec{p}_{1} \times \vec{p}_{2}).
\end{equation}
In the case of coplanarity $\theta_{k,1x2} = 90^{0}$.
An important advantage of the above definition is that
in order to calculate the angle $\theta_{k,1x2}$, one only needs to know the directions of the $\vec{p}_{1}$ and $\vec{p}_{2}$ vectors and not necessarily the magnitudes.
This is exactly our case, since due to the lack of the magnetic field in the Central Detector during the present
experiment, only the directions of the $\vec{p}_{1}$ and $\vec{p}_{2}$ vectors are measured in the MDC.

\begin{figure}[!ht]
      \centering
      \scalebox{\scaleFactor}
         {
         \includegraphics{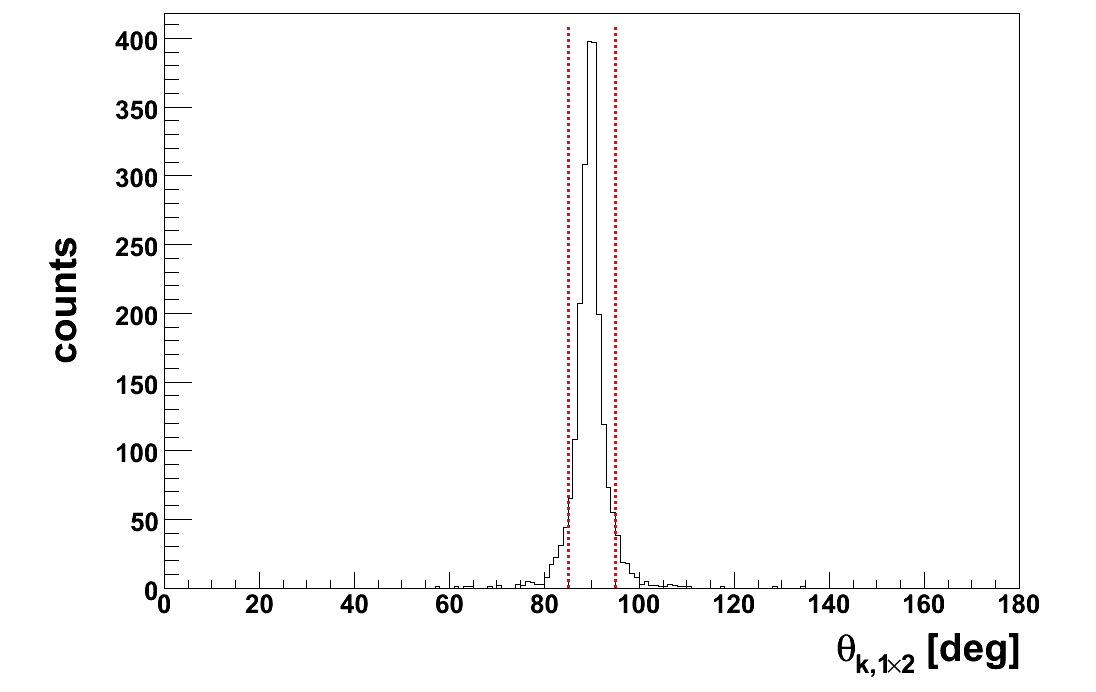}
	
  	 }
	 \scalebox{\scaleFactor}
         {
	 \includegraphics{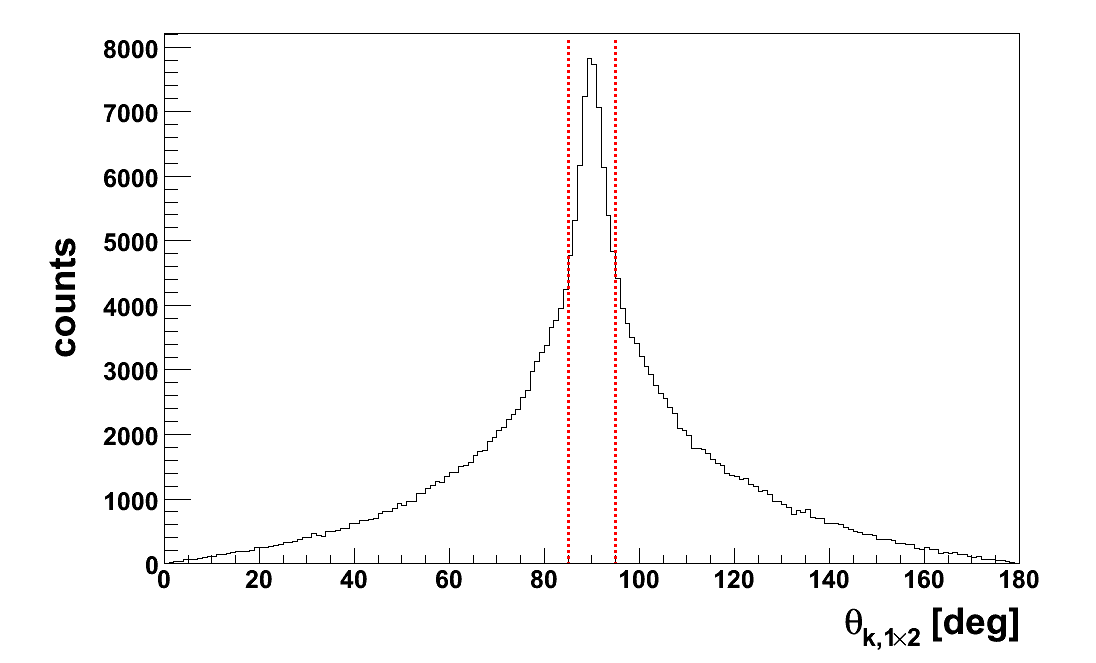}
	 }
         \caption[The coplanarity observable for 3-body reactions.]{Distribution of the coplanarity observable $\theta_{k,1x2}$ determined for the simulations of the $dd \rightarrow {^3\mbox{He}} p \pi^{-}$ reaction (upper panel)
         and for the experimental data (lower panel).
         The cut applied to the data - $\theta_{k,1x2} \in (90 \pm 5 ^{\circ})$ -
         is marked by the vertical dotted lines.}
         \label{coplanarityCut_data_p}
\end{figure}

The upper panel of Fig.~\ref{coplanarityCut_data_p} shows distribution of the angle $\theta_{k,1x2}$
obtained in simulations of the $dd \rightarrow {^3\mbox{He}} p \pi^{-}$ reaction.
The smearing of the angle around the expected value of $90^{\circ}$ is due to uncertainties
of reconstruction of particle tracks.
The experimental distribution of  $\theta_{k,1x2}$ presented in the lower part of Fig.~\ref{coplanarityCut_data_p}
has much longer tails which presumably originate from final states with more than three particles.

Based on the MC simulations we apply a cut $\theta_{k,1x2} \in (90 \pm 5 ^{\circ})$ which reduces
the background coming from reactions with more than three particles in the final state.
It eliminates also some part of three-particle events with the reconstructed vectors being far from coplanarity
due to experimental uncertainties.
The last case is important in the context of the $p$ -$\pi^{-}$ momentum determination method which is based on the assumption that the vectors are coplanar as it is discussed in the further part of the present chapter.
Elimination of events which are far from coplanarity essentially improves the resolution
of the momentum reconstruction.

\subsection{p and \texorpdfstring{$\pi^{-}$}{pi-} identification}

\begin{figure}[!ht]
\begin{center}

      \scalebox{\scaleFactor}
         {
         \includegraphics{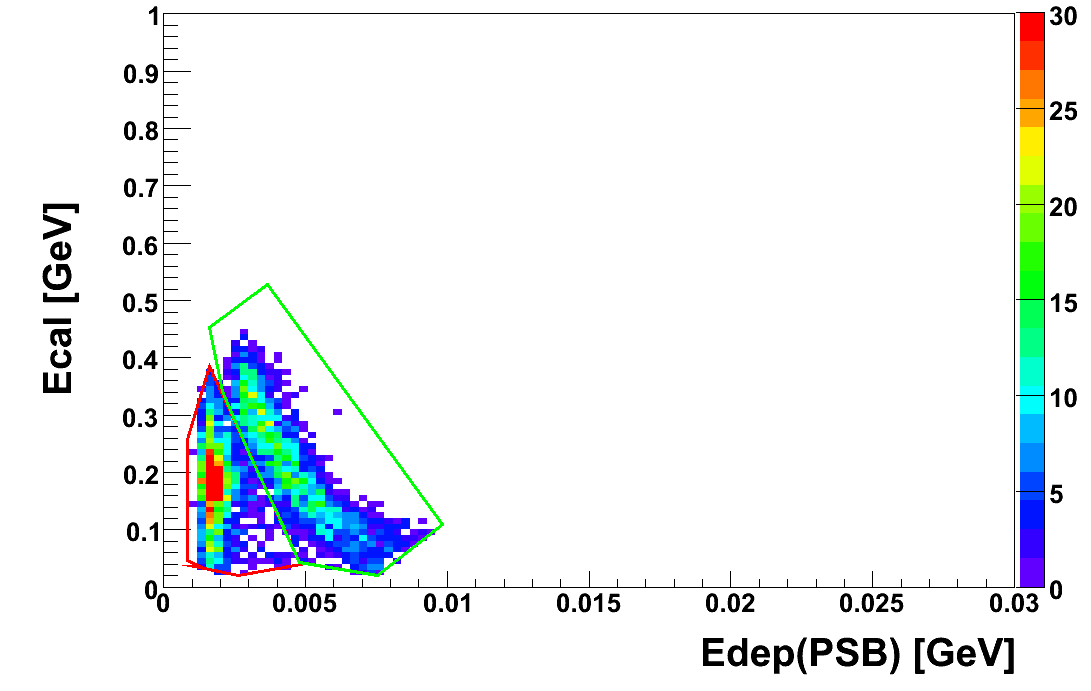}
	
  	 }
      \scalebox{\scaleFactor}
         {
         \includegraphics{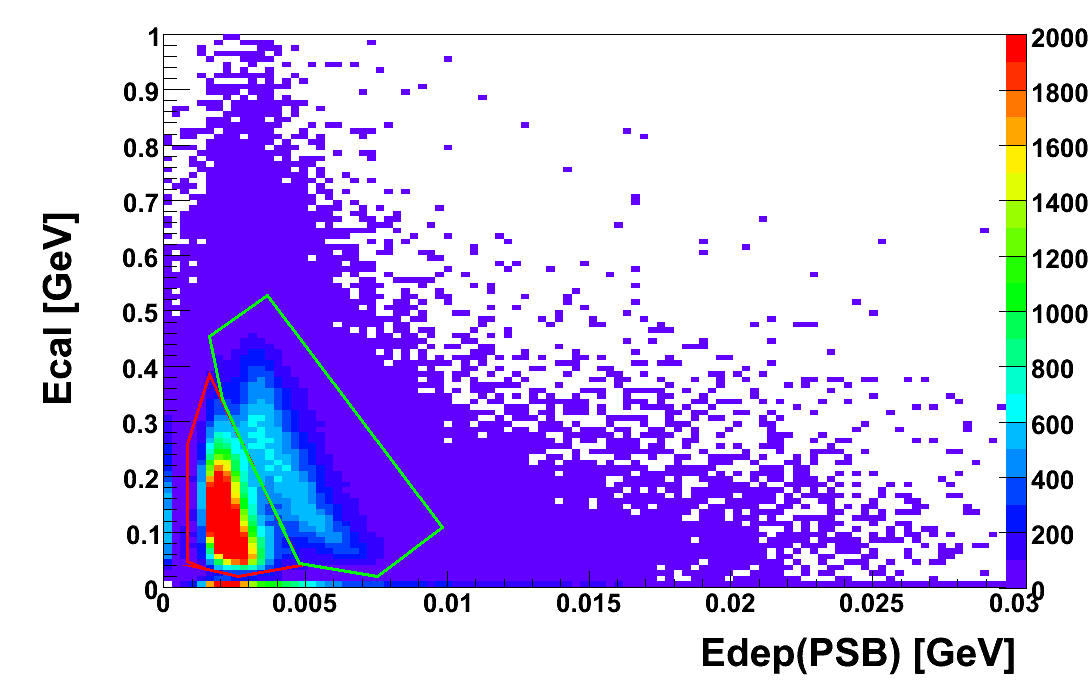}
	
  	 }
\caption[Identification of $p$ and $pi^{-}$.]{\label{ecal_vs_ps_p}Comparison of the Monte Carlo simulation (upper plot) and the experimental spectrum (bottom plot) of the energy loss in the Plastic Scintillator Barrel ($x$-axis) combined with the energy deposited in the Electromagnetic Calorimeter ($y$-axis). The green and red curve represents
the applied graphical cut to separate the protons and pions.}
\end{center}
\end{figure}

The standard method of identification of charged particles in the Central Detector is based on the measurement of
energy loss in the Plastic Scintillator Barrel and of momentum in the MDC ($\Delta E-p$ method).
However, during the present experiment the cooling of the solenoid was broken and, therefore, this method cannot be applied. Instead, the energy loss in the Plastic Scintillator Barrel was combined with the energy deposited in the Electromagnetic Calorimeter to identify protons and pions ($\Delta E-E$ method).
In the experimental $\Delta E-E$ spectrum pions and protons are well separated which is in line with the simulations
(see  Fig.~\ref{ecal_vs_ps_p}).

\subsection{p and \texorpdfstring{$\pi^-$}{pi-} momentum reconstruction}

As it was already mentioned, the failure of the solenoid cooling excluded the possibility of
a direct momentum determination for the charged particles registered in the Central Detector.
However, for the analysed reaction $dd \rightarrow {^3\mbox{He}} p \pi^{-}$ the momentum of the pions and protons
can be calculated using momentum conservation and the information about their emission angles measured
in the MDC.

\begin{figure}[!ht]
\begin{center}
  \scalebox{0.6}
  {
    \includegraphics{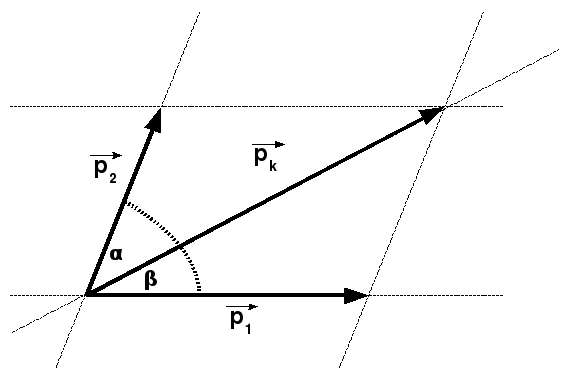}
  }
\caption[ $p$ and $\pi^{-}$ momentum reconstruction]{\label{sum_p} Relation between vectors and angles used
in the reconstruction of the $\pi^{-}$ and $p$ momenta.}
\end{center}
\end{figure}

The reconstruction of the $p$ an $\pi^{-}$ momenta can be reduced to a simple geometrical problem
of determination of the $\vec{p}_{1}$ and $\vec{p}_{2}$ magnitudes, assuming that we know their directions
and that the sum $\vec{p}_{1}+ \vec{p}_{2}$ is equal to a known vector $\vec{p}_k$ (see Fig.~\ref{sum_p}).

The magnitudes of $\vec{p}_{1}$ and $\vec{p}_{2}$ vectors are calculated using the relations
following from the geometrical law of sines:
\begin{equation}
 |\vec{p}_{1}|=|\vec{p}_k| \times \frac{\sin \beta}{\sin(\alpha + \beta)},
 |\vec{p}_{2}|=|\vec{p}_k| \times \frac{\sin \alpha}{\sin( \alpha + \beta)},
\end{equation}
where the angles $\alpha$ and $\beta$ are defined in Fig.~\ref{sum_p}.

In the calculations we implicitly assume that the vectors  $\vec{p}_{1}$, $\vec{p}_{2}$ and $\vec{p}_k$ are co-planar.
However, due to experimental uncertainties the directions of these vectors reconstructed on the basis of the data
from the tracking detectors might deviate from coplanarity. The applied cut described in the section~\ref{three_particle_coplanarity_cut_l} eliminates most of the  non-coplanar events and therefore improves the resolution.

\begin{figure}[!ht]
\begin{center}
  \scalebox{\scaleFactor}
  {
    \includegraphics{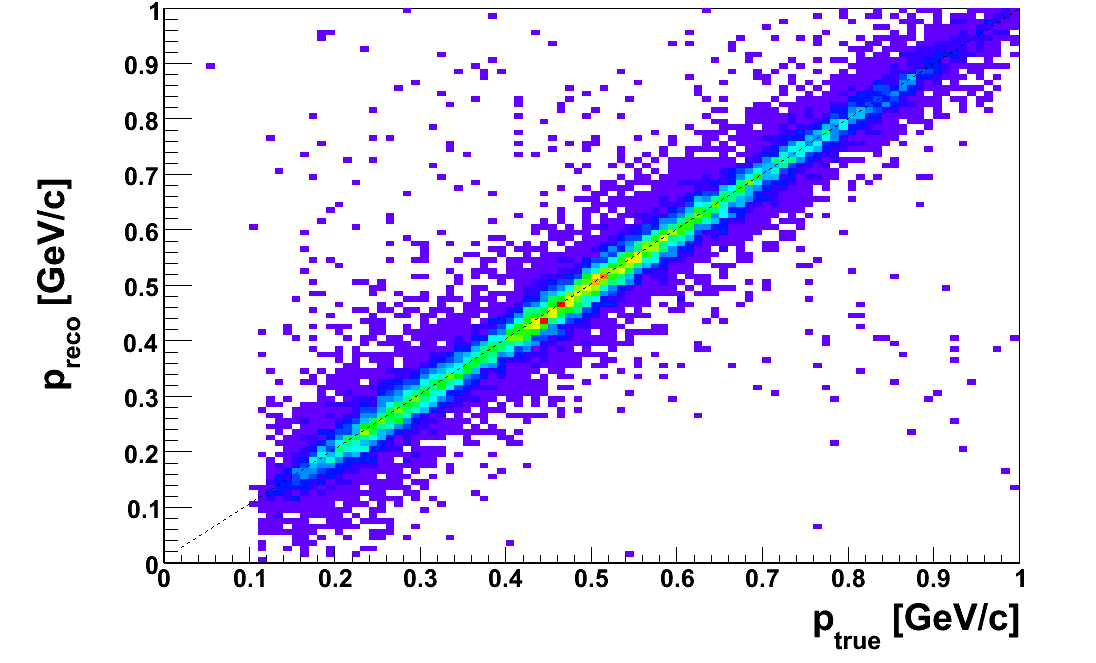}
  }
  \scalebox{\scaleFactor}
  {
  \includegraphics{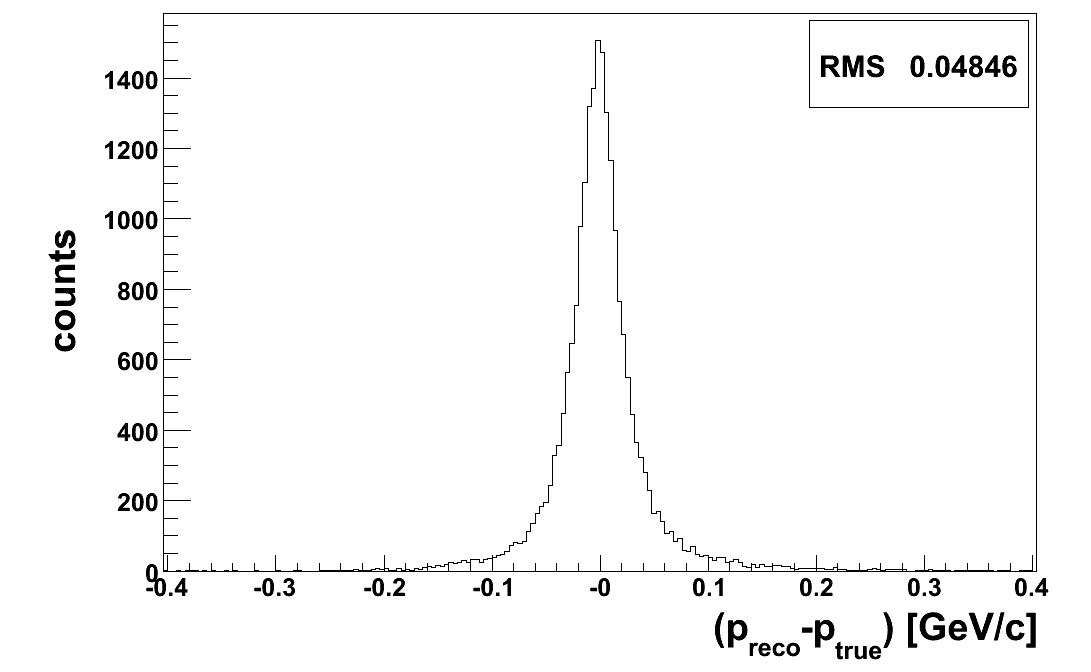}
  }
\caption[Test of the momentum reconstruction method of $p$ and $\pi$]{\label{momentum_reco_diff_true_LAB_direct_p} Distribution of reconstructed versus true values of  $p$ and $\pi$ momenta obtained for simulated events (top panel). Difference between reconstructed and true values of $p$ and $\pi$ momenta in the LAB system (bottom panel).}
\end{center}
\end{figure}

In order to estimate the precision of the presented method of the pion and proton momentum reconstruction,
we applied it to simulated $dd \rightarrow ({^4\mbox{He}} \eta)_{bound} \rightarrow {^3\mbox{He}} p \pi^{-}$ events.
Fig.~\ref{momentum_reco_diff_true_LAB_direct_p} presents difference between the reconstructed and generated momenta.
The obtained distribution is symmetric around zero, which means that the applied method does not introduce any shift in the reconstructed momenta, and the resolution can be estimated from the RMS value as equal to about 50\,MeV/c in the LAB frame.

\subsection{Cut on \texorpdfstring{$^3\mbox{He}$}{3He} momentum}
\label{cut_he3_momentum_sect}

\begin{figure}[!ht]
\begin{center}
  \scalebox{\scaleFactor}
  {
    \includegraphics{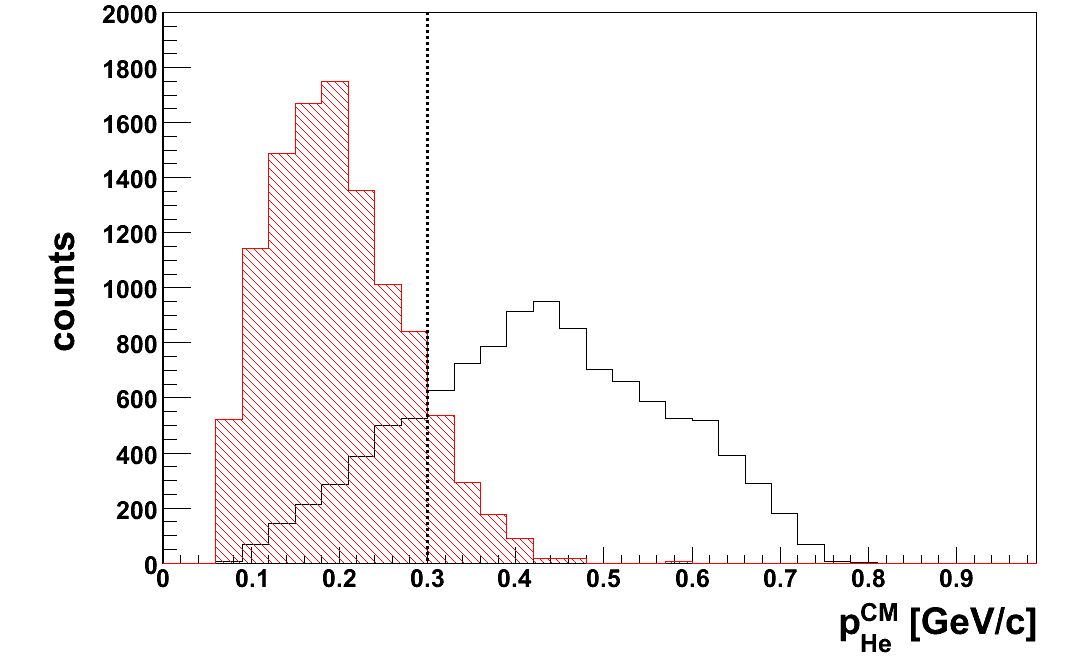}
  }
\caption[Helium momentum distribution in the CM frame (MC).]{\label{Momentum_he3_CM_comparision_direct_signal_v2_p} Distribution of the ${^3\mbox{He}}$ momentum seen in the CM system as simulated for the processes leading to the creation of the ${^4\mbox{He}}\eta$ bound state:  $dd\to({^4\mbox{He}}\eta)_{bound}\to{^3\mbox{He}}p\pi^{-}$ (red area)
and for the direct production $dd\to^3{\mbox{He}}p\pi^{-}$ (solid line). The blue dashed line shows the selection cut
applied in our analysis.}
\end{center}
\end{figure}

In the reaction  $dd \rightarrow ({^4\mbox{He}}-\eta)_{bound} \rightarrow {^3\mbox{He}} p \pi^{-}$
the $^3\mbox{He}$ plays the role of a spectator (see \ref{basic_idea_l}) and is expected to move with low momenta
in the CM frame corresponding to the Fermi momenta in the $^4\mbox{He}$ nucleus.
Hence, we use this feature to make a distinction between two regions in the CM momentum distribution: the low momentum region (up to 0.3 GeV), where the signal from the bound state decay is expected and the high momentum region
where the direct production $dd \rightarrow {^3\mbox{He}} p \pi^{-}$ dominates
(see Fig.~\ref{Momentum_he3_CM_comparision_direct_signal_v2_p}).

\begin{figure}[!ht]
\begin{center}
  \scalebox{\scaleFactor}
  {
    \includegraphics{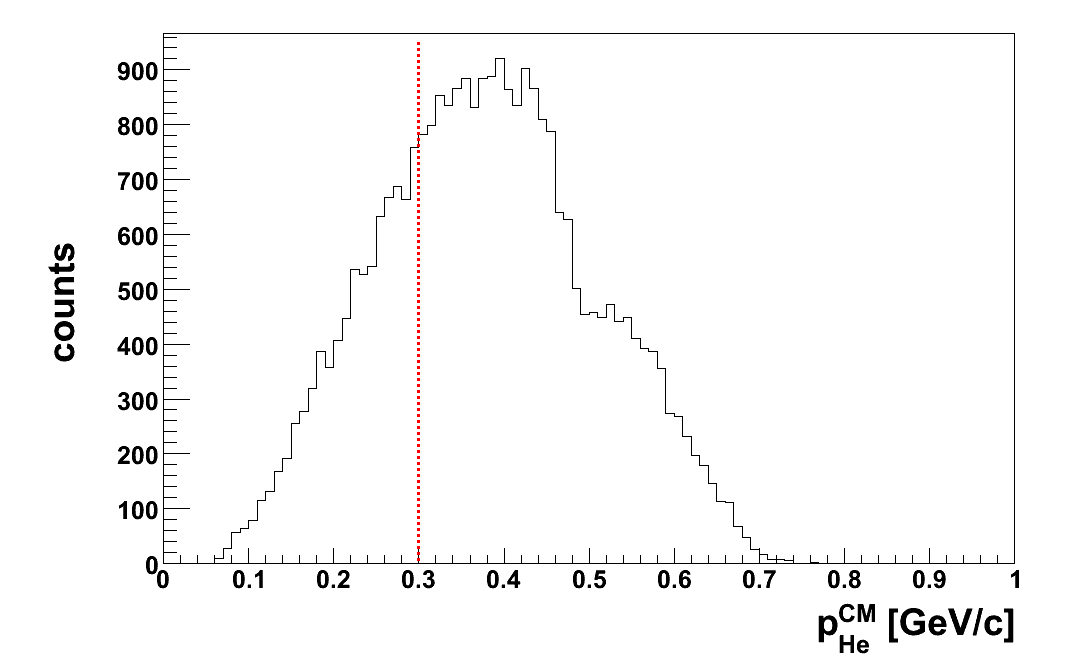}
  }
\caption[Helium momentum distribution in the CM frame (data).]{\label{Momentum_he3_CM_exp_p} Experimental distribution of the ${^3\mbox{He}}$ momentum seen in the CM system.  The red dashed line shows the selection cut.}
\end{center}
\end{figure}

The experimental distribution of the $^3\mbox{He}$ momenta in the CM system is shown in Fig.~\ref{Momentum_he3_CM_exp_p}. In the region below 0.3\,GeV/c, we see a monotonic rise
of the counts as a function of the momentum.
If the process $dd\to(^4{\mbox{He}}\eta)_{bound}\to^3{\mbox{He}}p\pi^{-}$
was dominating in this region then a structure similar to the Fermi momentum distribution
should appear with a maximum at around 0.2\,GeV
as visible for the Monte Carlo simulations.

\subsection{Cut on \texorpdfstring{$p$}{p} and \texorpdfstring{$\pi^{-}$}{pi-} opening angle and kinetic energies}
\label{cut_ekin_sec}
\begin{figure}[!ht]
\begin{center}
  \scalebox{\scaleFactor}
  {
    \includegraphics{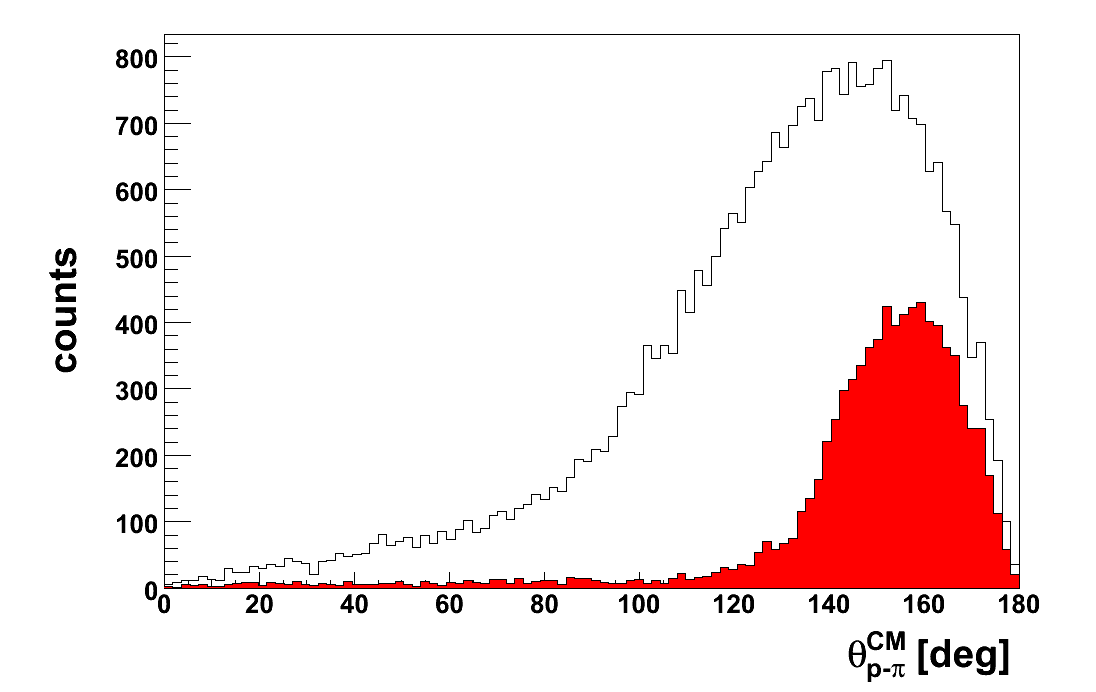}
  }
\caption[Opening angle in CM before and  after momentum cut.]{\label{opening_angle_cm_exp_comparison_momcut_v2_p} Experimental distribution of the  $p-\pi^{-}$ opening angle in CM frame before (white area) and after (red area) the selection of the low momentum $^3\mbox{He}$.}
\end{center}
\end{figure}

Fig.~\ref{opening_angle_cm_exp_comparison_momcut_v2_p} shows experimental distribution
of the $p-\pi^{-}$ opening angle in CM frame $\Theta_{p-\pi}^{CM}$ before and after the selection
of low momentum region of the $^3\mbox{He}$ ions.
The choice of the low momentum $^3\mbox{He}$ is strongly correlated with selection of high opening angles $\Theta_{p-\pi}^{CM}$.
In spite of it, we apply a cut on the opening angle corresponding to the region of  $(140^{\circ}-180^{\circ})$
as expected on the basis of the simulations (see Fig.~\ref{proton_pion_openAngleCM_p}).
This cut removes a small amount of events with the opening angle below $140^{\circ}$.

\begin{figure}[!hpt]
\begin{center}
  \scalebox{\scaleFactor}
  {
    \includegraphics{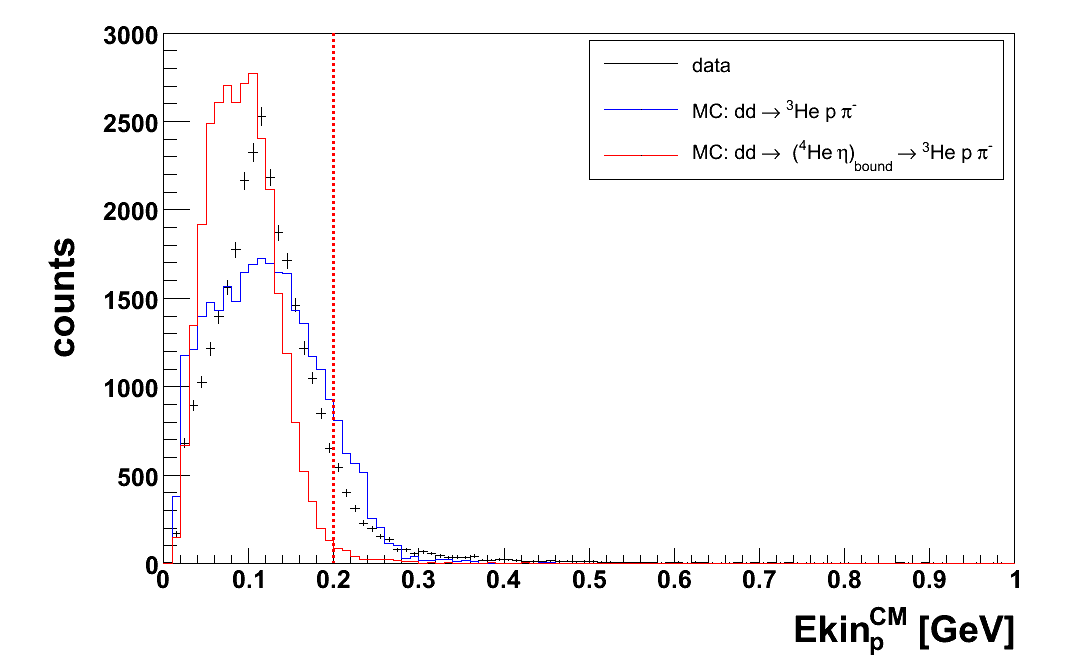}
  }
  \scalebox{\scaleFactor}
  {
    \includegraphics{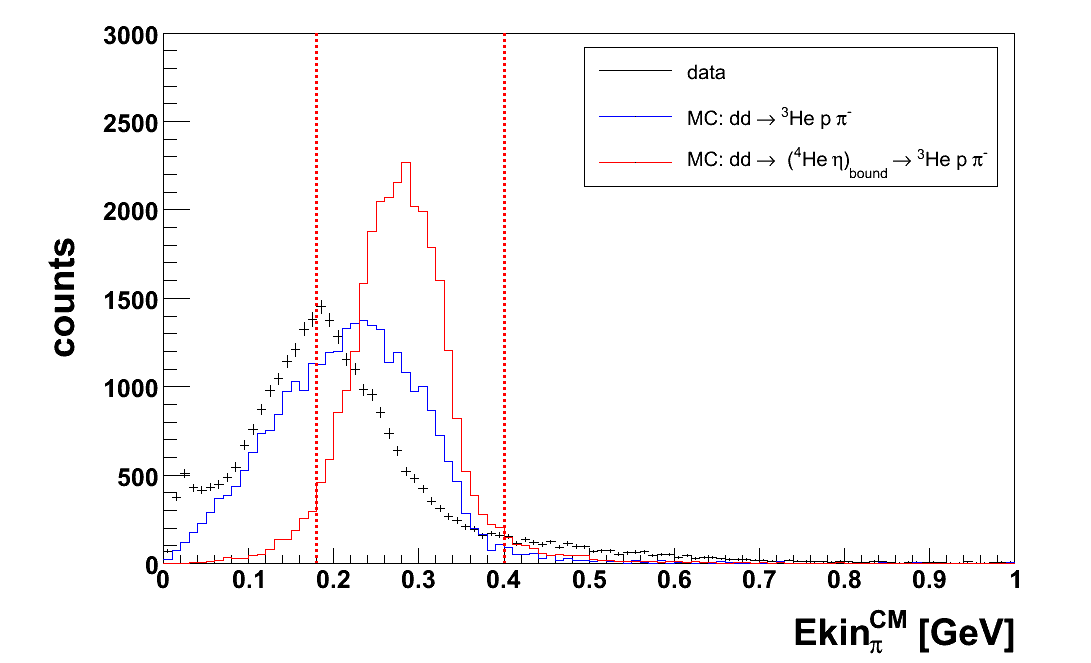}
  }
\caption[Kinetic energy distributions for $p$ and $\pi^{-}$ in the CM frame.]{\label{ekin_p_pim_p}Distribution of kinetic energy of $p$ (top plot) and $\pi^{-}$ (bottom plot) in the CM frame obtained from experiment and from the MC simulations. The red dashed line indicates a boundary of the applied cuts: ($Ekin^{CM}_{p} < 200$ MeV, $Ekin^{CM}_{\pi^{-}} \in (180,400)$ MeV).}
\end{center}
\end{figure}

An additional cut, which was used to improve  selection of events
from the  $dd \rightarrow (^4\mbox{He}\eta)_{bound} \rightarrow ^3\mbox{He} p \pi^{-} $ reaction,
was applied to the $p$ and $\pi^- $ kinetic energies in the CM system.
These kinetic energies  are released in the process of the $\eta$-meson conversion on a neutron
leading to creation of the $p-\pi^-$ pair and result from the mass difference $m_{\eta}-m_{\pi}$.
The kinetic energy spectra of protons originating from the bound state decay and
from the direct production are similar and are also close to the experimental distributions
(see Fig.~\ref{ekin_p_pim_p}-upper plot).
Therefore, the applied cut: $Ekin^{CM}_{p} < 200$\,MeV is not efficient in reduction of experimental background.
Larger differences are observed for the distribution of the pion kinetic energy which has a maximum at around
300\,MeV  for the bound state simulations compared to about 220\,MeV in the case of the direct production
(see Fig.~\ref{ekin_p_pim_p}-lower plot).
The applied cut: $Ekin^{CM}_{\pi^{-}} \in (180,400)$ MeV rejects more than 50\% of experimental events lying outside
the selected region.

\section{Luminosity determination}
\label{ch:luminosity}
The present measurements were performed using
the ramped beam technique. During an acceleration cycle the luminosity could vary due to beam losses
caused by the interaction with the target and with the rest gas in the accelerator beam line, as well as
due to displacement of the beam at the target position correlated with the variation of the momentum.
For this reason it was necessary to determine not only the total integrated luminosity
but also  its dependence on the beam momentum.
A precise knowledge of this dependence is important
for an accurate relative normalization of data points from the studied
excitation function of the reaction $dd \rightarrow {^3\mbox{He}} p \pi^-$,
since too large normalization uncertainties
can mask the searched signal from the decay of the ${^4\mbox{He}}-\eta$ bound state.
It is also essential, that the ${^4\mbox{He}}-\eta$ bound state does not decay
or decays with negligible probability in the channel used for the luminosity determination.
Otherwise, the normalization procedure can cancel the searched signal partly or, in the worst case, fully.

In the present experiment, the total integrated luminosity was determined by comparison
of experimental counts from the $dd \rightarrow {^3\mbox{He}} n$ reaction with corresponding
cross-sections taken from literature.
Because of the limited statistics of events registered for this reaction, the relative normalization of
luminosities corresponding to individual beam momentum intervals, was based on the quasi-elastic
proton-proton scattering which was observed with essentially higher rate.
In our analysis we assumed, that the energy dependence of the cross-sections for the quasi-elastic
proton-proton scattering in the deuteron-deuteron collisions is proportional
to the elastic proton-proton cross-sections.

In the following three subsections we respectively present:
basic formulas applied for the luminosity calculations,
the procedure for the determination of the integrated luminosity
and the calculations of integrated luminosities for individual beam momentum intervals.

\subsection{Basic formulas}

In scattering experiments, number of registered events per unit of time for a given process
can be expressed as:
\begin{equation} \label{eq:R}
\frac{dN}{dt}(t)= L(t) \cdot \sigma \cdot \varepsilon  ,
\end{equation}
where $\sigma$ is a cross-section for the process, $\varepsilon$ is a detection efficiency
and $L(t)$ is an instantaneous luminosity.
In experiments with a fixed target like WASA-at-COSY, the luminosity is equal to the product of
areal density of the target and the number of incident beam particles per unit time.
The detection efficiency $\varepsilon$ includes geometrical acceptance of the detector, efficiency
of detection system for registration of the investigated events and efficiency
of reconstructing events during the analysis.

The total number of events $N$ registered during the measurement can be obtained by
integrating  both sides of the Eq.~(\ref{eq:R}) over the time of measurement:
\begin{equation} \label{eq:N}
N=\mathcal{L} \cdot \sigma \cdot \varepsilon  ,
\end{equation}
where $\mathcal{L}=\int_{t1}^{t2} L(t) dt$ is the integrated luminosity.

In the case of the differential cross-section $\frac{d\sigma}{dx}$, the Eq.~(\ref{eq:N})
can be modified to:
\begin{equation} \label{eq:dN}
 \frac{dN}{dx}(x)= \mathcal{L} \cdot \frac{d\sigma}{dx}(x) \cdot \varepsilon(x) .
\end{equation}

The integrated luminosity can be determined on the basis of the measured counts $\frac{dN}{dx}$
for a process for which the differential cross-section is known:
\begin{equation} \label{eq:L}
\mathcal{L} = \frac{\frac{dN}{dx}(x)}{\frac{d\sigma}{dx}(x) \cdot \varepsilon(x)} .
\end{equation}

For the luminosity determination we use angular distributions of the cross-section
and, therefore, we substitute $x=\cos\theta$, $dx=d(\cos\theta)=d\Omega/2\pi$.
The previous equation takes the following form:
\begin{equation} \label{eq:Lth}
\mathcal{L} = \frac{\frac{dN}{d\Omega}(\cos\theta)}{\frac{d\sigma}{d\Omega}(\cos\theta)
\cdot \varepsilon(\cos\theta)} .
\end{equation}


The present measurements were performed with a ramped beam
and the integrated luminosity was calculated for each beam momentum interval separately.
For this purpose we used Eq.~(\ref{eq:Lth}) reformulated in the following way:
\begin{equation} \label{eq:Lthi}
\mathcal{L}_i = \frac{\frac{dN_i}{d\Omega}(\cos\theta)}{\frac{d\sigma}{d\Omega}(\cos\theta,p_i)
\cdot \varepsilon(\cos\theta,p_i)} ,
\end{equation}
where index $i$ numerates the beam momentum intervals and $p_i$ is the central momentum for $i$'th interval.

For determination of the integrated luminosities $\mathcal{L}_i$ we use the quasi-elastic
proton-proton scattering.
In our analysis we assume, that the cross-section for this process
is proportional to
the cross-section for the elastic proton-proton scattering:
\begin{equation} \label{eq:sq}
 \frac{d\sigma}{d\Omega}(\cos\theta,p_i)= c \cdot \frac{d\sigma^{elast}}{d\Omega}(\cos\theta,p_i) ,
\end{equation}
and $c$ is a proportionality coefficient which can be interpreted in terms of
shadowing of the protons by the neutrons in the deuteron-deuteron collisions.
The integrated luminosity for the $i$'th beam momentum interval can be written as:
\begin{equation} \label{eq:Lthic}
\mathcal{L}_i = \frac{\frac{dN_i}{d\Omega}(\cos\theta)}{c\cdot\frac{d\sigma^{elast}}{d\Omega}(\cos\theta,p_i)
\cdot \varepsilon(\cos\theta,p_i)} =\frac{1}{c}l_i.
\end{equation}
The quantities $l_i$ are proportional to the integrated luminosities $\mathcal{L}_i$ and are calculated
using the elastic proton-proton cross-sections.

The total integrated luminosity $\mathcal{L}$ can be expressed as a sum of partial luminosities
$\mathcal{L}_{i}$ over all momentum intervals:
\begin{equation} \label{eq:LumSum}
\mathcal{L}=\sum_{\substack{i}} \mathcal{L}_{i} .
\end{equation}
The unknown constant $c$ is calculated from the integrated luminosity $\mathcal{L}$ which was determined
on the basis of measurements of the $dd \rightarrow {^3\mbox{He}} n$ reaction:
\begin{equation} \label{eq:const}
c = \frac{\sum_{\substack{i}} l_{i}}{\mathcal{L}}.
\end{equation}

\subsection{Integrated luminosity}

\subsubsection{Cross-section for $dd \rightarrow {^3\mbox{He}} n$ reaction }
\label{param_l}
In order to determine the absolute value of the integrated luminosity
we used the experimental data on the $dd \rightarrow {^3\mbox{He}} n$
cross-sections measured at SATURNE for four beam momenta between 1.65
and 2.49 GeV/c \cite{bizard}.
As shown in Ref.\ \cite{bizard} the experimental cross-sections can be
very well described by a sum of three exponential functions of the
four momentum transfer $t$:
\begin{equation} \label{eq:sigt}
\frac{d\sigma}{dt}=\sum_{i=1}^{3}a_i \cdot e^{b_{i}(t-t_{max})} ,
\end{equation}
where $a_i$, $b_i$ are fit parameters and $t_{max}$ denotes the maximum kinematical value of $t$ out of the set of measurements for a given beam energy at SATURNE.

In our analysis we used a parametrization of the coefficients $
a_i$, $b_i$ with a hyperbolical function
of the total energy $\sqrt{s}$ as proposed in Ref.\ \cite{pricking}:
\begin{equation} \label{eq:hyperbol}
a_i(\sqrt{s}), b_i(\sqrt{s}) =\frac{k_{i}}{\sqrt{s}-q_i} +r_i.
\end{equation}
The parameters $k_i$, $q_i$, $r_i$ obtained from a fit to the SATURNE data
are listed in Table~\ref{tab:param}.
\renewcommand{\arraystretch}{1.2}
\begin{table}[hbt]

 \begin{center}
 \begin{tabular}{|c|ccc|}
    \hline
    &$k_i$ & $q_i$ & $r_i$ \\
    \hline
    $a_{1}$ &11.64 & 4.05 & -14.49 \\
    $b_{1}$ &0.78 & 3.92 & 9.04\\
    $a_{2}$ &2327.04 & -1.44 & -3.99.27\\
    $b_{2}$ &0.78 & 3.92 & 9.04\\
    $a_{3}$ &0.22 & 4.08 & 1.24\\
    $b_{3}$ &0.78 & 3.92 & 9.04\\
    \hline
 \end{tabular}
 \caption[Parameters of the total cross-section]{Parameters for the total energy $\sqrt{s}$ and
the four momentum transfer $t$ dependence of the cross-section
for the reaction $dd \rightarrow {^3\mbox{He}} n$.
In the applied parametrization,
the $\sqrt{s}$ and $t$ values are expressed in GeV and (GeV/c)$^2$, respectively,
and the cross-section is given in $\mu$b/(GeV/c)$^2$.}
\label{tab:param}
\end{center}
\end{table}

\begin{figure}[!htp]
\begin{center}
      \scalebox{\scaleFactor}
      {
         \includegraphics{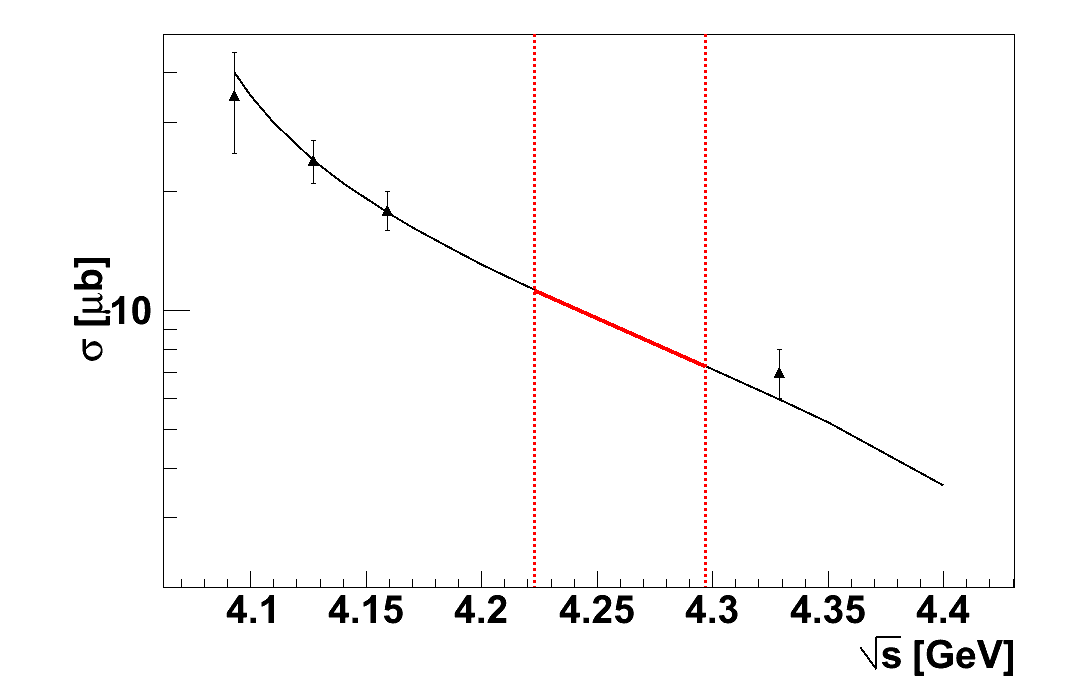}
      }
\caption[Parametrization for the total cross-section for  the $dd \rightarrow {^3\mbox{He}} n$ reaction based on the SATURNE data]{\label{SATURNEparam_p} Total cross-section as function of energy $\sqrt{s}$ available in the CM frame . The solid black line presents the applied parametrization. The SATURNE data are marked as black triangles. The red dashed lines indicate the range of the energy covered by our experiment ($\sqrt{s}\in[4.223, 4.297] $ GeV). }
\end{center}
\end{figure}

\begin{figure}[!htp]
\begin{center}
      \scalebox{\scaleFactor}
      {
         \includegraphics{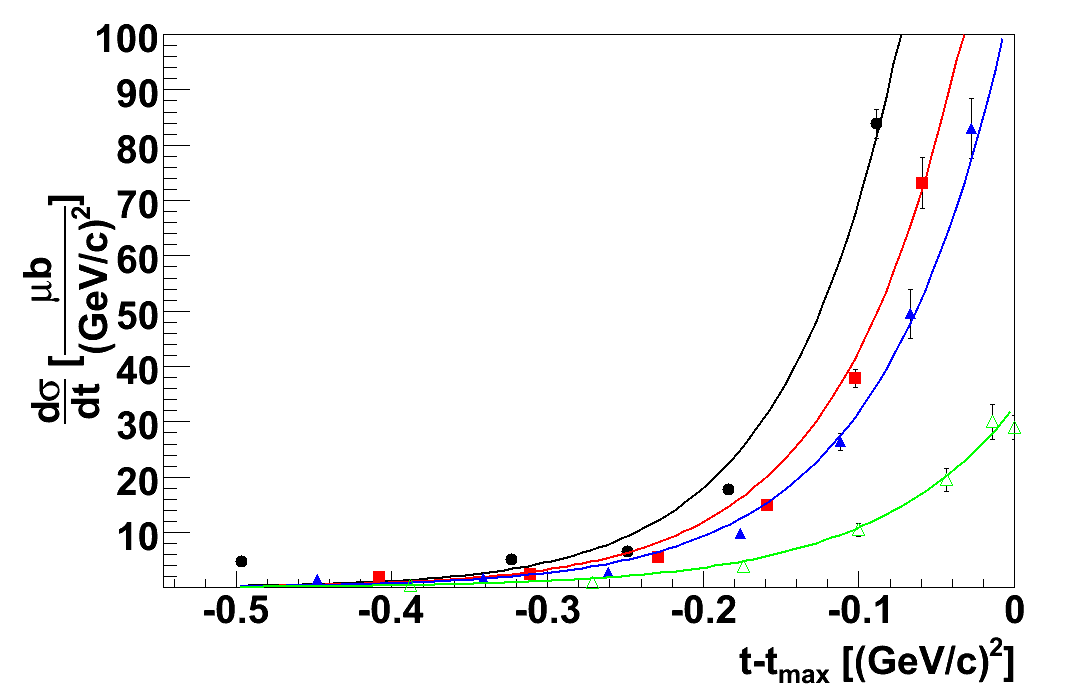}
      }
\caption[Differential cross-section from SATURNE]{\label{SATURNEparam2_p} Differential cross-section as a function of the four-momentum transfer. The lines present the applied parametrization. Four sets of the SATURNE data are plotted as: black full dots ($\sqrt{s}$=4.093 GeV), red full squares ($\sqrt{s}$=4.127 GeV), blue full triangles ($\sqrt{s}$=4.159 GeV) and green empty triangles ($\sqrt{s}$=4.329 GeV).}
\end{center}
\end{figure}

The SATURNE data and the applied parametrization are shown in Fig.~\ref{SATURNEparam_p} and \ref{SATURNEparam2_p}.

To determine the integrated luminosity we compared the angular distribution
of the measured counts with the parametrized cross-sections in the CM system.
In this system the four momentum transfer can be expressed as:
\begin{equation}
 t=(p_{beam}-p_{He})^2 = {m_d}^2 + {m_{He}}^2  -2 \cdot E_{beam} \cdot E_{He} + 2 \cdot |\vec{p}_{beam}|
\cdot |\vec{p}_{He}| \cdot \cos \theta_{CM} ,
\end{equation}
where ${\theta_{CM}}$ is the ${^3\mbox{He}}$ emission angle in the CM frame.

To calculate the parametrized cross-section in the CM system we use the transformation:
\begin{equation}
  \frac{d\sigma}{d\Omega}=\frac{1}{2\pi}\frac{d\sigma}{d(\cos\theta_{CM})}
=\frac{1}{2\pi}\frac{d\sigma}{d(t-t_{max})} \cdot \frac{d(t-t_{max})}{d(\cos\theta_{CM})} ,
\end{equation}
where the Jacobian term $d(t-t_{max})/d(\cos\theta_{CM})$  is equal to:
\begin{equation}
 \frac{d(t-t_{max})}{d(\cos\theta_{CM})}= -2 \cdot |\vec{p}_{beam}| \cdot |\vec{p}_{He}| .
\end{equation}

The statistical uncertainties of the SATURNE cross-sections are much smaller
than the overall normalization uncertainty equal to 7\%.


\subsubsection{Experimental counts}

\begin{figure}[!ht]
\begin{center}
      \scalebox{\scaleFactor}
         {
         \includegraphics{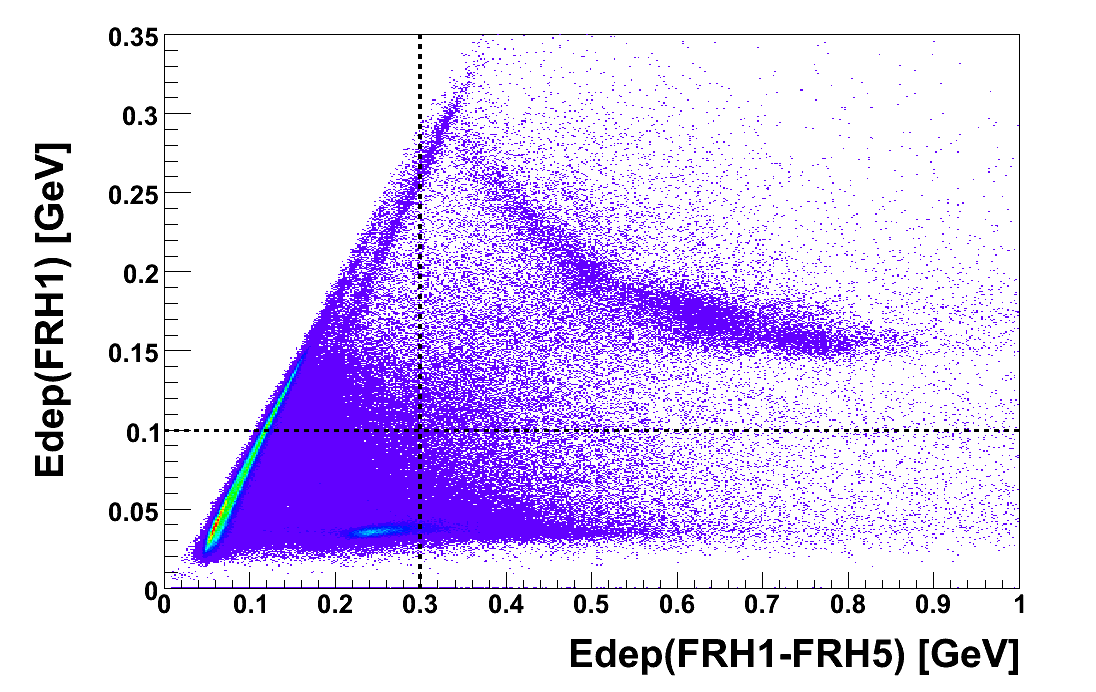}
	
  	 }
\caption[Identification of He]{\label{edepFRH1vsEdepTot_p}  Deposited energy in the first layer of the Forward Range Hodoscope ($y$-axis) versus deposited energy in all Hodoscope layers ($x$-axis).   The black dashed lines indicates the applied cut: $Edep(FRH1)>0.1 $\,GeV AND $Edep(FRH_{tot})>0.3 $\,GeV.}
\end{center}
\end{figure}

The measurement of the $dd \rightarrow {^3\mbox{He}} n$ reaction was based on the registration
of the ${^3\mbox{He}}$ ions in the Forward Detector.
The applied hardware trigger required at least one charged particle in the Forward Detector and,
in addition, a high energy threshold in the Window Counter Detector was set to reduce the background
from fast protons and pions.
At low beam momenta, the ${^3\mbox{He}}$ ions
from the $dd \rightarrow {^3\mbox{He}} n$ reaction were stopped in the third  layer
of the Forward Range Hodoscope and at high momenta they were stopped in the fourth layer.
Their identification was based on the $\Delta E- E$ information from the Forward Range Hodoscope
as it is shown in Fig.~\ref{edepFRH1vsEdepTot_p}.


\begin{figure}[!ht]
\begin{center}
      \scalebox{\scaleFactor}
         {
         \includegraphics{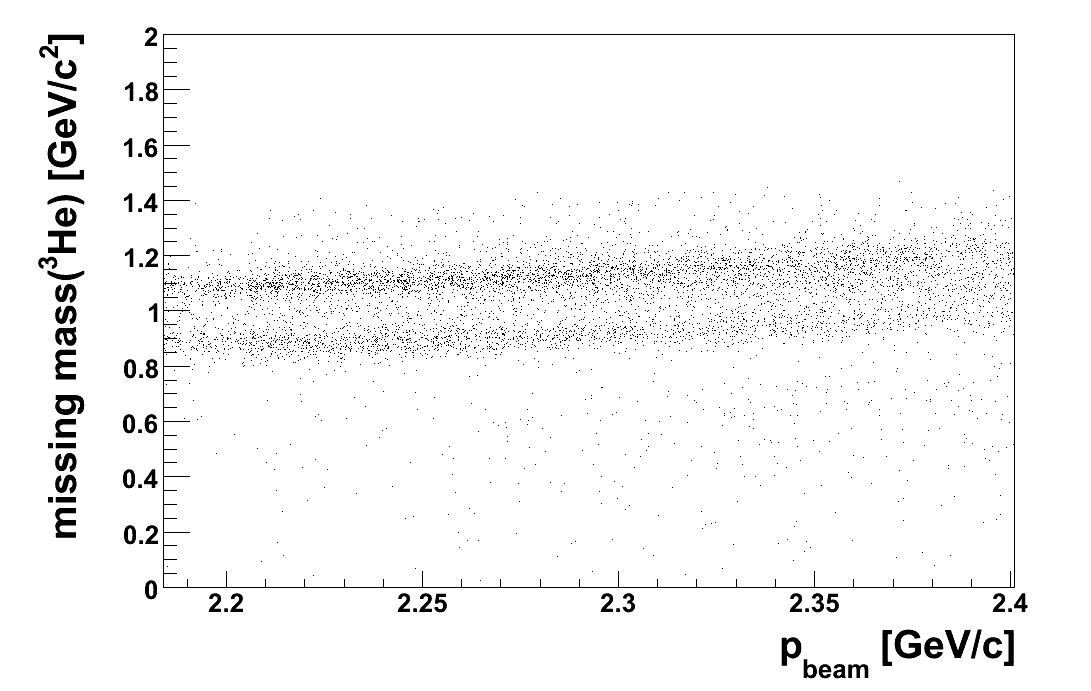}
  	 }

\caption[Missing mass vs beam momentum]{\label{he3MM_vs_beam_p}  Missing mass of the ${^3\mbox{He}}$ as a function of the beam momentum. }
\end{center}
\end{figure}

The outgoing neutrons were identified using the missing mass technique.
Fig.~\ref{he3MM_vs_beam_p} presents the missing mass of the ${^3\mbox{He}}$ as a function of the beam momentum.
Two distinct bands are visible. The lower one corresponds to the $dd \rightarrow {^3\mbox{He}} n$ events,
whereas the upper one results from pion production in processes
like $dd \rightarrow {^3\mbox{He}} n \pi^0$.
For the beam momenta above circa 2.3~GeV/c the missing mass resolution is worse
than at lower momenta.
This might be due to the fact, that at higher momenta the ${^3\mbox{He}}$ ions are stopped
in the fourth layer of the FRH,
for which the energy calibration is less accurate than for the first three layers (see Chapter \ref{calib_ch}).
Also, the visible increase of the missing mass as a function of the beam momentum can result from systematic
uncertainty of the energy calibration for the fourth layer of the FRH.

\begin{figure}[~h]
\begin{center}
      \scalebox{\scaleFactor}
         {
         \includegraphics{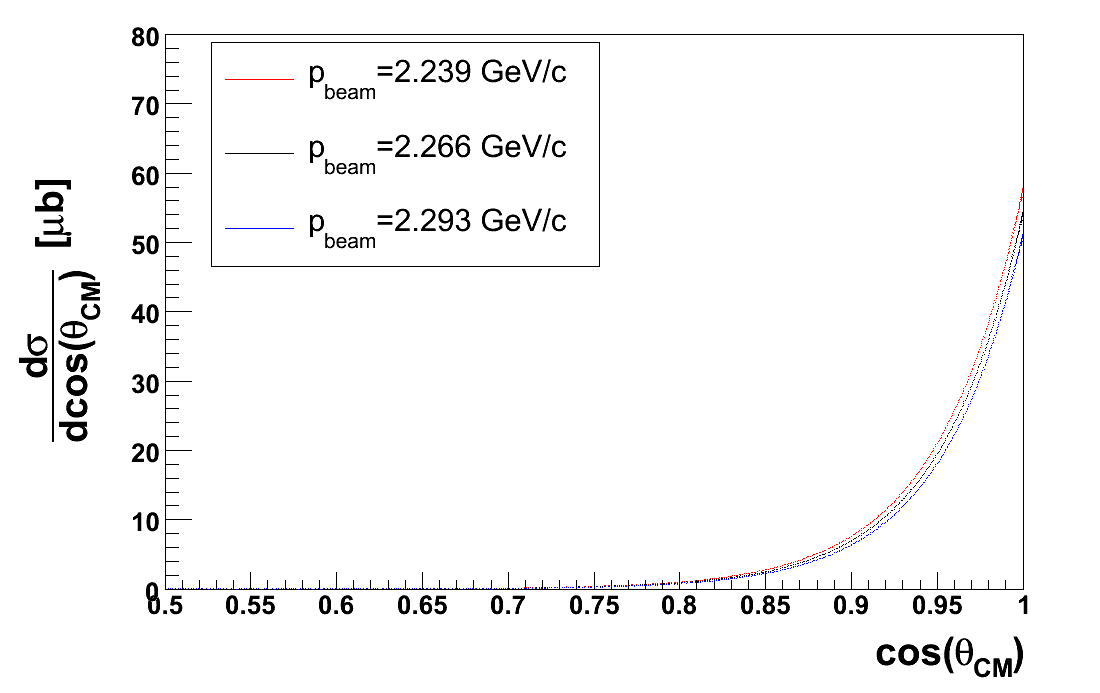}
  	 }
\caption[3Hen cross-section]{\label{xsectionfDiffMom_p} Parametrization of the $dd \rightarrow {^3\mbox{He}} n$
cross-section as a function of $\cos(\theta_{CM})$ for three different beam momenta from the analysed beam momentum range: the lowest value $p_{beam}=$ 2.239\,GeV/c (red line), the middle value 2.266\,GeV/c (blue line) and the highest values 2.2293\,GeV/c (black line).}
\end{center}
\end{figure}

For determination of the integrated luminosity we chose
the beam momentum range from 2.239 to 2.293 GeV/c, corresponding
to about one quarter of the total beam momentum range.
On the one hand, the selected interval is wide enough to determine the integrated luminosity
with statistical accuracy much smaller than the systematic uncertainty of 7\%
of the $dd \rightarrow {^3\mbox{He}} n$ cross-section (see previous subsection).
On the other hand, it is sufficiently narrow to neglect uncertainties connected with
averaging the cross-section as a function of beam momentum, so that for the luminosity
determination the cross-section for the central momentum can be taken
(see Fig.~\ref{xsectionfDiffMom_p}).

\begin{figure}[!ht]
\begin{center}
\scalebox{\scaleFactor}
         {
         \includegraphics{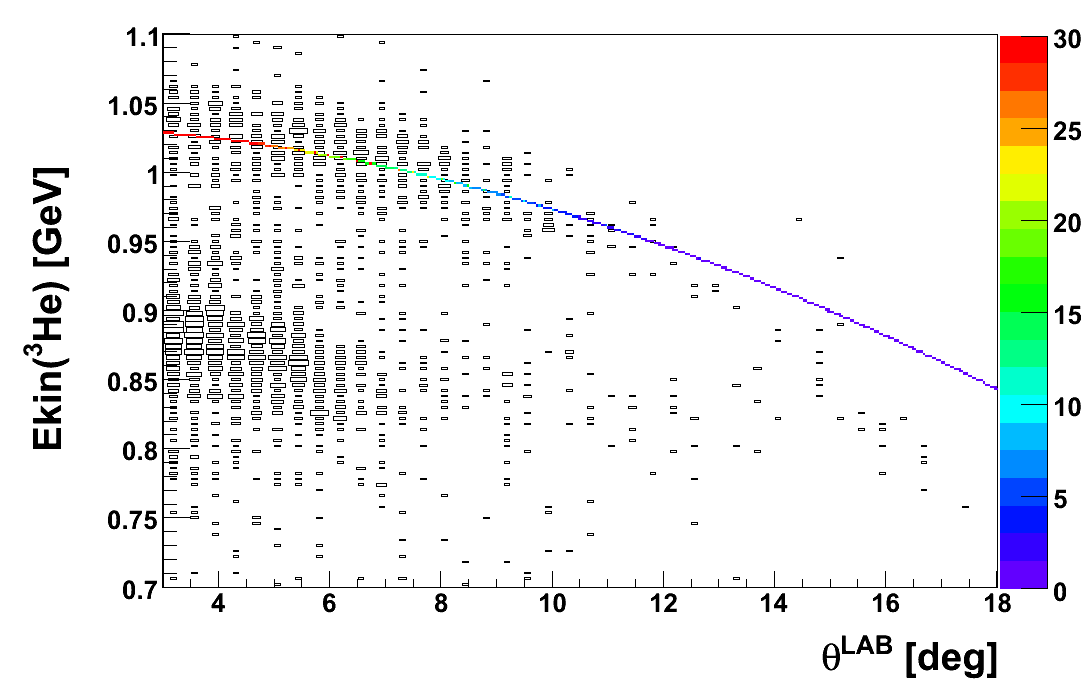}
         }
\caption[Ekin vs theta for He]{\label{Ekin_vs_thetaLAB_he3_p} Experimental distribution of kinetic energy versus the scattering angle in the LAB frame. The simulated kinematic dependence for the $dd \rightarrow {^3\mbox{He}} n$  is marked as a colour curve. The red colour indicates the area with the cross-section two orders of magnitude higher than in other parts of the curve according to the applied parametrization. }
\end{center}
\end{figure}

Fig.~\ref{Ekin_vs_thetaLAB_he3_p} shows distribution of the kinetic energy versus the scattering angle in the LAB frame
for the identified ${^3\mbox{He}}$ ions.

\begin{figure}[!ht]
\begin{center}
\scalebox{\scaleFactor}
         {
         \includegraphics{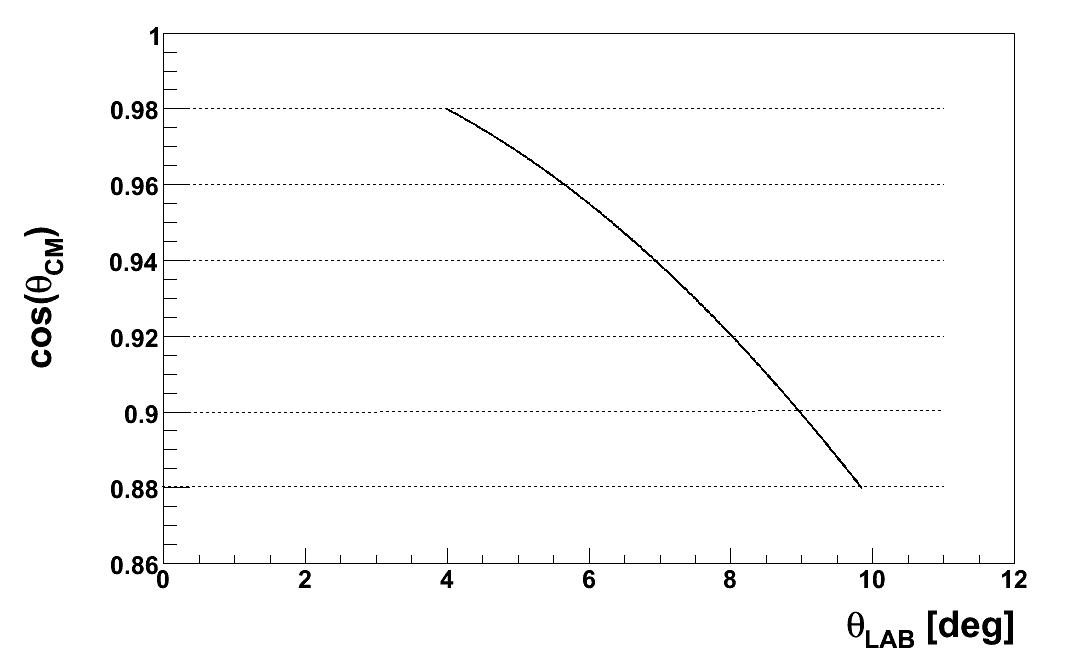}
         }
\caption[$cos(\theta_{CM})$ versus $\theta_{LAB}$ ]{\label{cosThetaCMvsThetaLAB_p} Kinematic
correspondence between the scattering angle $\theta_{LAB}$ and $\cos\theta_{CM}$
for the $dd \rightarrow {^3\mbox{He}} n$ reaction.
The dashed lines indicate the selected intervals in  $\cos\theta_{CM}$. \label{fig:cosThetaCM_vs_ThetaLAB}}
\end{center}
\end{figure}

The counts for the $dd \rightarrow {^3\mbox{He}} n$ reaction cover the angular range
from about 4$^{\circ}$ to 10$^{\circ}$.
As shown in Fig.~\ref{fig:cosThetaCM_vs_ThetaLAB}, in the CM frame this range corresponds
roughly to the $\cos\theta_{CM}$ interval from 0.88 to 0.98.
For comparison of angular distribution of  the experimental counts with the cross-sections we divided the above $\cos\theta_{CM}$ range into five bins of equal width as indicated in Fig.~\ref{fig:cosThetaCM_vs_ThetaLAB}.

\begin{figure}[!ht]
\begin{center}
\scalebox{0.32}
         {
         \includegraphics{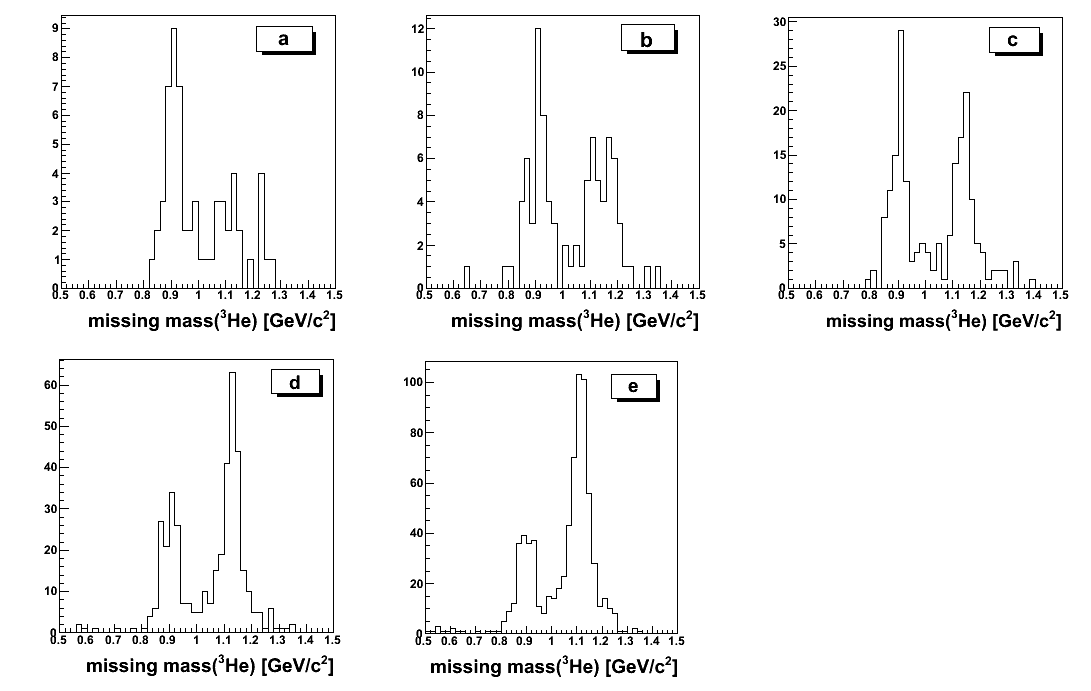}
         }
\caption[Missing mass for different $\cos(\theta_{CM})$]{\label{he3MM_differentCosTheta_p}
 Missing mass spectra of ${^3\mbox{He}}$ for five $\cos\theta_{CM}$ intervals: a) 0.88-0.9, b) 0.9-0.92, c) 0.92-0.94, d) 0.94-0.96, e) 0.96-0.98. A clear peak around the neutron mass is visible. The second peak around 1.1~GeV/c$^2$
originates from pion production reactions.}
\end{center}
\end{figure}

\begin{figure}[!ht]
\begin{center}
\scalebox{\scaleFactor}
         {
         \includegraphics{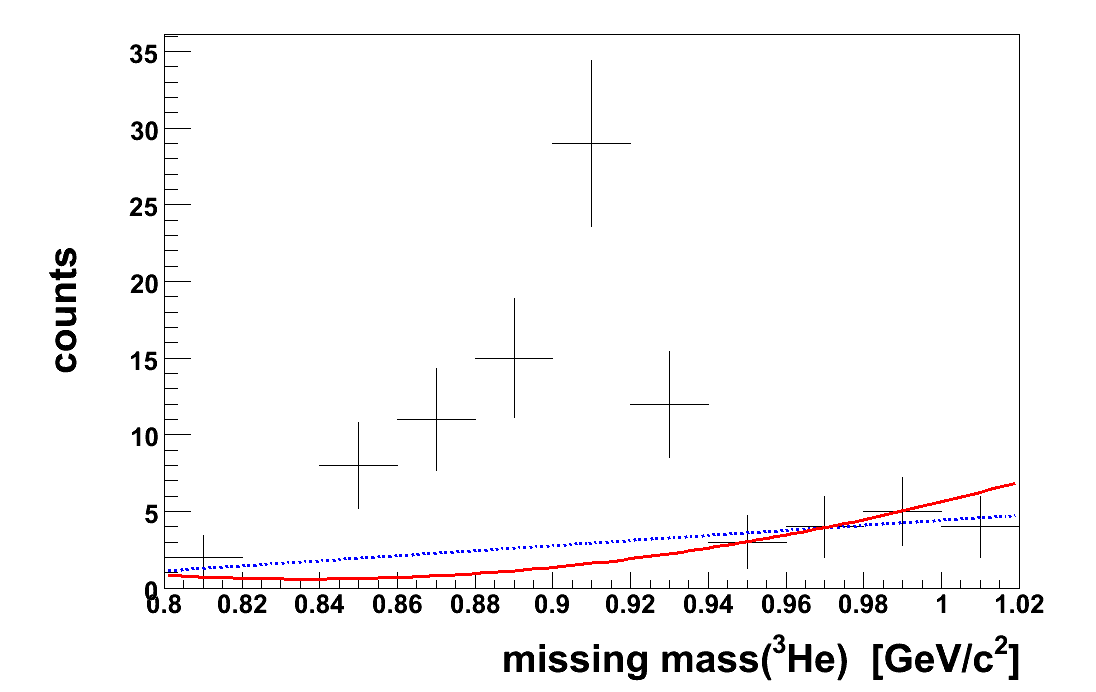}
         }
\caption[Background subtraction]{\label{backg_subtraction_p} Distribution of missing mass of ${^3\mbox{He}}$ for $cos\theta_{CM}\in[0.92,0.94]$. The background was fitted with a linear function (blue dashed line)
and a quadratic function (red solid line).}
\end{center}
\end{figure}

The numbers of $dd \rightarrow {^3\mbox{He}}n $ events  in the chosen $cos\theta_{CM}$ intervals
were obtained from counts in the neutron peak in the corresponding missing mass spectra
shown in Fig.~\ref{he3MM_differentCosTheta_p}.
The number of counts in the neutron peak  was corrected for the background under the peak
approximated with a quadratic function which was fitted to a few points selected on the left
and on the right side of the peak (see Fig.~\ref{backg_subtraction_p}).
The uncertainty of the background subtraction was estimated by comparison of fits with
a linear and quadratic function.
This uncertainty is equal to 8\% and is included as a systematic error in the luminosity determination.

\begin{figure}[!ht]
\begin{center}
\scalebox{\scaleFactor}
         {
         \includegraphics{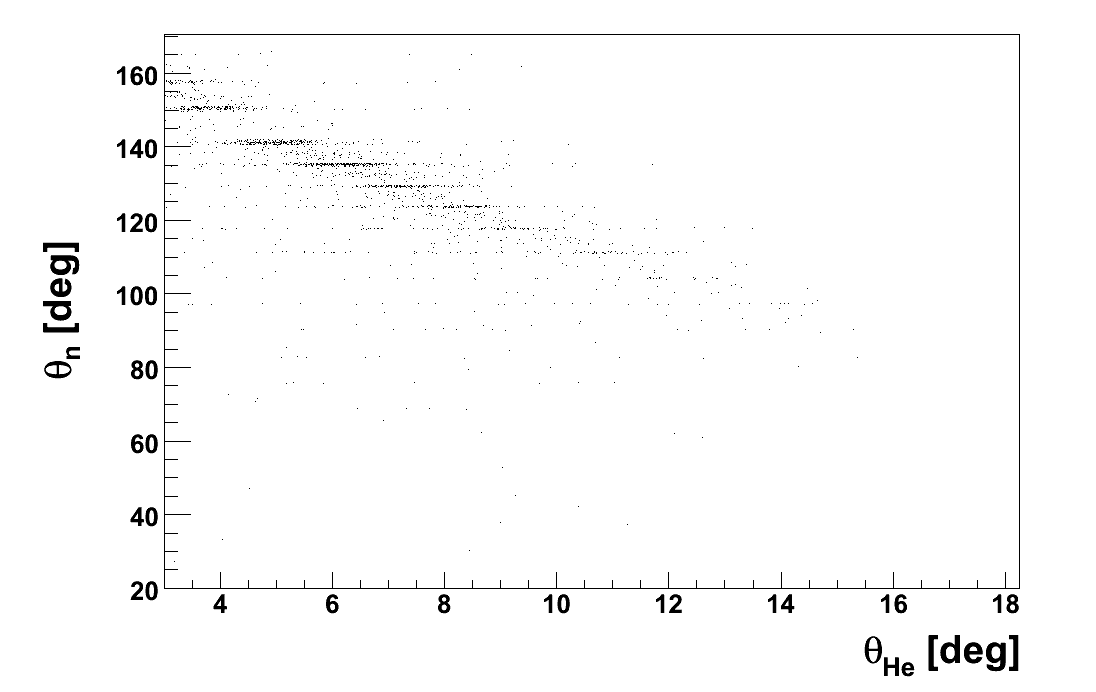}
         }
\scalebox{\scaleFactor}
         {
         \includegraphics{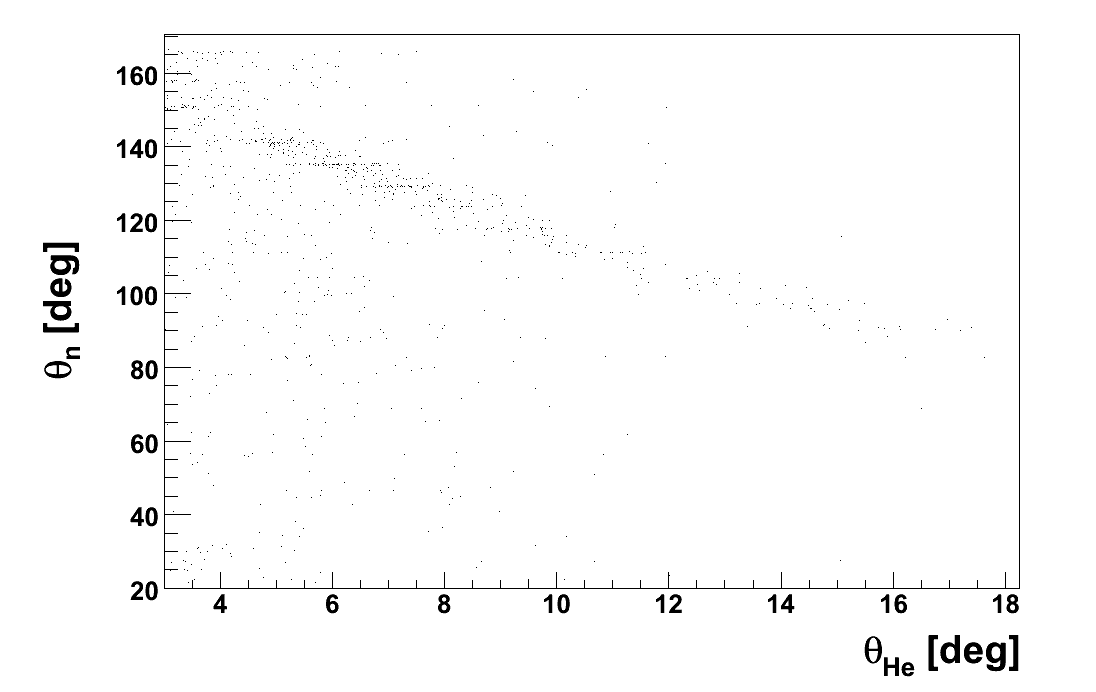}
         }
\caption[$^{3}\mbox{He}$ and neutrons]{\label{ThetaHe3vsThetaN_p} Scattering angle of charged particle detected in the Forward Detector versus scattering angle of neutral particle registered in the Calorimeter. The cuts to select the $^{3}\mbox{He}$ candidates were applied as explained in the text. In addition exactly one neutral particle registered in the Central Detector was demanded. Finally the co-planarity condition was applied, namely: $|\phi_{fd}-\phi_{cd}|<10 ^{\circ}$. The upper plot is the $dd \rightarrow {^3\mbox{He}}n $ MC simulation according to the phenomenological parametrization (see \ref{param_l}). The lower plot shows the experimental data. The visible signal from the ${^3\mbox{He}}n$ is  seen in both plots.}
\end{center}
\end{figure}

In order to check the applied selection procedure for the $dd \rightarrow {^3\mbox{He}}n$ events,
we investigated the response of the Scintillating Electromagnetic Calorimeter to the outgoing neutrons.
Because of relatively small thickness of the SEC from the point of view of the nuclear interactions
 the probability that a neutron will interact in the calorimeter material
is only about 40\%.
In order to choose the neutron candidates, in addition to the ${^3\mbox{He}}$ selection cuts, we demanded exactly one neutral particle registered in the Central Detector.
Besides, we request that the difference between the azimuthal angles for the outgoing $^{3}\mbox{He}$ ion
and the neutron is close to 180$^\circ$ as expected for a two body reaction:
$|\phi_{fd}-\phi_{cd}-180^{\circ}|<10 ^{\circ}$.
As it is presented in the Fig.~\ref{ThetaHe3vsThetaN_p}, a clear correlation in the scattering angle distribution of the forward and the central tracks is visible. The experimental distribution agrees nicely
with results of simulations.
This result confirms correctness of the selection procedure for ${^3\mbox{He}}$, however due to the poor description of the nuclear processes in the SEC
the coincidence with the neutrons registered in the SEC was not used for the luminosity determination.

\subsubsection{Results for the integrated luminosity}

The efficiency term $\varepsilon(\cos\theta_{CM})$  for the analysed reaction was determined on the basis Monte Carlo simulations.
It turned out, that the calculated efficiency does not depend on the scattering angle $\theta_{CM}$ and is equal to 93\%.
The remaining 7\% corresponds to events which couldn't be reconstructed mostly due to nuclear interactions
of the $ ^3\mbox{He}$ ions in the FRH.

The obtained number of counts $\Delta N$, as well as the determined luminosities for each $\cos(\theta_{CM})$ interval, are given in Tab.~(\ref{TabLum}).
{
\renewcommand{\arraystretch}{1.2}
\begin{table}[hbpt]
 \begin{center}
 \begin{tabular}{|l|l|l|}
        \hline
        $\cos(\theta_{CM})$ & $\Delta N \pm stat $ & $L_{k} [{nb}^{-1}] \pm stat$  \\
        \hline
        0.88-0.9  & $ 37 \pm 6 $ & $29.5 \pm 4.9$ \\
        \hline
	0.9-0.92  & $ 43 \pm 6  $ & $21.6 \pm 3.3$ \\
	\hline
	0.92-0.94 &$ 93  \pm 10 $ & $31.7 \pm 3.3$ \\
        \hline
	0.94-0.96 &$ 143 \pm 12 $  & $32.3 \pm 2.8$\\
	\hline
	0.96-0.98 &$ 222 \pm 15 $ & $34.0 \pm 2.4$\\
        \hline
 \end{tabular}
 \caption{Number of counts and integrated luminosity values for different $\cos(\theta_{CM})$ bins.}
 \label{TabLum}
 \end{center}
\end{table}
}

\begin{figure}[!ht]
\begin{center}
\scalebox{\scaleFactor}
      {
         \includegraphics{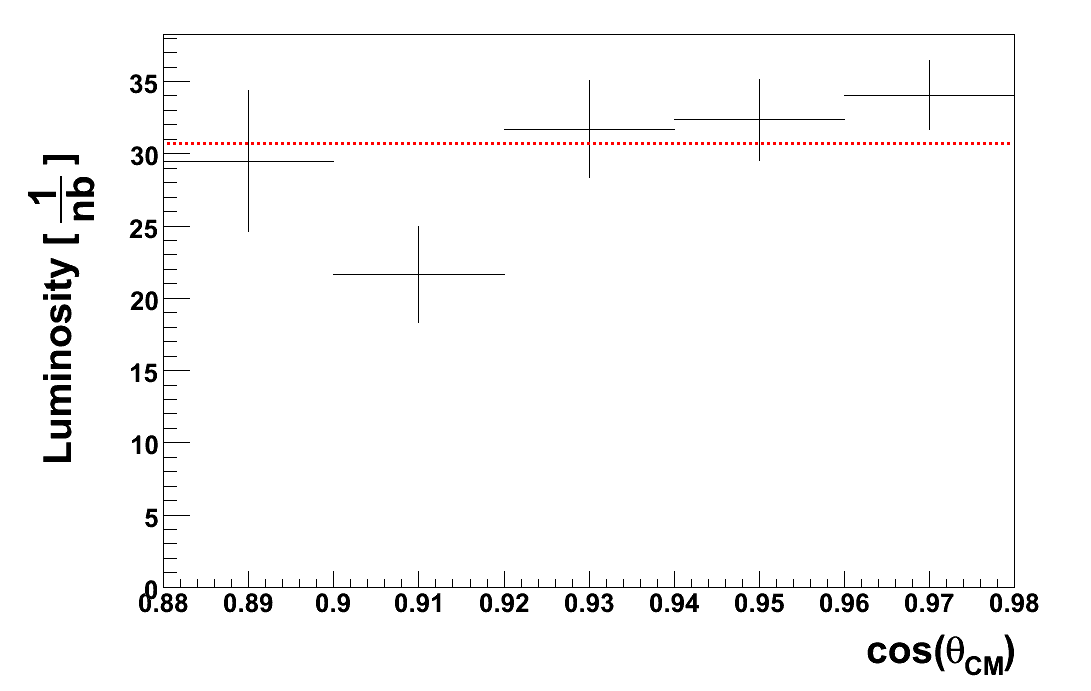}
      }
\caption[Final luminosity]{\label{luminosity2_v3_p} Estimated luminosities $L_{i}$ for different $\cos\theta_{CM}$
bins. The statistical uncertainties are marked as a vertical bars. The integrated luminosity equal to the weighted average is marked as a dashed red line.}
\end{center}
\end{figure}

The integrated luminosity was calculated as a weighted
average of the luminosities determined on the basis of 5 individual $\cos\theta_{CM}$ intervals:
\begin{equation}
\overline{L}=\frac{\sum_{k=1}^{=5} L_{k} \cdot \frac{1}{(\sigma_{k})^2}}{\sum_{k=1}^{5} \frac{1}{(\sigma_{k})^2}}.
\end{equation}
This average integrated luminosity equals 30.72\,nb$^{-1}$. It is marked in the Fig.~\ref{luminosity2_v3_p}
with dashed line.
The uncertainty of the integrated luminosity equal to 11.54\%
was calculated by adding in quadrature the statistical uncertainty ($\pm$4.5\%), the systematic error due to the background subtraction ($\pm$8\%) and the normalization uncertainty of the SATURNE data for the
$dd \rightarrow {^3\mbox{He}} n$ cross-sections \cite{bizard} ($\pm$7\%).

\begin{figure}[!ht]
\begin{center}
\scalebox{\scaleFactor}
      {
         \includegraphics{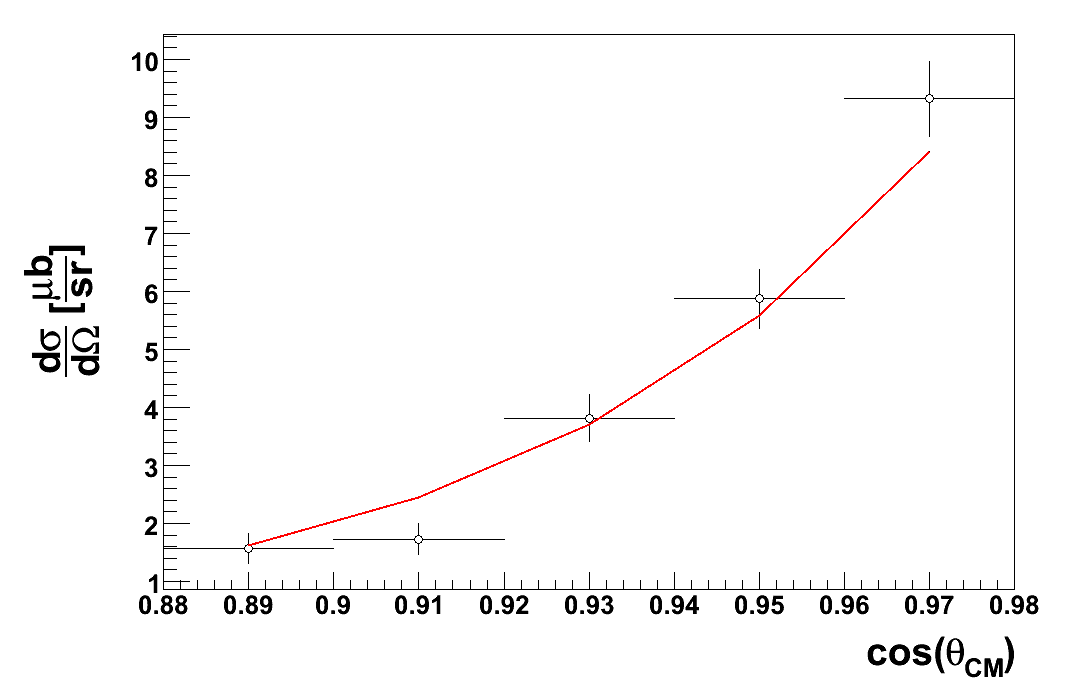}
      }
\caption[Angular distribution of $dd \rightarrow {^3\mbox{He}} n$ cross-section]{\label{xsection_vs_data_p}
Angular distribution of the $dd \rightarrow {^3\mbox{He}} n$ cross-section at beam momentum of 2.266\,GeV/c
determined assuming the final result for the integrated luminosity.
The vertical error bars represent the statistical uncertainties.
The cross-section parametrization of the SATURNE data is shown with red line.}
\end{center}
\end{figure}

In order to illustrate the capability of the present experiment for determination
of angular distributions of the $dd \rightarrow {^3\mbox{He}} n$ cross-section
in Fig.~\ref{xsection_vs_data_p}  we show a comparison between the parametrization of the SATURNE data for
the beam momentum equal to 2.266 GeV/c and the angular distribution based on the present data assuming
the integrated luminosity equal to the weighted average. Only the statistical
uncertainties of the present data points are indicated.
They can be reduced by extending the measurement time.
More serious are the systematic errors due to the background subtraction.
Only careful study of the shape of the background can help to reduce this error.

The above integrated luminosity corresponds to the range of the beam momenta from 2.239 to 2.293~GeV/c.
In order to calculate the integrated luminosity in the full range of the beam momenta which
was used in the search for the $\eta$-mesic ${^4\mbox{He}}$ (2.192-2.400~GeV/c)
we calculated the corresponding correction according to the Eq.~(\ref{eq:const})
using the relative normalization of luminosities determined in various
intervals of the beam momentum as described in the next section.
The total integrated luminosity is equal to $117.9 \pm 13.6$\,nb$^{-1}$.
The obtained result is in line with a rough estimation of the total integrated luminosity of 170\,nb$^{-1}$, 
based on the beam intensity and the target thickness (see Appendix~\ref{app:lum_estimate}).

\clearpage

\subsection{Dependence on the beam momentum}

\subsubsection{Coincident spectra of two charged particles}
In order to determine the luminosity dependence on the beam momentum
we used the  quasi-elastic proton-proton scattering.
In the deuteron-deuteron collisions, the protons from the deuteron beam scatter
on the protons in the deuteron target.
The neutrons from the colliding deuterons play the role of spectators, which means that before
and after the collision they move with the Fermi momentum in the CM system of their parent deuteron.
We use the following notation for this process: $dd \rightarrow pp n_{sp} n_{sp}$.

The hardware trigger applied for the registration of the quasi-elastic scattering required at least one charged particle detected in the Forward Detector and another one in the Central Detector.
In the analysis, a more selective condition of exactly one charged particle in the FD
and one particle in the CD was set.

\begin{figure}[!ht]
\begin{center}
      \scalebox{\scaleFactor}
      {
         \includegraphics{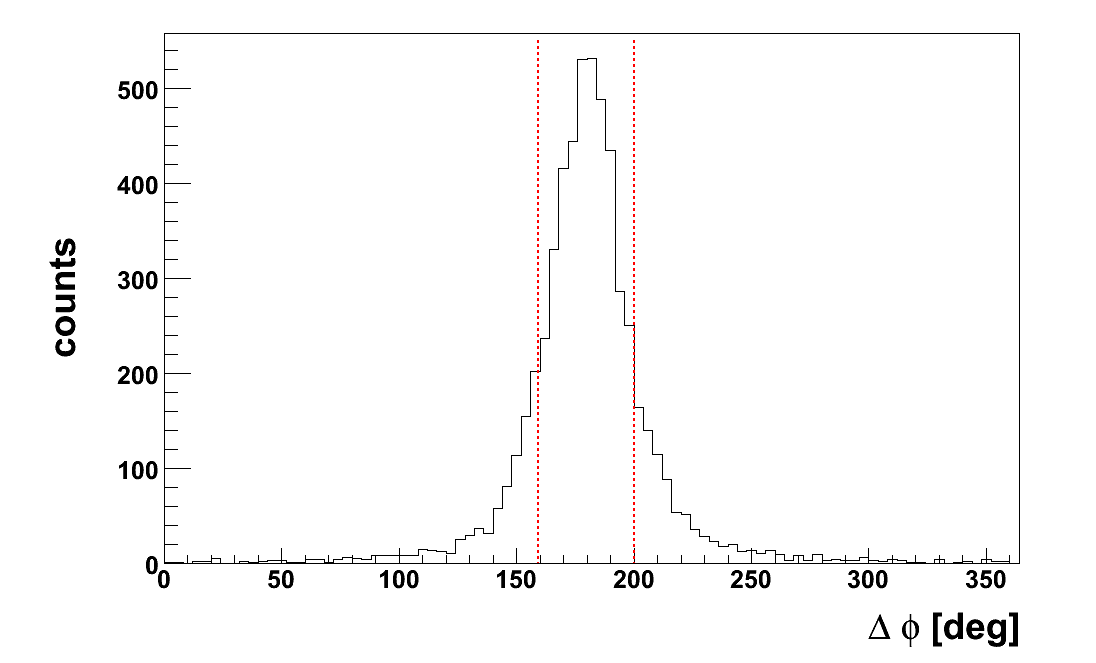}
      }
      \scalebox{\scaleFactor}
      {
         \includegraphics{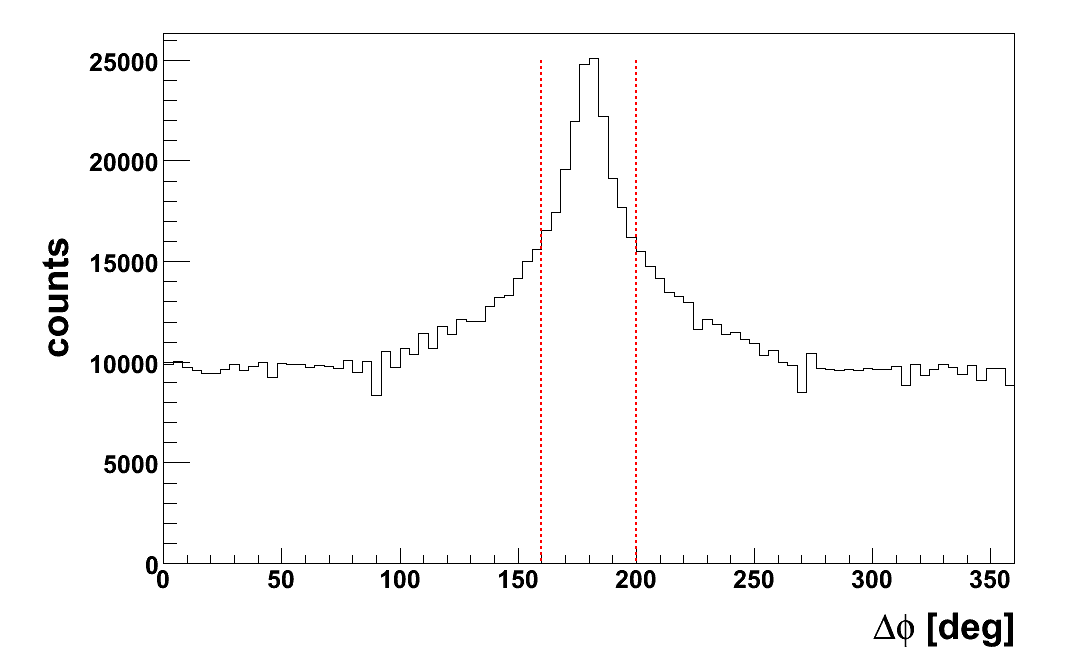}
      }
\caption[Distribution of difference of azimuthal angles $\Delta\phi=\phi_{FD}-\phi_{CD} - 180^{\circ}$]{\label{deltaPhi_pp_p} Distribution of the difference of azimuthal angles $\Delta\phi=\phi_{FD}-\phi_{CD} - 180^{\circ}$ for MC simulation of the quasi-elastic $dd \rightarrow pp n_{sp} n_{sp}$ reaction (upper plot) and for the experimental events (lower plot). The applied cut is marked by  red dashed lines.}
\end{center}
\end{figure}

In an ideal case of lack of Fermi motion in the deuteron, the sum of momentum vectors of the scattered protons in the LAB system is oriented along the beam direction.
In this case the azimuthal angles for the protons registered in the FD
and in the CD fulfil the condition:  $\Delta\phi=\phi_{FD}-\phi_{CD} - 180^{\circ} = 0^{\circ}$.
The Fermi motion leads to smearing of $\Delta\phi$
as shown in the upper plot in Fig.~\ref{deltaPhi_pp_p} representing results of simulations
of the quasi-elastic proton-proton scattering.
The lower plot in Fig.~\ref{deltaPhi_pp_p} represents the difference of the azimuthal angles in FD and in CD determined for the experimental data. The central peak with maximum at $\Delta\phi=180^{\circ}$ lies on a high level of background
originating from many body processes.

For further analysis we set the condition $|\Delta\phi|<20^{\circ}$ as expected for the $p-p$
quasi-elastic scattering on the basis of the simulations.

Events selected according the above conditions are presented on a plot showing the correlation
between the polar angles $\theta_{FD}$ and $\theta_{FD}$ of
particles registered in the FD and in the CD, respectively.

In this plot we identified events corresponding to the following quasi-free processes:
\begin{itemize}
\item $dd \rightarrow pp n_{sp} n_{sp}$, \\
\item $dd \rightarrow d_{b} p_{t} n_{sp}$, \\
\item $dd \rightarrow p_{b} d_{t} n_{sp}$, \\
\item $dd \rightarrow d\pi^{+}  n_{sp} n_{sp}$, \\
\end{itemize}
where the subscripts ${b}$ and ${d}$ denote the particles from the beam and from the target, respectively.

\begin{figure}[!ht]
\begin{center}
 \scalebox{0.38}
      {
      \includegraphics{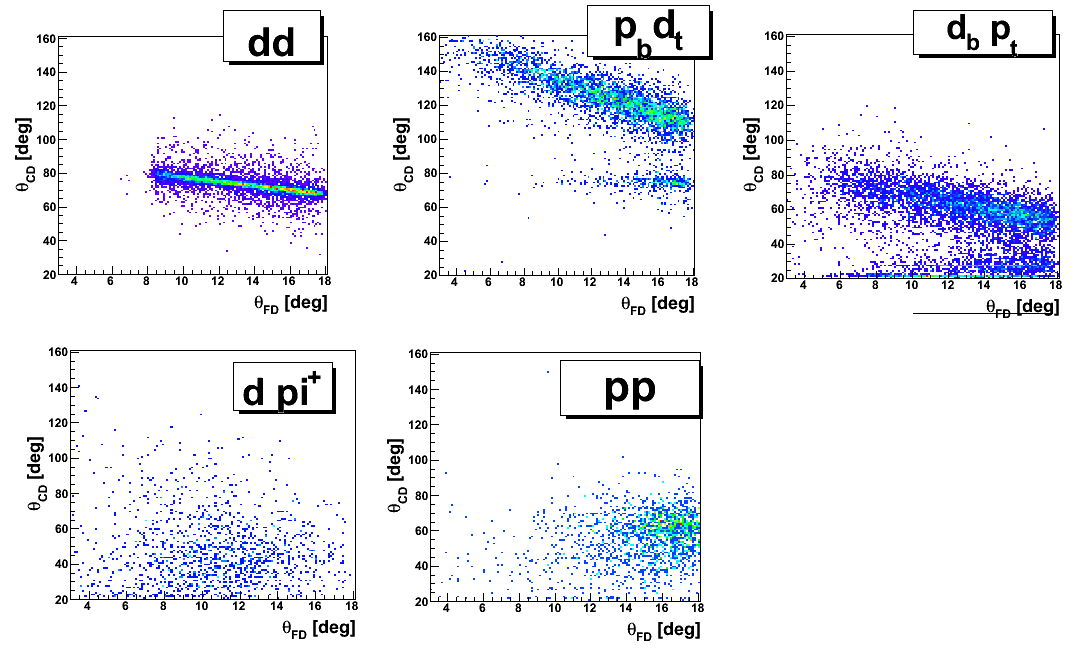}
      }
\caption[$\theta_{FD}$ vs $\theta_{CD}$ MC]{\label{all_theta_vs_theta_mc_2_2275_v2_p} Correlations between the polar angles $\theta_{FD}$ and $\theta_{CD}$ obtained in simulations for the different quasi-elastic processes: 1) $dd \rightarrow dd$, 2) $dd \rightarrow p_{b} d_{t} n_{sp}$, 3) $dd \rightarrow d_{b} p_{t} n_{sp}$. 4) $dd \rightarrow d\pi^{+}  n_{sp} n_{sp}$, 5) $dd \rightarrow pp n_{sp} n_{sp}$.}
\end{center}
\end{figure}

\begin{figure}[!ht]
\begin{center}
 \scalebox{0.32}
      {
         \includegraphics{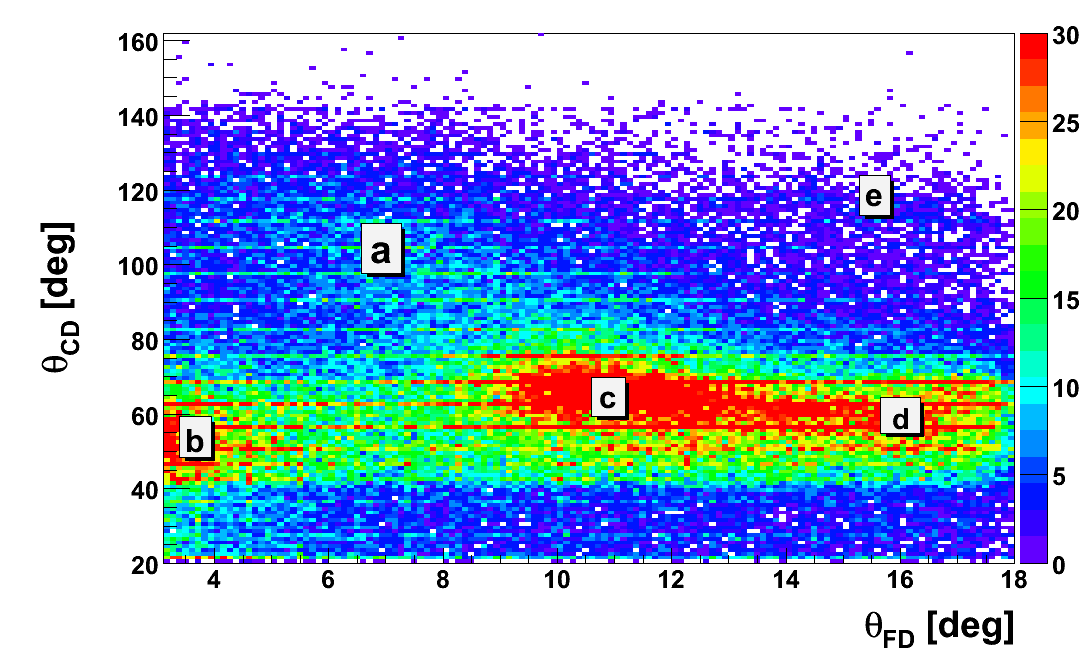}
      }
\caption[Distribution $\theta_{FD}$ vs $\theta_{CD}$ (data)]{\label{thetaFDvsthetaCD_noPScut_v2_p}Correlations between the polar angles $\theta_{FD}$ and $\theta_{CD}$ from the experiment. The indicated areas correspond to the: a)$dd \rightarrow pp n_{sp} n_{sp} \rightarrow d\pi^{+}  n_{sp} n_{sp}$,b) $dd \rightarrow pn p_{sp} n_{sp}$ ,c) $dd \rightarrow d_{b} p_{t} n_{sp}$ and $dd \rightarrow pp n_{sp} n_{sp}$, d) $dd \rightarrow pp n_{sp} n_{sp}$ e) $dd \rightarrow p_{b} d_{t} n_{sp}$ .}
\end{center}
\end{figure}

The identification of these processes was based on a comparison with simulations taking into account the Fermi motion of nucleons inside the deuteron.
The correlations between the polar angles $\theta_{FD}$ and $\theta_{CD}$
obtained in simulations of the studied processes shown in Fig.~\ref{all_theta_vs_theta_mc_2_2275_v2_p} coincide with ones
for the experimental counts presented in Fig.~\ref{thetaFDvsthetaCD_noPScut_v2_p}.

In Fig.~\ref{thetaFDvsthetaCD_noPScut_v2_p}, besides the events from quasi-free processes  discussed above,
there is a pronounced concentration of events in the area $\theta_{FD}\in[3-4 ^{\circ}]$ and $\theta_{CD}\in[40-60 ^{\circ}]$.
We presume, that these events originate from
the quasi-free scattering of the neutron from the deuteron beam on the proton inside the target:
$dd \rightarrow pn p_{sp} n_{sp}$.
In the FD we register the spectator proton from the deuteron beam
and in the CD  the recoil proton from the $n-p$ quasi-elastic scattering.
For this process we expect no correlation between the azimuthal angles of particles
registered in the FD and CD, and indeed, in the experimental data we see no correlation between
these two angles for the discussed events.

\begin{figure}[!ht]
\begin{center}
\scalebox{0.32}
      {
         \includegraphics{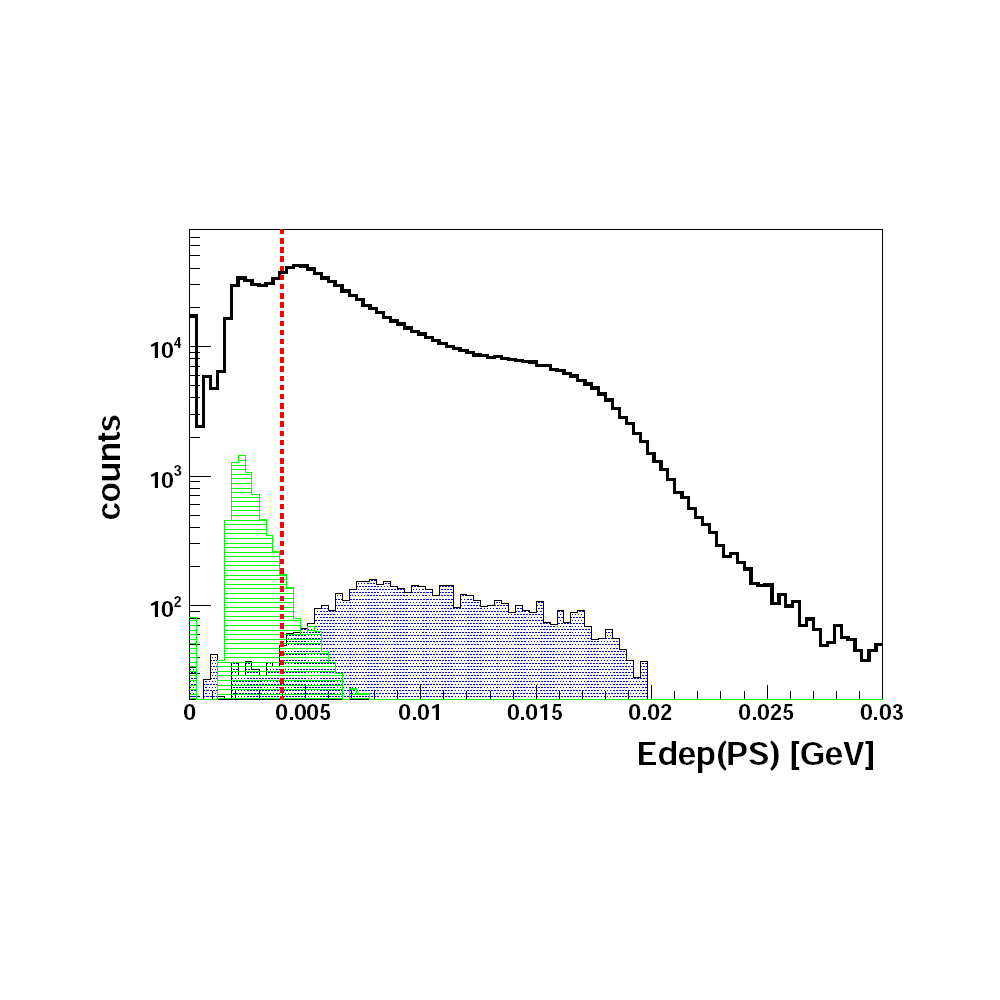}
      }
\caption[ Distribution of the energy loss in PSB]{\label{PScut_p} Distribution of the energy loss in the Plastic Scintillator Barrel. The experimental data is shown with black line.
The results from MC simulation of the reaction $dd \rightarrow d\pi^{+}  n_{sp} n_{sp}$ are marked as  green area, whereas the results from the  $dd \rightarrow pp n_{sp} n_{sp}$ are marked as blue area. The MC counts are arbitrarily normalized. The applied cut ($Edep(PS) >0.004$\,GeV) is shown as a red dashed line. Please note that the $y$ axis is in logarithmic scale.}
\end{center}
\end{figure}

In order to suppress the events which contain the charged pions registered in the Central Detector, we applied an additional cut on the energy deposited in the Plastic Scintillation Barrel as showed in the Fig.~\ref{PScut_p}.

\begin{figure}[!ht]
\begin{center}
\scalebox{0.32}
      {
         \includegraphics{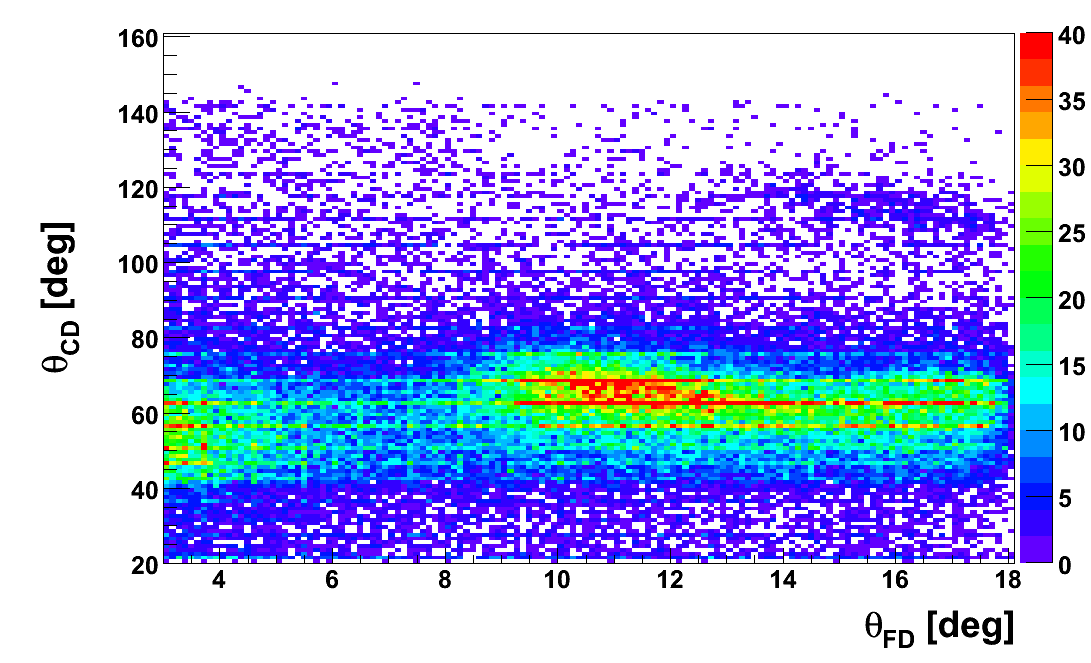}
      }
\caption[Distribution $\theta_{FD}$ vs $\theta_{CD}$ after the application of the cut $Edep(PS) >0.004$\,GeV (data)]{\label{thetaFDvsthetaCD_withPScut_p}Correlations between the polar angles $\theta_{FD}$ and $\theta_{CD}$ from the experiment after the application of the cut $Edep(PS) >0.004$\,GeV.}
\end{center}
\end{figure}

As one could expect, after this cut the $dd \rightarrow  d\pi^{+}  n_{sp} n_{sp}$
events disappear from the plot showing the correlation between the polar angles
$\theta_{FD}$ and $\theta_{CD}$ (see Fig.~\ref{thetaFDvsthetaCD_withPScut_p})

From the reactions discussed above, for the luminosity determination we decided to use
the quasi-elastic $p-p$  scattering.
This process is registered with relatively high statistics.
Although we can not separate it from the quasi-elastic $d-p$ scattering,
we show in the next subsection, that it is absolutely dominating in the angular region $\theta_{FD}\in 15-17 ^{\circ}$.

\begin{figure}[!ht]
\begin{center}
      \scalebox{\scaleFactor}
      {
         \includegraphics{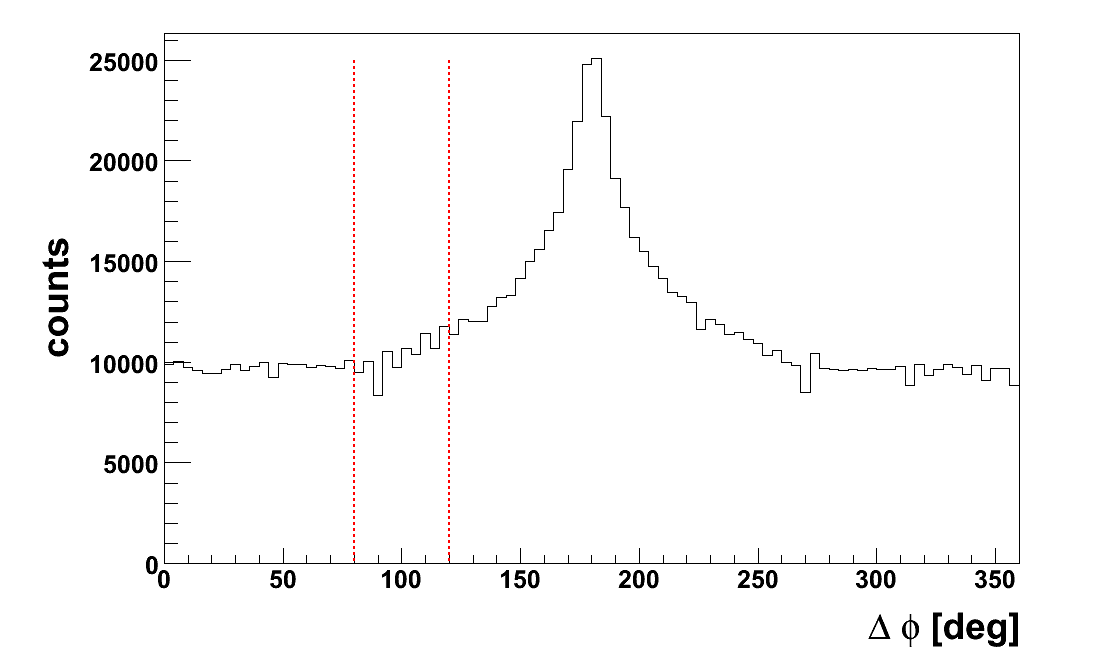}
      }
\caption[Distribution of difference of azimuthal angles $\Delta\phi=\phi_{FD}-\phi_{CD} - 100^{\circ}<20^{\circ}$ (background)]{\label{deltaPhi_background_p} Selection of events 
in the $|\phi_{FD}-\phi_{CD} - 100^{\circ}|<20^{\circ}$ range marked with red dashed lines for estimation of background
from many-body reactions.}
\end{center}
\end{figure}

\begin{figure}[!ht]
\begin{center}
\scalebox{0.32}
      {
         \includegraphics{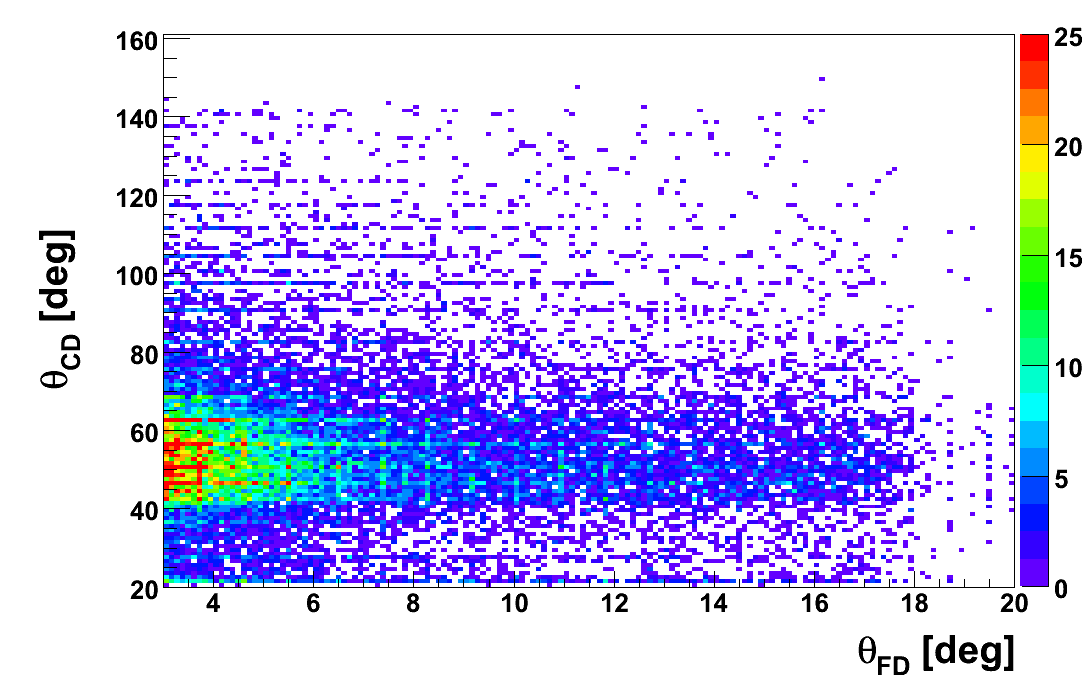}
      }
\caption[Distribution $\theta_{FD}$ vs $\theta_{CD} $ for background reactions data)]{\label{thetaFDvsthetaCD_background_p}Correlations between the polar angles $\theta_{FD}$ and $\theta_{CD}$ from the experiment after the choice of the background area.}
\end{center}
\end{figure}

We also, investigated the background from many-body reactions.
For this we chose the $|\phi_{FD}-\phi_{CD} - 100^{\circ}|<20^{\circ}$ area (see Fig.~\ref{deltaPhi_background_p})
and we checked the distribution of counts in the resulting $\theta_{FD}$-$\theta_{CD}$ plot shown in Fig.~\ref{thetaFDvsthetaCD_background_p}.
The counts in the angular range $\theta_{FD}\in 15-17 ^{\circ}$, where the quasi-elastic $p-p$ scattering dominates, 
is about 20 times smaller than the $p-p$ signal.

\subsubsection{Cross sections for the elastic $p-p$ and $d-p$ scattering}

For estimation of the cross-sections for the quasi-free proton-proton scattering
we used the data for the elastic proton-proton scattering at the proton beam momentum equal to one-half of the momentum of the deuteron beam.
For the analysed beam momentum range of 2.192-2.400 GeV/c, the corresponding proton momenta
change between 1096 and 1.200 GeV/c.
The cross-sections for the elastic proton-proton scattering are taken from the SAID partial-wave analysis
\cite{SAID}. We used the energy dependent "Current Solution".
We checked that its predictions, describes very well the experimental cross-sections measured close to our energy range
by the EDDA collaboration \cite{EDDA}.

\begin{figure}[!htp]
\begin{center}
 \scalebox{\scaleFactor}
      {
         \includegraphics{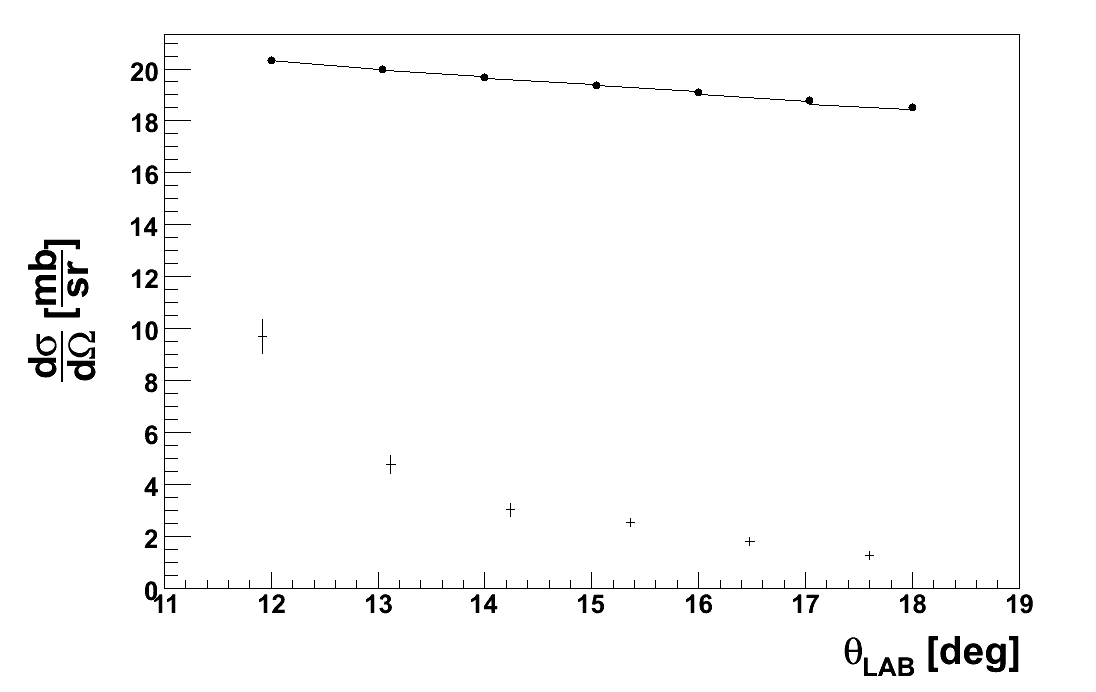}
      }
 \scalebox{\scaleFactor}
      {
         \includegraphics{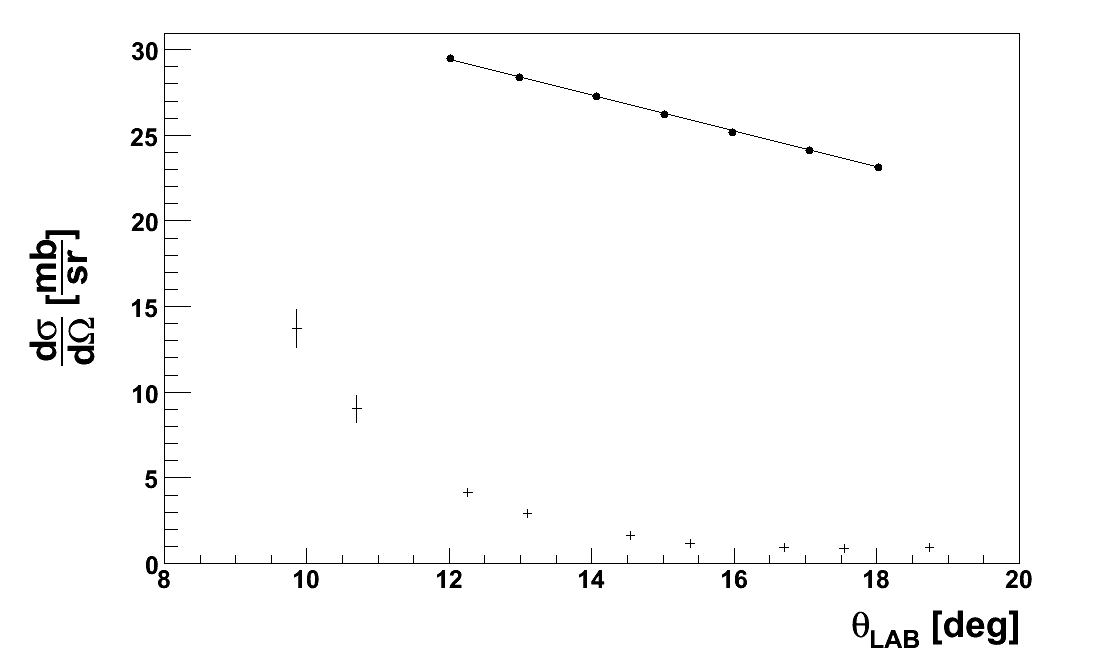}
      }
\caption[Angular distributions of $p-p$ and $d-p$  cross-section]{\label{xsection_dp_plab_p}
Experimental data for the $d-p$ elastic cross-section for deuteron beam momentum of
$p_{LAB}=1979$\,GeV/c (upper plot) and $p_{LAB}=2391$ GeV/c (lower plot) from \cite{Booth} and  \cite{Boschitz}, respectively.
For comparison, the elastic $p-p$ cross-sections calculated using SAID for proton beam momenta of 0.989\,GeV/c (upper plot)
and 1.195\,GeV/c (lower plot) are shown with solid line.}
\end{center}
\end{figure}

To estimate the contributions from the quasi-free $d-p$ scattering,
which in our measurements can not be distinguished from the quasi-free $p-p$ scattering,
we assumed, that it is described by the elastic $d-p$ cross-sections.
For this, we made use of the experimental $p-d$ cross-sections of Booth et al.~\cite{Booth}
and of Boschitz et al.~\cite{Boschitz}, measured with deuteron target and proton beam with momentum
of 990~MeV/c and 1196.2~GeV/c, respectively.
The equivalent momenta for a deuteron beam scattered on a proton target are equal to 1.979~GeV/c
and 2.391~GeV/c, respectively.
These two momenta are very close to the lower and the upper limit
of the present beam momentum range.
In Fig.~\ref{xsection_dp_plab_p}, the experimental $d-p$ cross-sections are compared with the elastic $p-p$
cross-sections calculated using the SAID parametrization.
For the forward scattering angles of about $\theta_{FD} = 17 ^{\circ}$, the $d-p$ cross-sections are about
20 times smaller than the $p-p$ cross-sections.
Therefore, we neglect the contributions from the quasi-elastic $d-p$ scattering in the current analysis.

We also checked that the cross-sections for the elastic $d-d$ scattering \cite{alberi},
which in our experiment can not be distinguished from the $p-p$ quasi-elastic scattering,
within the investigated angular region are smaller than the elastic $p-p$ cross-sections by a factor of approximately 50.
Therefore, we can also neglected contributions from the $d-d$ scattering in our analysis.

\subsubsection{Integrated luminosities for beam momentum intervals}
\label{integrated_luminosities_l}
In order to calculate the integrated luminosity for each beam momentum interval we used Eq.~(\ref{eq:Lthic}).
The number of the quasi-elastic $p-p$ events was corrected for the detection efficiency
determined in the simulations and equal to 53\%.
Also, the prescaling factor of the applied experimental trigger equal to $\frac{1}{4000}$ was taken into account.
The integrated luminosities for the individual beam momentum intervals were calculated
using the $p-p$ elastic cross-sections.
The sum of these luminosities was then normalized to the absolute value of the integrated luminosity
determined on the basis of the  $dd \rightarrow {^3\mbox{He}} n$ measurements using Eq.~(\ref{eq:const}).
The resulting value of the correcting factor $c$ equals 0.4
and could be understood as a result of shadowing of the proton by the neutron
inside the deuteron (shadowing effect) which reduces the probability of the quasi-elastic scattering. 
This result should be further investigated in measurements of $d-d$ collisions with the WASA-at-COSY
making use of the momentum analysis of particles registered in the CD for a cleaner identification
of the $p-p$ quasi-elastic scattering.

\begin{figure}[htp]
\begin{center}
 \scalebox{\scaleFactor}
      {
         \includegraphics{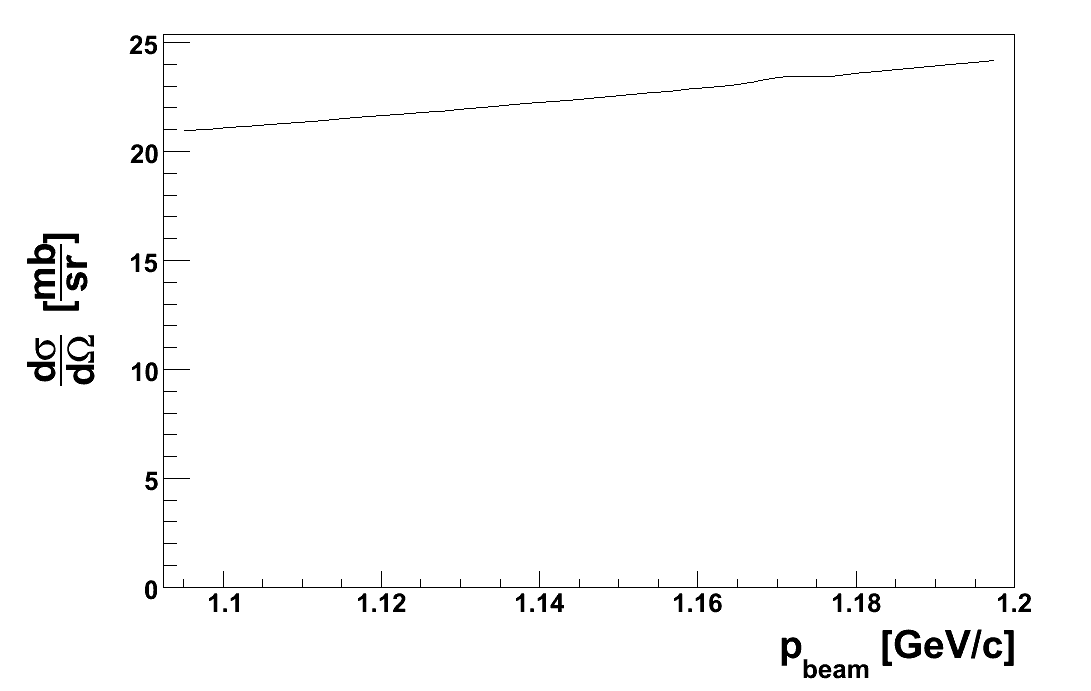}
      }
 \scalebox{\scaleFactor}
      {
         \includegraphics{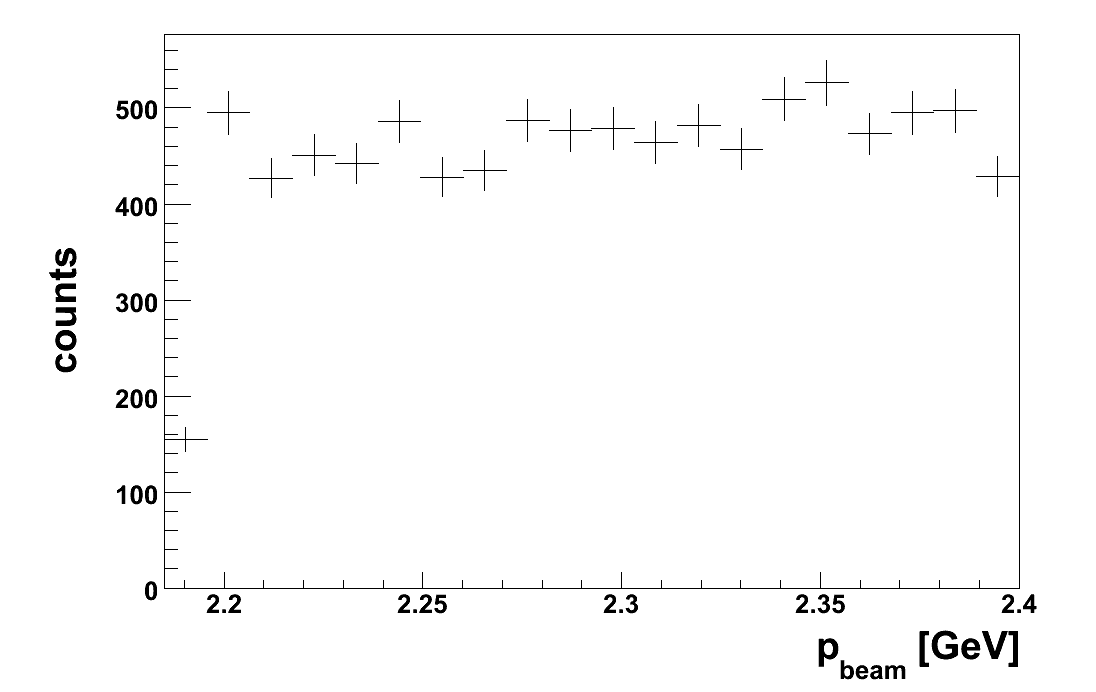}
      }
 \scalebox{\scaleFactor}
      {
         \includegraphics{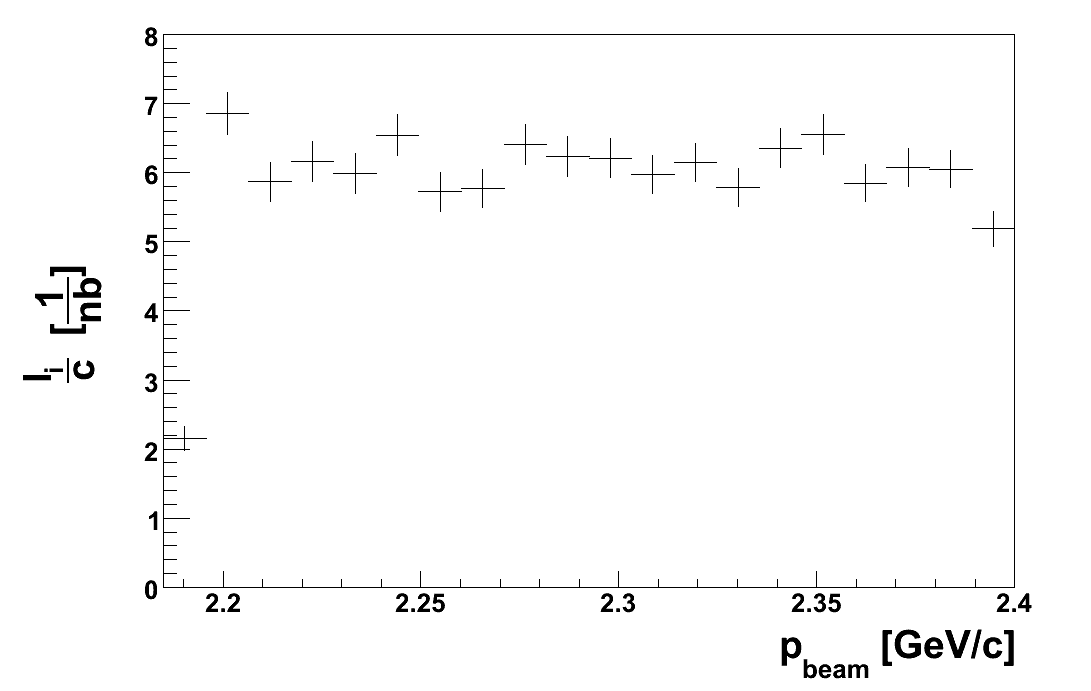}
      }
\caption[The integrated luminosity calculated for individual beam momenta bins]{\label{weights_p} Differential cross-section for the $pp \rightarrow pp$  elastic scattering at $\theta=17^{\circ}$ based on the SAID parametrization (upper plot); the number of counts of selected $dd \rightarrow pp (nn)$ events(middle plot) and the integrated luminosity calculated for individual beam momentum bins (lower plot).}
\end{center}
\end{figure}

The integrated luminosities calculated for individual beam momentum intervals
are shown in Fig.~\ref{weights_p} (bottom). The statistical uncertainty of each point is about 4.7\%.

\cleardoublepage




\chapter{Discussion of results}
\label{ch:disscusion}

In the present chapter we describe determination of the excitation function
for the $dd \to (^4\mbox{He}\eta)_{bound} \rightarrow {^3\mbox{He}} p \pi^-$ reaction.
A fit of the excitation function using the Breit-Wigner curve
is used for a search of a signal from decay of the ${^4\mbox{He}}-\eta$ bound state.
We investigate also the experimental background
on the basis of angular and momentum distributions of the final state particles.

\section{Excitation function}

\begin{figure}[!hp]
\begin{center}
  \scalebox{\scaleFactor}
  {
    \includegraphics{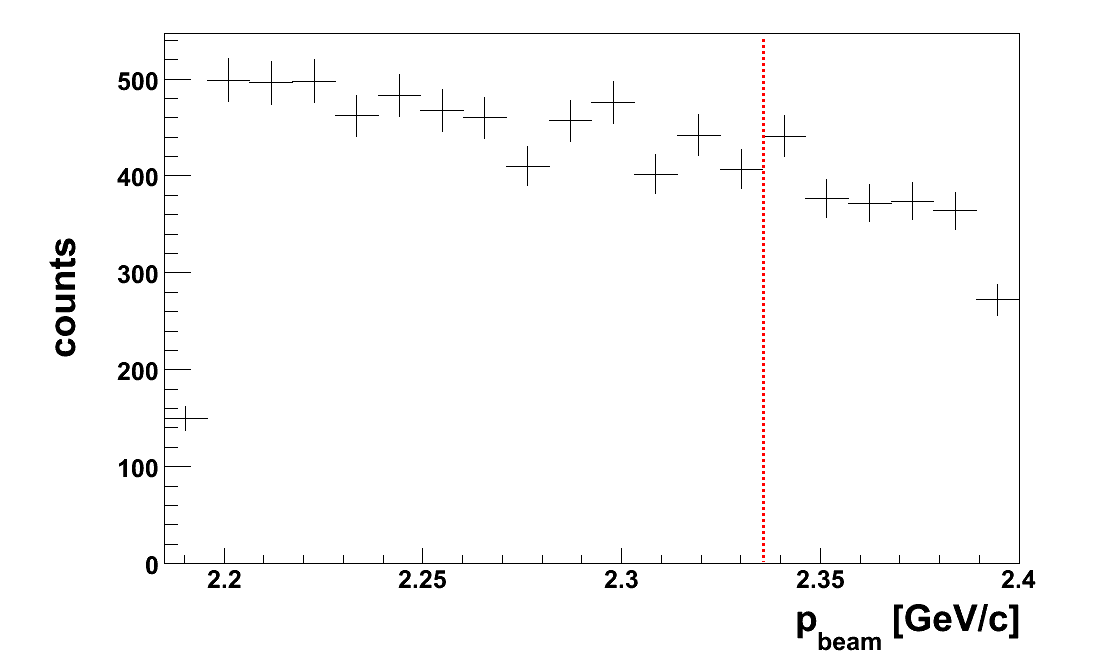}
  }
  \scalebox{\scaleFactor}
  {
    \includegraphics{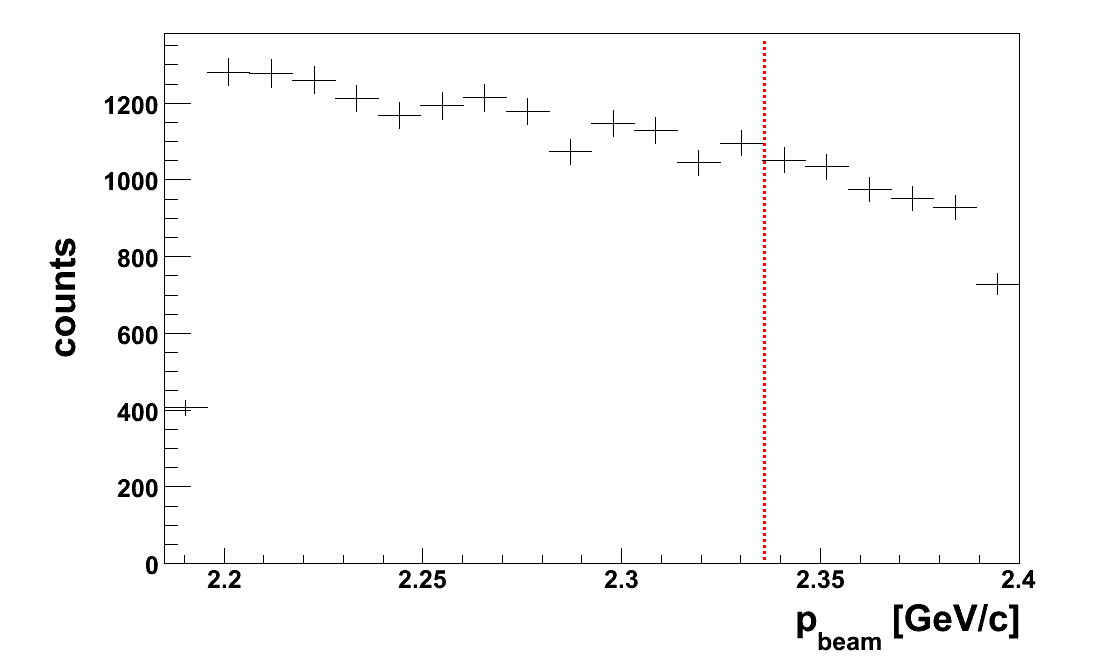}
  }
  \scalebox{\scaleFactor}
  {
    \includegraphics{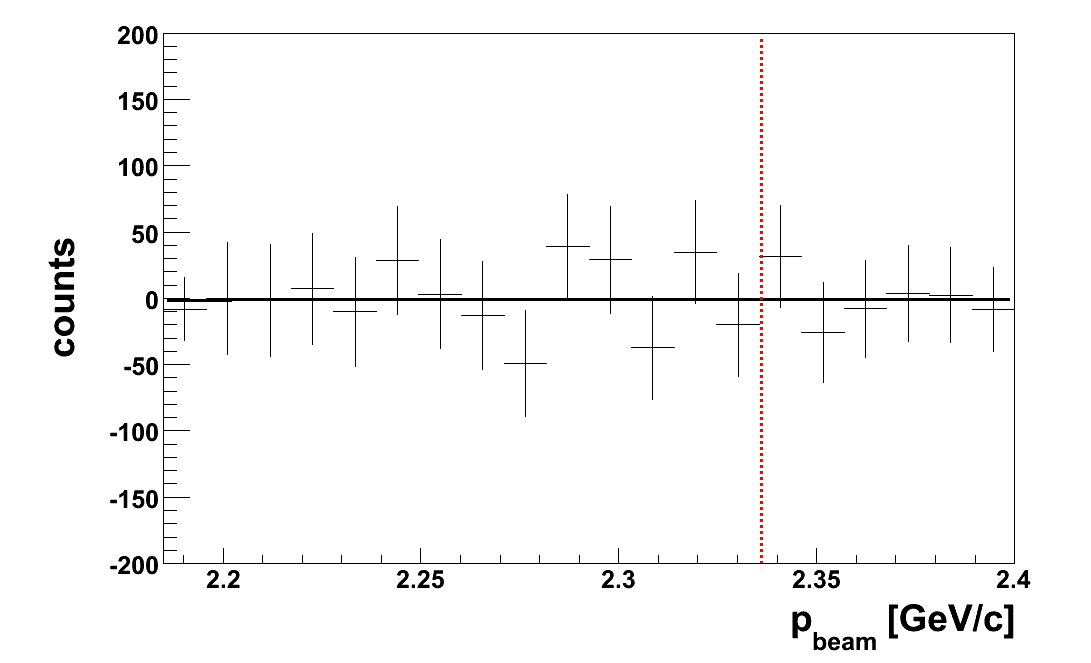}
  }
\caption[Not normalized excitation functions]{\label{hExcitFuncCM_mom_exp_bad_p}Excitation function for the $dd \rightarrow {^3\mbox{He}} p \pi^{-}$ reaction for the "signal-rich" region corresponding to ${^3\mbox{He}}$ momentum below 0.3\,GeV/c (upper panel) and the "signal-poor" region with  ${^3\mbox{He}}$ momentum above 0.3\,GeV/c (middle panel).
Difference of the excitation functions for the "signal-rich" and "signal-poor" regions after the normalization
to the lowest beam momentum bin is shown in the lower panel. The black solid line represents a straight line fit.
The beam momentum corresponding to the ${^4\mbox{He}}-\eta$ threshold is marked by the vertical red line.}
\end{center}
\end{figure}

We constructed two types of excitation function for the $dd \rightarrow {^3\mbox{He}} p \pi^{-}$ reaction.
They differ in the selection of the events and in the way of normalizing the data points.
The first excitation function uses events from the  "signal-rich" region
corresponding to the ${^3\mbox{He}}$ CM momenta below 0.3\,GeV/c, as it was discussed in Sect.~\ref{cut_he3_momentum_sect}.
The counts are plotted as a function of the beam momentum as it is shown Fig.~\ref{hExcitFuncCM_mom_exp_bad_p}(top).
The obtained function is smooth an no clear signal, which could be interpreted as a resonance-like
structure, is visible.
A similar dependence is obtained for events originating from the "signal-poor" region
corresponding to ${^3\mbox{He}}$ CM momenta above 0.3\,GeV/c (see Fig.~\ref{hExcitFuncCM_mom_exp_bad_p}(middle)).
We checked also for possible structures in the difference between the discussed functions
for the "signal-rich" and "signal-poor" region.
For this, we multiplied the function for the "signal-poor" region by a factor chosen in such a way,
that the difference of the two functions for the lowest beam momentum bin is equal to zero.
This difference is presented in Fig.~\ref{hExcitFuncCM_mom_exp_bad_p}(bottom).
The obtained dependence is flat and is consistent with zero. No resonance structure is visible.

\begin{figure}[!ht]
\begin{center}
      \scalebox{\scaleFactor}
      {
         \includegraphics{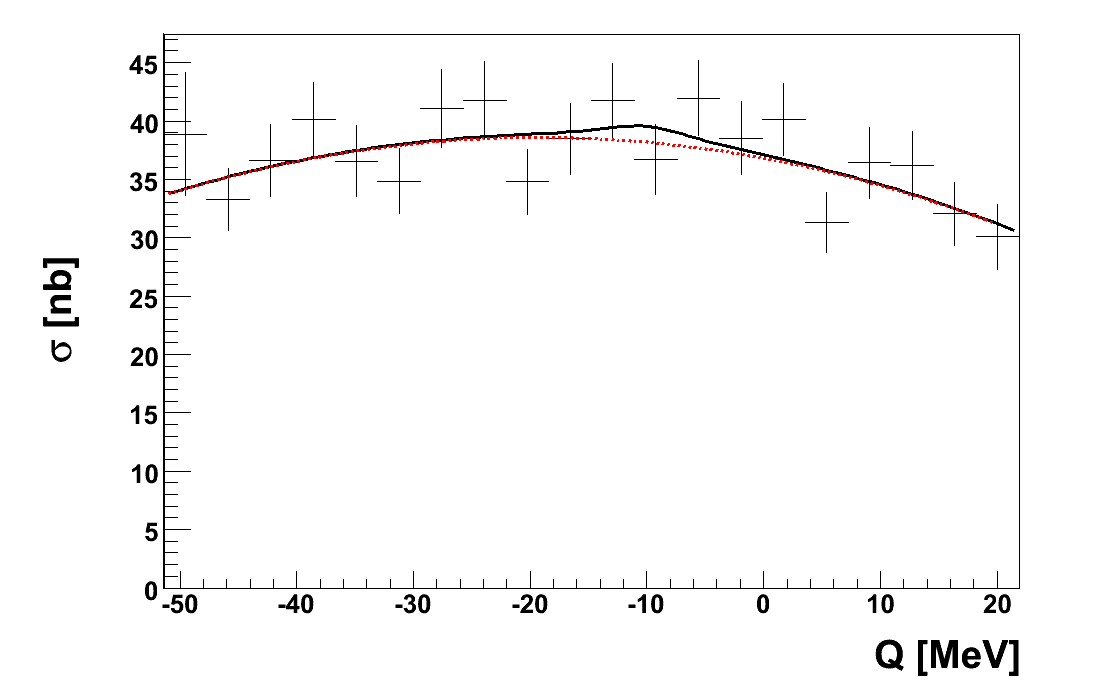}
      }
\caption[Breit-Wigner fit to the normalized excitation function]{\label{fit_10_10_allcuts_p} 
Excitation function for the $dd \rightarrow {^3\mbox{He}} p \pi^{-}$ reaction obtained by normalizing the events
selected in individual excess energy intervals by the corresponding integrated luminosities.
The solid line represents a fit with second order polynomial combined with a Breit-Wigner function 
with  fixed binding energy and width equal to -10 and 10 MeV, respectively. 
The dotted line corresponds to the contribution from the second order polynomial in the performed fit. The $\sigma$ values are not corrected for acceptance and efficiency cuts.}
\end{center}
\end{figure}

However, we do not treat this result as a final conclusion of non observation
of the ${^3\mbox{He}}-\eta$ bound state,
since one can apply further cuts to reduce the background.
Additional cuts on the $p$ and $\pi^-$ kinetic energy distributions
and the $p-\pi^-$ opening angle in the CM system lead us to the construction of a second excitation curve.
The CM kinetic energies of protons and pions originate
from the mass deficit $m_{\eta}-m_{\pi}$ are around 50\,MeV and 350\,MeV, respectively.
We select the kinetic energy of protons smaller than 200 MeV  and of pions from the interval (180, 400) MeV
as it is presented in Sect.~\ref{cut_ekin_sec}.
We apply also a cut on the relative $p-\pi^-$ angle in the CM system in the range of (140$^{\circ}$-180$^{\circ}$) (see~\ref{cut_he3_momentum_sect}).
The number of selected events in each beam momentum interval is divided
by the corresponding integrated luminosity.
In order to use the Breit-Wigner distribution for the description of a possible resonance structure
in the excitation function, we translate the beam momentum intervals into intervals of the excess energy with respect
to the ${^4\mbox{He}} \eta$ production threshold.
The excitation function is presented in Fig.~\ref{fit_10_10_allcuts_p}.
It can be  well described with a quadratic fit resulting in the chi-squared value per degree of freedom of 0.98.
In the excitation function we observe no structure which could be interpreted
as a resonance originating from decay of the $\eta$-mesic ${^4\mbox{He}}$.

\section{Upper limit for the \texorpdfstring{$dd \to (^4\mbox{He}\eta)_{bound} \rightarrow {^3\mbox{He}} p \pi^-$}{dd -> (4He-eta)bound -> 3He p pi-} cross-section}

We assume, that a signal from the bound state in the excitation curve  determined as a function of the excess energy $Q$ 
with respect to the ${^4\mbox{He}}-\eta$ threshold, can be described by the Breit-Wigner shape:
\begin{equation} \label{bw_eq}
\sigma(Q,E_{BE},\Gamma,A)= \frac{A \cdot (\frac{\Gamma}{2})^2}{(Q-E_{BE})^2 +(\frac{\Gamma}{2})^2},
\end{equation}
where $E_{BE}$ is the binding energy,  $\Gamma$ is the width and $A$ is the amplitude.
The value of the Breit-Wigner function for the central energy ($Q=E_{BE}$)  corresponds to the maximum cross-section
for the decay of the  $\eta$-mesic ${^4\mbox{He}}$ into the ${^3\mbox{He}} p \pi^{-}$ channel.
As it was shown in the previous section the analysis of the excitation functions for the $dd \rightarrow {^3\mbox{He}} p \pi^{-}$ reaction does not indicate any statistically significant enhancement for energies below the $\eta$ production threshold.
Therefore, we can only determine an upper limit for the cross-section for formation of the ${^4\mbox{He}}-\eta$ bound state and its decay into the ${^3\mbox{He}} p \pi^{-}$ channel.
For this, we fitted the excitation function 
with quadratic function describing the background combined with the Breit-Wigner function.
In the fit we adjusted the quadratic background and the amplitude $A$ of the Breit-Wigner distribution.
The binding energy $E_{BE}$ and the width $\Gamma$ were fixed during the fit.
An example of the fit with $E_{BE}$=-10\,MeV and $\Gamma$=10\,MeV is shown in Fig.~\ref{fit_10_10_allcuts_p}.
The fit was performed for various values of the binding energy and the width representing different hypothesis of the bound state
properties.
Obtained results comprising the amplitude $A$ and its uncertainty $\sigma_{A}$ for the binding energies
of -10, -20, -30\,MeV and the widths of 10, 20, 30\,MeV, are gathered in the Table~\ref{TabBW}.
In each case, the value of the amplitude $A$ is consistent with zero within the uncertainty $\sigma_{A}$, 
which confirms the null hypothesis of non-observation of the signal.
{
\renewcommand{\arraystretch}{1.2}
\begin{table}[!ht]
 \begin{center}
 \begin{tabular}{|l|l|l|l|l|}
	\hline
	$E_{BE}$ [MeV] & $\Gamma$ [MeV] & A [nb]&  $\sigma_{A}$ [nb] & $\sigma_{max}$ [nb] \\

 \hline
-10 & 10 & 1.38 & 3.39 & 29.3\\
 \hline
-10 & 20 & 2.36 & 3.95 & 34.2\\
 \hline
-10 & 30 & 3.06 & 5.24 & 45.4\\
 \hline
-20 & 10 & -3.36 & 3.20 & 27.7\\
 \hline
-20 & 20 & -2.59 & 3.69 & 31.9\\
 \hline
-20 & 30 & -2.85 & 4.77 & 41.3\\
 \hline
-30 & 10 & -1.94 & 3.07 & 26.6\\
 \hline
-30 & 20 & -1.92 & 3.32 & 28.7\\
 \hline
-30 & 30 & -2.51 & 4.10 & 35.5\\
 \hline

 \end{tabular}
 \caption{Amplitude of Breit-Wigner function and its uncertainty obtained from fit with different fixed values 
of binding energy $E_{BE}$ and width $\Gamma$. The last column contains the maximum cross section for 
the $dd \to (^4\mbox{He}\eta)_{bound} \rightarrow {^3\mbox{He}} p \pi^-$ process calculated with Eq.~\ref{sig_max}.} \label{TabBW}
 \end{center}
\end{table}
}

\begin{figure}[h]
\begin{center}
  \scalebox{\scaleFactor}
  {
    \includegraphics{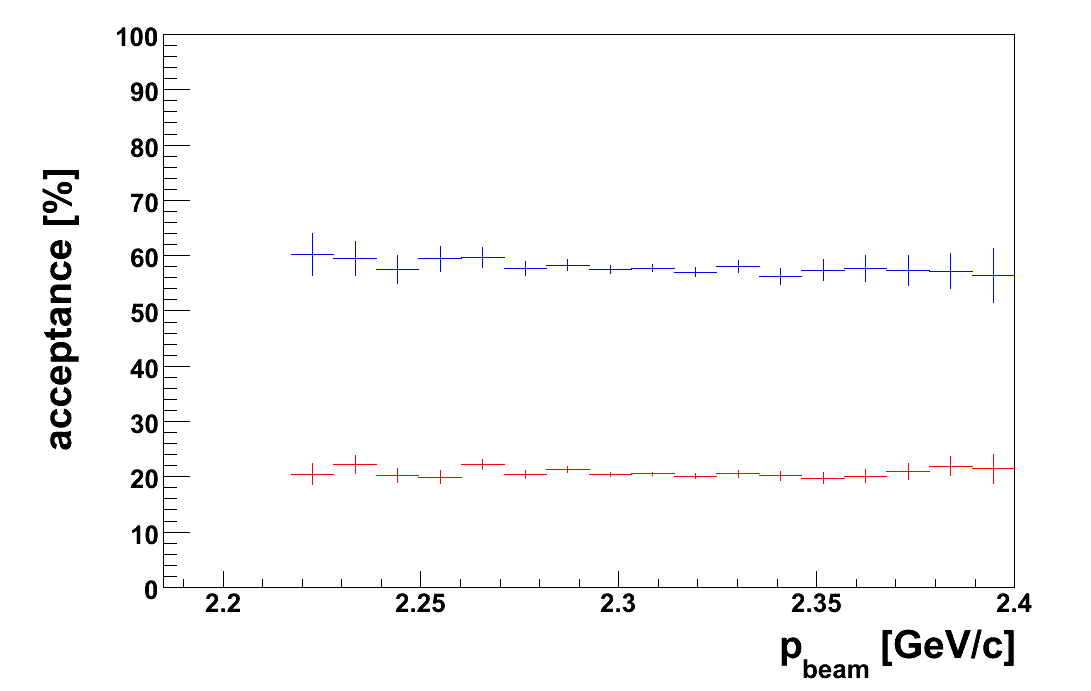}
  }
  \caption[Acceptance as a function of the beam momentum]{ \label{acceptance_mom_100000_p} Dependence on the beam momentum 
of acceptance for the $dd \to (^4\mbox{He}\eta)_{bound} \rightarrow {^3\mbox{He}} p \pi^-$ reaction.
  The geometrical acceptance is shown with blue points whereas the full efficiency including  detection and reconstruction 
efficiency is shown with red points.}
\end{center}
\end{figure}

In order to calculate the cross-sections corresponding to obtained values of the amplitude $A$ we still need to take
into account the detector acceptance for the process
$dd \to (^4\mbox{He}\eta)_{bound} \rightarrow {^3\mbox{He}} p \pi^-$.
We performed  simulations allowing to determine the acceptance  as a function of the beam momentum.
The study of the beam momentum dependence is of high importance, in order to exclude the possibility of creation artificial signals in the excitation curve due to variation of the acceptance.
Obtained geometrical acceptance is about 60\% and the full efficiency including all cuts applied in the analysis is about 19\% and it varies by only about 1\% along the whole beam momentum range as it is shown in Fig.~\ref{acceptance_mom_100000_p}.

In order to  calculate  an upper limit for the $dd \to (^4\mbox{He}\eta)_{bound} \rightarrow {^3\mbox{He}} p \pi^-$
cross-section, the $\sigma_{A}$ values obtained in the above described fit
have to be corrected for the efficiency $\varepsilon$ (equal to 19\%) and multiplied by the statistical factor $k$ equal to 1.64485 corresponding to the probability confidence level of 90\%:
\begin{equation}
\sigma_{max}=\frac{\sigma_{A} \cdot k}{\varepsilon}.
\label{sig_max}
\end{equation}

The obtained upper limits are given in the last column of Table~\ref{TabBW}.
One can notice that these limits depend mainly on the width of the bound state and not on the binding energy.
Therefore, we take average of $\sigma_{max}$ corresponding to different binding energies 
and we obtain the upper limits for the cross-sections of 28, 32 and 41\,nb for production and decays of the bound state 
with a width of 10, 20 and 30\,MeV, respectively.




Predictions of the cross-section for creation of the $\eta$-mesic nucleus is a very difficult task since
it involves computations of behaviour of the many-body system and, besides,
it depends on the strength of the relatively poorly known $\eta-N$ interaction.
A theoretical model for formation of ${^4\mbox{He}}-\eta$ bound state in the $d+d$ collisions is at present being developed
by Hirenzaki~{\em et al.}~\cite{hirenzaki}, however, quantitative predictions for the production cross-sections are not available yet.

In order to estimate if the present upper limit
for the $dd \to (^4\mbox{He}\eta)_{bound} \rightarrow {^3\mbox{He}} p \pi^-$
cross-section constitutes an essential constrain for existence of the  $\eta$-mesic ${^4\mbox{He}}$
we constructed a simple model of formation and decay of such state.
In our model we introduce a hypothesis
that the probability of the production of the $\eta$ meson in continuum and its absorption
on the helium nucleus is of the same order.
Based on this hypothesis we assume that the cross-section for the creation of ${^4\mbox{He}}-\eta$ the bound state
in the $d+d$ collisions in the maximum of the Breit-Wigner distribution is about 15\,nb as measured
for the $dd \to {^4\mbox{He}} \eta$ reaction in the vicinity of the kinematic threshold (see Fig.~\ref{4Heeta}).
Further on, we assume, that the probability of decay of the ${^4\mbox{He}}-\eta$ bound state
into the ${^3\mbox{He}} p \pi^-$ channel
is equal to $\frac{1}{4} \cdot \frac{1}{2}=\frac{1}{8}$.
The factor $\frac{1}{4}$ takes into account the fact that there are four possible $\eta$ absorption channels:
\begin{itemize}
\item $\eta p \rightarrow p \pi^0$   (corresponding to  $dd \rightarrow {^3\mbox{H}} p \pi^0$)
\item $\eta p \rightarrow n \pi^+$   (corresponding to  $dd \rightarrow {^3\mbox{H}} n \pi^+$)
\item $\eta n \rightarrow n \pi^0$   (corresponding to  $dd \rightarrow {^3\mbox{He}} n \pi^0$)
\item $\eta n \rightarrow p \pi^-$   (corresponding to  $dd \rightarrow {^3\mbox{He}} p \pi^-$)
\end{itemize}
In turns, the factor $\frac{1}{2}$ represents our guess of the probability that the three observer nucleons ($ppn$),
in the process of the $\eta$ absorption on the neutron in ${^4\mbox{He}}$, form  ${^3\mbox{He}}$ in the final state.
Estimation of this effect needs, in our understanding of the problem, projection of the ${^4\mbox{He}}$ wave function
on the wave function of the ${^3\mbox{He}}-n$ pair and inclusion of the pion and proton rescattering on the observer nucleons.
Since, this kind of estimations isn't at present in our reach, we justify our guess using an analogy between the decay
of the ${^4\mbox{He}}-\eta$ nuclei and the ${^4_{\Lambda}\mbox{He}} $-hypernuclei. For the later case it was observed
namely that in the $\pi^-$ decay channel the decay mode ${^4_{\Lambda}\mbox{He}} \rightarrow \pi^- p {^3\mbox{He}}$
is dominant ~\cite{Fet}.

The above considerations imply that the cross-section for the\\
$dd \to (^4\mbox{He}\eta)_{bound} \rightarrow ^3\mbox{He} p \pi^-$ reaction should be in the order of 2\,nb.
This number is by an order of magnitude smaller than the upper limits for this process determined in the present
experiment.
Therefore, the present experimental result does not exclude the existence of $\eta$-mesic  ${^4\mbox{He}}$.

\section{Angular and momentum distributions for background}

The background observed in the experiment limits the sensitivity of the search for the $\eta$-mesic ${^4\mbox{He}}$.
Therefore, understanding of reactions contributing to the background can be important
for the conducted search.
We performed a study of the background on the basis of the momentum and angular distributions of the outgoing protons,
pions and ${^3\mbox{He}}$ ions in the measured process $dd \rightarrow {^3\mbox{He}}p\pi^{-}$.
For this, we compared the experimental distributions with results of simulations of the following three processes:
\begin{itemize}
\item direct production corresponding to the uniform distribution over the phase space available for the reaction products,
\item formation and decay of the ${^4\mbox{He}}-\eta$ bound state,
\item reaction $dd\rightarrow {^3\mbox{He}} N^{\star} \rightarrow{^3\mbox{He}}p\pi^{-}$
proceeding with excitation of an intermediate $N^{\star}$ resonance which subsequently decays in the $p \pi^{-}$ pair.
\end{itemize}
The first two processes were already discussed.
In the following, we present some details of simulations of the third process.
In these simulations, the mass of the  intermediate $N^{\star}$ resonance was selected assuming the Breit-Wigner distribution
with the resonance energy $E_{0}=1535$\,MeV
and the width $\Gamma=150$\,MeV in accordance with PDG~\cite{pdg2010}.
For a given available energy in the CM frame $\sqrt{s}$ calculated from the beam momentum,
the selection of the $N^{\star}$ mass is limited by two conditions:
\begin{equation}\label{cond_one_eq}
m_{N^{\star}} +m_{^3\mbox{He}} \le \sqrt{s},
\end{equation}
\begin{equation}\label{cond_two_eq}
m_{N^{\star}}\ge  m_{\pi^{-}}+m_p .
\end{equation}
Eq.~\ref{cond_one_eq} with the equality condition can be considered as a limiting case
in which the whole available energy $\sqrt{s}$ goes to the creation of $N^{\star}$
and the ${^3\mbox{He}}-N^{\star}$ pair rests in the CM frame.
In turns, Eq.~\ref{cond_two_eq} expresses the fact that the mass of the resonance must be large enough
for its decay into the $p\pi^{-}$ pair.
After the selection of the resonance mass  the sum of kinetic energies 
of  $N^{\star}$ and of ${^3\mbox{He}}$ in the CM system can be determined in a following way:
\begin{equation}Ekin=\sqrt{s} -m_{N^{\star}}-m_{^3\mbox{He}} .
\end{equation}
The direction of the ${^3\mbox{He}}$ momentum vector is selected assuming
its isotropic distribution in the CM frame.
Next, the simulation of the $N^{\star}$ decay
into a $p-\pi^{-}$ pair is performed under assumption of its isotropic
angular distribution in the $N^{\star}$ rest frame.

One should stress, that for study of the background reactions we use momentum and angular distributions
of the outgoing protons, pions and ${^3\mbox{He}}$ ions constructed before application of cuts
on the  ${^3\mbox{He}}$ momenta and on the $p$ and $\pi^-$ kinetic energy and opening angle.
Application of these cuts to predictions of various simulation models enforces similarities of the investigated distributions
and makes the studies of background inconclusive.

\begin{figure}[!ht]
\begin{center}
      \scalebox{\scaleFactor}
      {
         \includegraphics{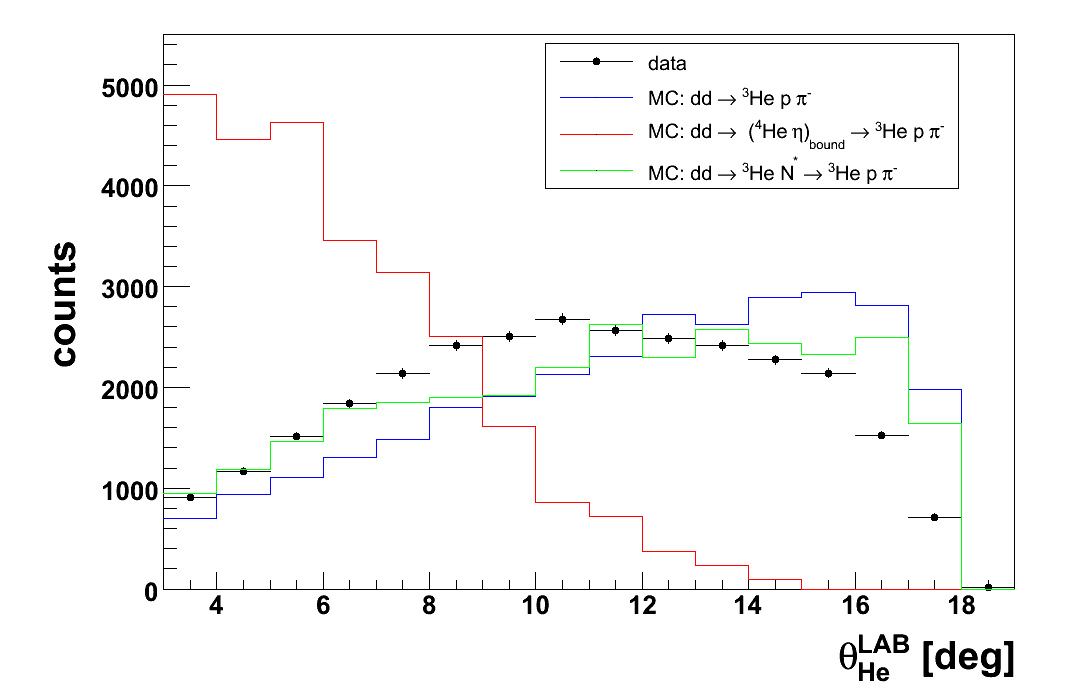}
      }
\caption[Distribution of $\theta$ scattering angle of $^3\mbox{He}$ in the LAB frame.]{\label{theta_lab_he3_compar_v2_p} Distribution of $\theta$ scattering angle of $^3\mbox{He}$ in the LAB frame obtained from experiment and from the MC simulations. }
\end{center}
\end{figure}

\begin{figure}[!ht]
\begin{center}
      \scalebox{\scaleFactor}
      {
         \includegraphics{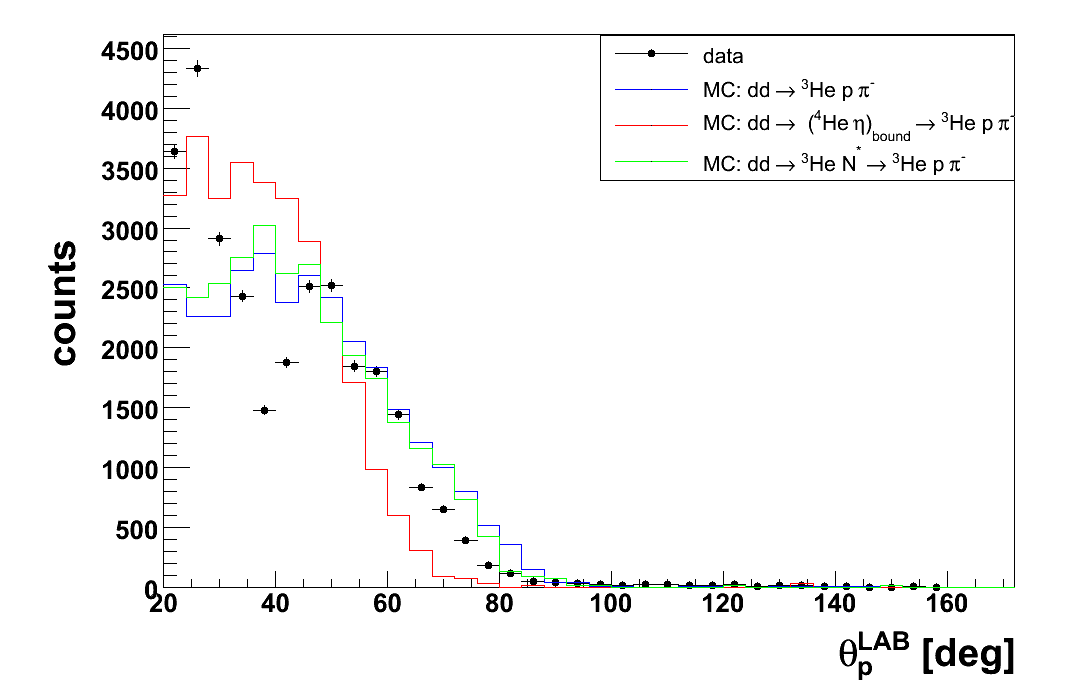}
      }
\caption[Distribution of $\theta$ scattering angle of protons in the LAB frame.]{\label{theta_lab_p_compar_v2_p} Distribution of $\theta$ scattering angle of protons in the LAB frame obtained from experiment and from the MC simulations.}
\end{center}
\end{figure}

\begin{figure}[!ht]
\begin{center}
      \scalebox{\scaleFactor}
      {
         \includegraphics{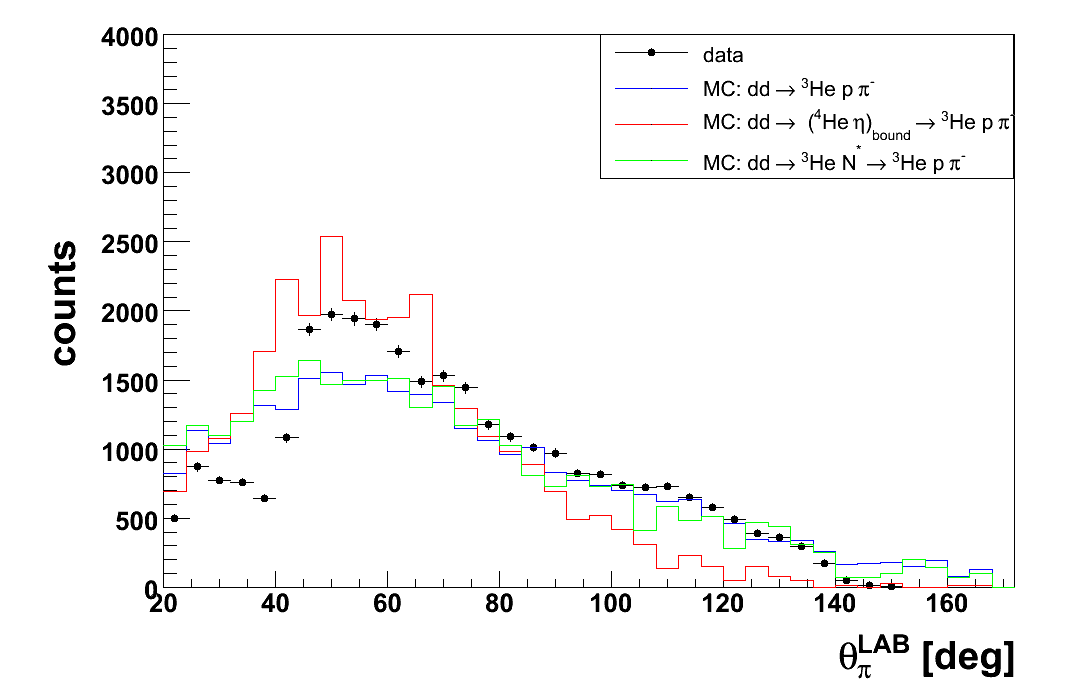}
      }
\caption[Distribution of $\theta$ scattering angle of $\pi^{-}$ in the LAB frame.]{\label{theta_lab_pi_compar_v2_p} Distribution of $\theta$ scattering angle of $\pi^{-}$ in the LAB frame obtained from experiment and from the MC simulations.}
\end{center}
\end{figure}

\begin{figure}[!ht]
\begin{center}
      \scalebox{\scaleFactor}
      {
         \includegraphics{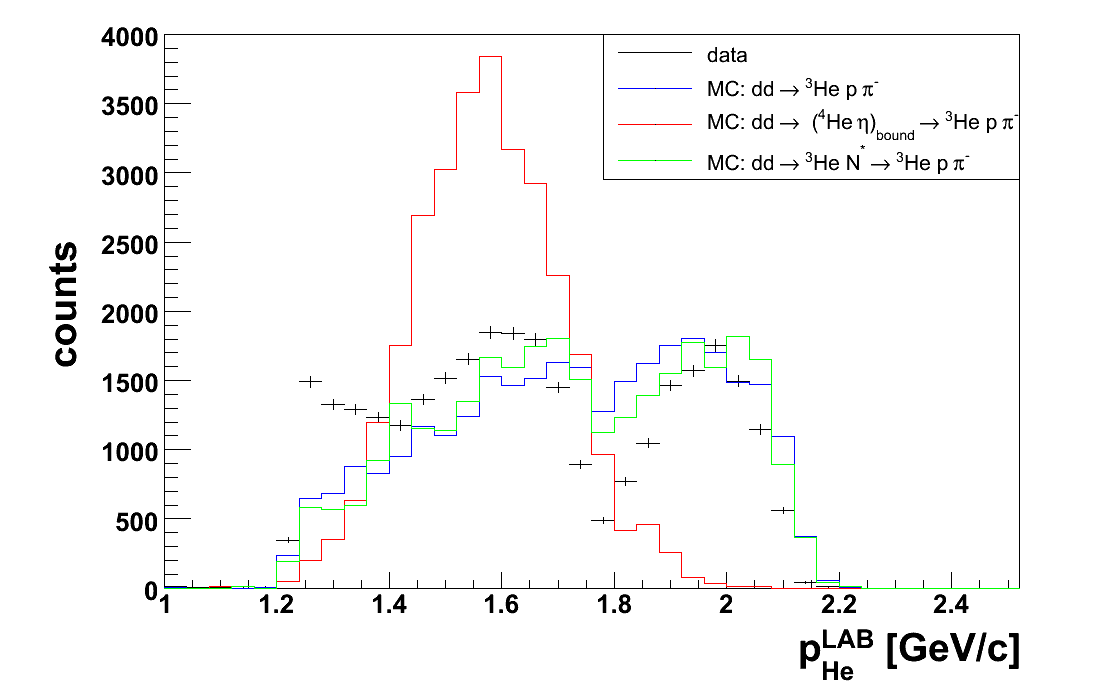}
      }
\caption[Distribution of momentum of $^3\mbox{He}$ in the LAB frame.]{\label{mom_lab_he3_compar_v2_p} Distribution of momentum of $^3\mbox{He}$ in the LAB frame obtained from experiment and from the MC simulations.}
\end{center}
\end{figure}

\begin{figure}[!ht]
\begin{center}
      \scalebox{\scaleFactor}
      {
         \includegraphics{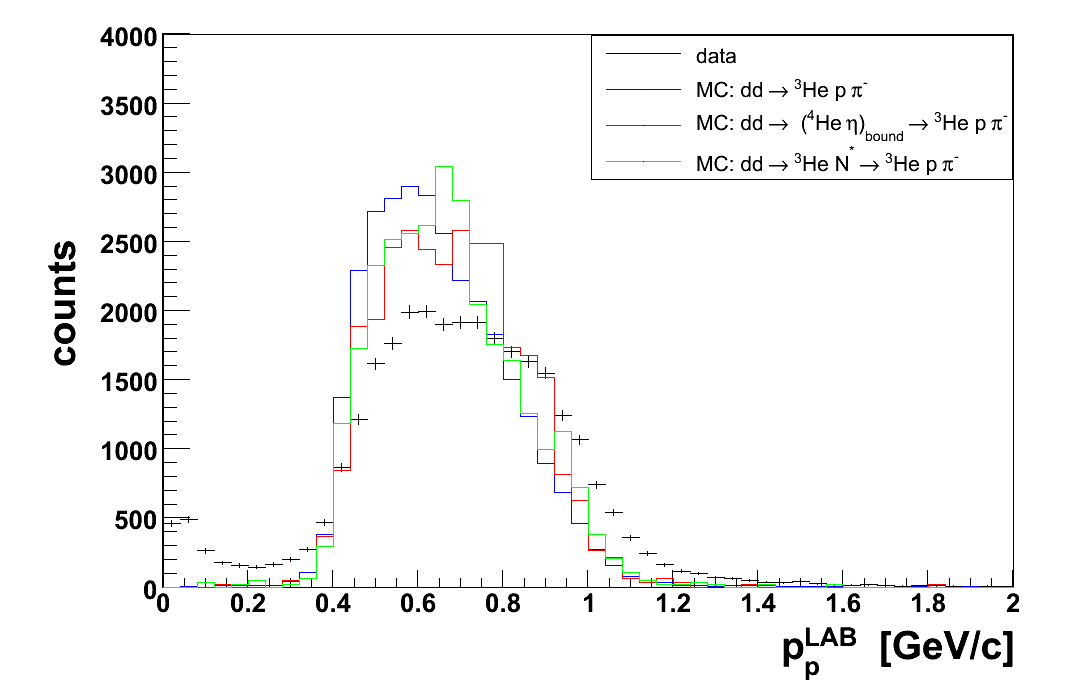}
      }
\caption[Distribution of momentum of $p$ in the LAB frame.]{\label{mom_lab_p_compar_v2_p} Distribution of momentum of $p$ in the LAB frame obtained from experiment and from the MC simulations.}
\end{center}
\end{figure}

\begin{figure}[!ht]
\begin{center}
      \scalebox{\scaleFactor}
      {
         \includegraphics{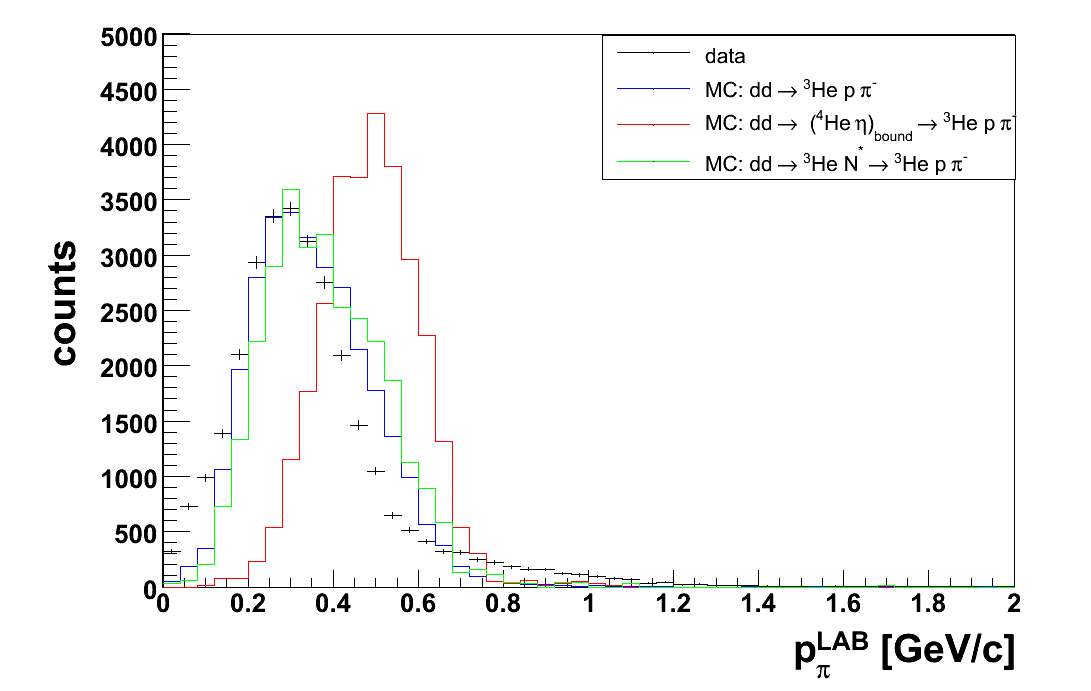}
      }
\caption[Distribution of momentum of $\pi^{-}$ in the LAB frame.]{\label{mom_lab_pi_compar_v2_p} Distribution of momentum of $\pi^{-}$ in the LAB frame obtained from experiment and from the MC simulations.}
\end{center}
\end{figure}

The experimental and simulated angular distributions in the LAB frame of ${^3\mbox{He}}$, protons and pions
are presented in Figs.~\ref{theta_lab_he3_compar_v2_p}, \ref{theta_lab_p_compar_v2_p} and \ref{theta_lab_pi_compar_v2_p}, respectively.
In turns the momentum distributions in the LAB system for ${^3\mbox{He}}$, protons and pions are shown
in Figs.~\ref{mom_lab_he3_compar_v2_p}, \ref{mom_lab_p_compar_v2_p} and \ref{mom_lab_pi_compar_v2_p}, respectively.
Study of these distributions led us to the following conclusions:\\
\begin{itemize}
\item Simulations of the direct production and of the excitation of the intermediate N$^{\star}$ resonance
provide very similar results. One can understand it when taking into account that the width of $N^{\star}$
is larger than the range of excess energy available for the $p-\pi^-$ pair in the CM system.
\item Results of simulations of the bound state differ essentially from ones for the direct
production in the case of the ${^3\mbox{He}}$ angular and momentum distribution and the pion momentum distribution.
\item The experimental spectra are reasonably well described by the direct production. A very good agreement
is observed in the case of the pion momentum distributions.
In turns there is a problem with description of the experimental angular distributions of protons and pions
for the LAB angles below 40$^{\circ}$. In the experimental spectra, a rapid drop of counts for this
angular region is visible.
We presume, that this could result from an inefficiency of the forward end-cap of the Plastic Scintillator Barrel,
which covers the angles below  40$^{\circ}$.
\end{itemize}

\chapter{Conclusions and outlook}
\label{ch:conclusions}

This thesis was devoted to a  search for the ${^4\mbox{He}}-\eta$ bound state via  exclusive measurement
of the excitation function for the $dd \rightarrow {^3\mbox{He}} p \pi^-$ reaction.
Theoretical and experimental background of the search for the $\eta$-mesic nuclei,
the applied method of measurement, the analysis of data and obtained results have been presented.

The measurement was performed with the internal deuteron beam of the COSY accelerator scattered on a deuteron target of 
the pellet type and with the WASA-at-COSY detection system applied for registration of the reaction products.
During the experimental run the momentum of the deuteron beam  was varied continuously within each acceleration cycle
from  2.185~GeV/c to 2.400~GeV/c, crossing the kinematical threshold for the $\eta$ production 
in the $dd \rightarrow {^4\mbox{He}}\,\eta$ reaction at 2.336~GeV/c.
This range of the beam momenta corresponds to an interval of the excess energy in the $^4\mbox{He}-\eta$
system from -51.4~MeV to 22~MeV.

In the off-line data analysis, events corresponding to decays of the $\eta$-mesic ${^4\mbox{He}}$ were selected using cuts
on the ${^3\mbox{He}}$ momentum, $p$ and $\pi^-$ kinetic energies and the relative $p-\pi^-$ angle
in the center of mass system.
The range of the applied cuts was inferred from simulations
based on assumption that the decay of the $^4\mbox{He}-\eta$ bound state proceeds via absorption
of the $\eta$ meson on one of neutrons in the ${^4\mbox{He}}$ nucleus leading to creation of the $p-\pi^-$ pair,
and that the outgoing ${^3\mbox{He}}$ plays a role
of spectator and it moves with the Fermi momentum in the CM frame.
The integrated luminosity in the experiment was determined using the $dd \rightarrow {^3\mbox{He}} n$ reaction 
and equals 117.9 $\pm$13.6\,$nb^{-1}$.
The relative normalization of points of the $dd \rightarrow {^3\mbox{He}} p \pi^-$ excitation
function was based on the quasi-elastic proton-proton scattering.

The obtained excitation function is well described with a second order polynomial an no resonance-like structure, 
which could originate from decay of the ${^4\mbox{He}}-\eta$ bound state, is visible.
A fit of the excitation function using the Breit-Wigner curve combined with a quadratic term
was performed and an upper limit for the $dd \to (^4\mbox{He}\eta)_{bound} \rightarrow {^3\mbox{He}} p \pi^-$ cross-section
was determined as equal to 28, 32 and 41\,nb for the bound state width of 10, 20 and 30\,MeV, respectively.

Sensitivity of the present search for $\eta$-mesic ${^4\mbox{He}}$ is limited by experimental background
corresponding to $dd \rightarrow {^3\mbox{He}} p \pi^-$ reaction proceeding without formation of the intermediate
${^4\mbox{He}}-\eta$ bound state.
The measured angular and momentum distributions of the reaction products are close
to those simulated under the assumption of uniform phase-space distribution.

We constructed a simple model of formation and decay of  the ${^4\mbox{He}}-\eta$ bound state
based on assumption that the formation cross section is equal to 15\,nb as measured for the $dd \to {^4\mbox{He}} \eta$ reaction close to kinematical threshold.
Estimation results in the $dd \to (^4\mbox{He}\eta)_{bound} \rightarrow {^3\mbox{He}} p \pi^-$ cross section of 2\,nb
being by an order of magnitude smaller than the determined experimental upper limits for this cross section.
Therefore, one can conclude that the present measurement does not exclude existence of $\eta$-mesic
${^4\mbox{He}}$.
Above estimation might be simplistic and more realistic theoretical calculations are needed.
Indeed, the  group of Hirenzaki are developing a theoretical model for formation of ${^4\mbox{He}}-\eta$ bound state~\cite{hirenzaki} and quantitative predictions for the $dd \to (^4\mbox{He}\eta)_{bound} \rightarrow {^3\mbox{He}} p \pi^-$ are expected to be available soon.

Also experimental search for the $\eta$-mesic ${^4\mbox{He}}$ is being continued with WASA-at-COSY~\cite{prop1861,prop1862}.
In November 2010 a new two-weeks measurement
of the $dd \rightarrow {^3\mbox{He}} p \pi^-$ channel and also of the $dd \rightarrow {^3\mbox{He}} n \pi^{0}$ reaction was
performed. Analysis of the collected data is in progress.
It is planned to extend the present study to the  ${^3\mbox{He}}-\eta$ system.
The search for $\eta$-mesic nuclei is part of the research program at J-PARC~\cite{fujioka} and $\eta$'-mesic nuclei have being proposed recently to be included in the research program of GSI ~\cite{gsi-loi}.

\appendix
\chapter{Pseudoscalar mesons}
In the quark model, mesons are  built of quark-antiquark pairs ($q\bar{q'}$)\footnote{We used the $'$ notation to underline the fact that the flavour of quark $q$ and antiquark $\bar{q'}$ could be different.
In the further part of this appendix we will omit $'$ symbol.}.
Quarks and antiquarks are fermions with spin $s=\frac{1}{2}$.
Consequently, the total intrinsic spin $S$ of the $q\bar{q}$ state can be either 0  or 1.
By convention, quarks have positive intristic parity $P$  and antiquarks have negative intristic parity.
If $L$ is the orbital angular momentum of the $ q\bar{q}$ system, then the parity $P$ of the state  equals
$(-1)* (-1)^{L} =(-1)^{L+1} $ where the first term $(-1)$ is related to opposite $ q\bar{q}$ intristic  parities
and $(-1)^{L}$ term arises from the space invertion transformation.
The total spin $J$ of meson is the sum of the intrinsic spin $S$ and the angular momentum $L$.
Each quark is assigned additive baryon number $B=\frac{1}{3}$
and antiquarks are assigned baryon number with opposite sign $B=-\frac{1}{3}$.
This implies that mesons have  baryon number equal to 0.
By convention, the sign of the flavour quantum number assigned to quark is the same as its charge $Q$.
It implies that the flavour carried by charged mesons has the same sign as its charge e.g.
the strangness of $K^{+}$ is $S=1$ and the third component of the isospin of $\pi^{-}$ is $I_{3}$=$-1$.

Lets consider only the set of the three lightest quarks ($u,d,s$) and antiquarks ($\bar{u}$,$\bar{d}$,$\bar{s}$).
Combining these fundamental triplets we can construct nine possible combinations of $q\bar{q}$ states
which are decomposed into SU(3) flavour-octet states and SU(3) flavour-singlet state.
This decomposition is symbolically denoted as:

\begin{equation}
3\otimes\bar{3}=8\oplus1
\end{equation}


If we consider only the ground state (relative angular momentum $L=0$) then the total meson spin $J$ is equal
to the intristic spin $S$ and the parity $P$ equals -1. Based on the $J^{P}$ values we can distinguish two nonets
of SU(3) mesons: pseudoscalar mesons with $J^{P}= O^{-}$ and vector mesons with $J^{P}=1^{-}$.
The quark composition of the nonet of the light pseudoscalar mesons is presented in the Table~\ref{pseudoscal_t}.

{
\renewcommand{\arraystretch}{2}
\begin{table}[hbpt]
 \begin{center}
 \begin{tabular}{|l|c|}
        \hline
        $\pi^{-}$ & $d\bar{u}$ \\
        \hline
        $\pi^{+}$ & $u\bar{d}$ \\
        \hline
        $K^{-}$ & $s\bar{u}$ \\
        \hline
        $K^{+}$ & $u\bar{s}$ \\
        \hline
	$K^{0}$ & $d\bar{s}$ \\
        \hline
        $\bar{K}^{0}$ & $s\bar{d}$ \\
        \hline
        $\pi^{0}$ & $\frac{1}{\sqrt{2}} (u\bar{u} - d\bar{d})$  \\
        \hline
        $\eta_{1}$ & $\frac{1}{\sqrt{3}} (u\bar{u} + d\bar{d} + s\bar{s})$ \\
        \hline
        $\eta_{8}$ & $\frac{1}{\sqrt{6}} (u\bar{u} + d\bar{d} - 2 s\bar{s})$ \\
        \hline
 \end{tabular}
 \caption{Quark composition of the nonet of the light pseudoscalar mesons.}\label{pseudoscal_t}
 \end{center}
\end{table}
}

\label{app:pseudoscal}
\chapter{Estimate of the luminosity from the target density and the beam intensity}
Presented in this appendix  calculations of the luminosity are based on several assumptions
concerning the pellet target parameters and the beam properties,
which can vary significantly during the acceleration cycle and also from cycle to cycle.
Hence, the result should be treated as a rough estimate with an order-of magnitude precision.

In experiments with a fixed target like the WASA-at-COSY, luminosity is equal to the product of
areal target density  $\rho_t$ and of number of incident beam particles per unit time $j_b$:
\begin{equation}
L=\rho_t \cdot j_b.
\end{equation}
The areal density can be expressed in the following form:
\begin{equation}
\rho_t= \frac{N_t}{A},
\end{equation}
where $N_t$ is the number of atoms in one  pellet and $A$ is an area of lateral beam extension
at the target position.
The value $N_t$ can be calculated from the following equation:
\begin{equation}
N_t= \frac{4}{3} \cdot \pi \cdot {R_{t}}^3 \cdot \rho_{D} \cdot \frac{n}{M} \cdot N_{A},
\end{equation}
where $R_t$ is average radius of pellets, $\rho_{D}$ is density of pellet material,
$n$ is number of atoms per one molecule,  $M$ is a moll mass and $N_{A}$ is the Avogadro constant.
For the WASA-at-COSY pellets produced of deuterium ($\mbox{D}_2$ molecules),
the following values for the areal density calculations
are assumed:\\
$R_t=15$\, mm,\\
$\rho_{D}=0.162$\,$\frac{\text{g}}{{\text{cm}}^3}$,\\
$n=2$,\\
$M=4$\,$\frac{\text{g}}{\text{mol}}$.

The number of incident beam particles per unit time $j_b$ can be estimated from the relation:
\begin{equation}
j_b= f \cdot N_{b} ,
\end{equation}
where $f$ is the revolution frequency of the accelerator and $N_{b}$ is the number of beam particles stored
in the accelerator ring.
The values of $f$ and $N_{b}$ were monitored during the experiment, and are about  1.2 MHz
and  3.7$ \cdot 10^9$ particles, respectively.

Taking all the aforementioned relations and numbers into account, we obtain a value
for the luminosity of about  $10^{31}$ cm$^{-2}$s$^{-1}$.
However, in the calculation we have assumed an ideal 100\%  overlap of beam and pellet.
The typical overlap factor in the WASA-at-COSY for a pellet rate of 8~kHz is about 0.5.
In addition, one needs to take into account the accelerator beam losses resulting in another factor of 0.5.
Hence, the final estimate of the luminosity is $L \sim 2.7 \cdot 10^{30}$ cm$^{-2}$s$^{-1}$.
Taking into account the effective time of measurement $T=62760$ s,
we obtained a value for the integrated luminosity $\mathcal{L}$ of about 170\,nb$^{-1}$.

\label{app:lum_estimate}
\chapter*{Acknowledgements}
I had the pleasure of meeting and working with many amazing people without whom I  would have never  accomplished this dissertation.
It is impossible to list all people  who contributed to this work. I would like to express my gratitude to all of them.

I wish to thank my supervisor Prof. Jerzy Smyrski for introducing me to this interesting topic  
and for all his help. Without his remarks and many corrections this dissertation would have been much shorter and of much lower quality.

This thesis would not have been possible without the help of Prof. Pawel Moskal who has been encouraging me through all the years of my struggle with data analysis. I am very grateful for all fruitful discussions, suggestions  and especially for the support in the hard period of my student's life.

I would like to acknowledge Prof. Jim Ritman for allowing me to work in the Research Centre Juelich.

I am grateful to Prof. Boguslaw Kamys for allowing me to prepare this dissertation in the Faculty of Physics, Astronomy and Applied Computer Science of the Jagiellonian University.

I would like to thank my colleagues from Krak\'{o}w: Tomasz Bednarski, Dr.~Eryk Czerwi\'{n}ski, Damian Gil, Dr.~Piotr Hawranek, Malgorzata Hodana, Ayeh Jowzaee,Grzegorz Korcyl, Beata Michalska-Trebacz, Szymon Nied\'{z}wiecki, Iryna Ozerianska, Andrzej Pyszniak, Dr.~Witold Przygoda, Michal Silarski, Magdalena Skurzok, Radoslaw Trebacz,  Dr.~Aleksandra Wro\'{n}ska, Marcin Zieli\'{n}ski.
for the nice and scientific atmosphere over the years of work.
  
I am thankful to Dr.~Jacek Otwinowski who has been always ready to give me an explanation concerning Hydra, C++ or to play a basketball game.

I wish to thank my roommate Adrian Dybczak for many discussions about physics, life, and for his cooperation in our projects.

Special thanks to Jaroslaw Zdebik for discussions, his friendship and his support in many situations. Also, I am grateful for the corrections to this dissertation.

I am very grateful to inz.~Andrzej Misiak who always found time to explain me the secrets of electronics.

I would like to thank Tsitohaina Hary Randriamalala for many discussions in IKP.

I am indebted to Dr.~Pawel Klaja and Dr.~Joanna Klaja who were always hospital and helpful every time I went back to Juelich.
 
I would like to express my gratitude to my colleagues from WASA collaboration.

I wish  to thank  Dr.~Annette Pricking and  Dr.~Volker Hejny for providing parametrizations which saved me a lot of work.

Especially, I would like to thank Patrik Adlarson for his friendship and for our endless discussions about physics and philosophy during the long night shifts in Juelich and in many different places in Europe. 

Finally, I would like to thank my family for their support and the belief that one day I will finish the dissertation.

\addcontentsline{toc}{chapter}{Acknowledgements}

\listoffigures
\addcontentsline{toc}{chapter}{List of Figures}
\listoftables
\addcontentsline{toc}{chapter}{List of Tables}
\printnomenclature
\addcontentsline{toc}{chapter}{List of Abbreviations}
\bibliographystyle{plain}
\bibliography{thesis}

\addcontentsline{toc}{chapter}{Bibliography}

\end{document}